\documentclass[aip,apl,twocolumn,reprint]{revtex4-1}
\usepackage{graphicx}
\usepackage{dcolumn}
\usepackage{bm}
\usepackage{color}
\usepackage{tabularx}
\usepackage{array}
\usepackage{amsmath}
\usepackage{stmaryrd}  

\begin{document}

\title{Advances in coherent coupling between magnons and acoustic phonons}


\author{Yi Li}
\affiliation{Materials Science Division, Argonne National Laboratory, Argonne, IL 60439, USA}

\author{Chenbo Zhao}
\affiliation{Materials Science Division, Argonne National Laboratory, Argonne, IL 60439, USA}

\author{Wei Zhang}
\affiliation{Department of Physics, Oakland University, Rochester, MI 48309, USA}

\author{Axel Hoffmann}
\affiliation{Department of Materials Science and Engineering, University of Illinois at Urbana-Champaign Urbana, IL 61801}

\author{Valentyn Novosad}
\email{novosad@anl.gov}
\affiliation{Materials Science Division, Argonne National Laboratory, Argonne, IL 60439, USA}

\date{\today}

\begin{abstract}

The interaction between magnetic and acoustic excitations have recently inspired many interdisciplinary studies ranging from fundamental physics to circuit implementation. Specifically, the exploration of their coherent interconversion enabled via the magnetoelastic coupling opens a new playground combining straintronics and spintronics, and provides a unique platform for building up on-chip coherent information processing networks with miniaturized magnonic and acoustic devices. In this Perspective, we will focus on the recent progress of magnon-phonon coupled dynamic systems, including materials, circuits, imaging and new physics. In particular, we highlight the unique features such as nonreciprocal acoustic wave propagation and strong coupling between magnons and phonons in magnetic thin-film systems, which provides a unique platform for their coherent manipulation and transduction. We will also review the frontier of surface acoustic wave resonators in coherent quantum transduction and discuss how the novel acoustic circuit design can be applied in microwave spintronics.

\end{abstract}

\maketitle

\section{Introduction}

The interconversion between distinct excitations is a fundamental physical process when two media hosting different excitations are interacting with each other. These interconversions allow us to study material properties, explore physics and engineer excitation transduction. In particular, their coherent interactions, usually involving the excitation of the hybrid dynamics, have found application potential in advanced information processing, communications and sensing \cite{KurizkiPNAS2015,ClerkNphys2020}, and created many new interdisciplinary research directions with different combinations, such as cavity spintronics \cite{HuPhysCanada2016}, cavity optomechanics \cite{AspelmeyerRMP2014}, cavity optomagnonics \cite{KusminskiyarXiv2019}, cavity magnomechanics \cite{ZhangScienceAdv2016} and cavity piezomechanics \cite{HanNComm2020}.

Among various excitations, magnetic excitations, or magnons, show unique features because of their frequency tunability and the ease to reach strong coupling with microwave photons \cite{HueblPRL2013,TabuchiPRL2014,ZhangPRL2014,GoryachevPRApplied2014,BhoiJAP2014,BaiPRL2015,LiPRL2019_magnon,HouPRL2019}. Magnons describe the collective dynamics of exchange coupled spins in magnetically ordered materials. The dynamics is described by the Landau-Lifshitz-Gilbert equation, which yields a field dependence of precession frequency in the gigahertz range, similar to the Zeeman splitting of individual spins. \textcolor{black}{Due to the high spin density in magnetic materials, magnons can achieve strong magnetic dipolar interaction with photons, which can be orders of magnitude larger than what spin ensembles can achieve}. Furthermore, magnetic excitations exist in solid-state materials, are compatible with modern fabrications, and allow electric manipulation and detection, making them promising for device and circuit integrations. Acoustic excitations are also fundamental solid-state excitations which have found wide applications such as nanoelectromechanical systems (NEMS) and surface acoustic wave (SAW) devices for sensing and communication. Recently mechanical excitations have been exploited in quantum optics and quantum information by coupling with optical and microwave electromagnetic radiations. Thus the exploration of diverse interactions with phonons will extend their functionality in quantum applications.

\textcolor{black}{The interaction between magnons and mechanical excitations has a long history of research, particularly in the excitation of magneto-acoustic modes \cite{KittelPhysRev1958,SpencerPRL1958,BommelPRL1959,SchlomannJAP1960,PomerantzPRL1961,MatthewsPRL1962,SchlomannJAP1964,SeaveyProcIEEE1965,ComstockProcIEEE1965,KobayashiPRB1973,BelyaevaSov1992}. Recently, a rapid growth has been witnessed on this topic in the micro- and nano-systems owing to the advances in thin-film materials growth and structure fabrication, which provide a remarkable capability of controlling dimensions, orientations and wavelengths associated to the dynamics for engineering the magnetoacoustic effects. For example, spin waves in magnetic films can be directly stimulated by SAWs that are electrically excited by integrating lithographically defined interdigital transducers on piezoelectric substrates \cite{WeilerPRL2011,DreherPRB2012,ChangPRApplied2018,LisenkovPRB2019,RuckriegelPRB2014,LiXJAP2017,PueblaJPD2020}. In addition, new techniques for detecting responses of magnetic excitations, such as Brillouin light scattering and spin Hall effects, have provided unique insights in analyzing coupled magnetoelastic excitations and band structures, with examples of Bose-Einstein condensations \cite{HickPRB2012,BozhkoPRL2017,SchneiderNNano2020} and angular momentum of phonons \cite{SasakiPRB2017,HolandaNPhys2018}. Those studies have stimulated interests in exploring novel physics and finding applications in microwave signal processing \cite{BozhkoLTPhys2020}.}

\textcolor{black}{In this perspective, we focus on the recent advances in coherent interactions between magnons and acoustic phonons \cite{KikkawaPRL2016,BozhkoPRL2017,HayashiPRL2018,BerkNComm2019,AnPRB2020,YahiroPRB2020,GodejohannPRB2020}. In particular, the capability of exciting and characterizing short-wavelength magnons and phonons have provided a new playground for studying and engineering their interactions, included the achievement of mode anticrossing \cite{BerkNComm2019,AnPRB2020,GodejohannPRB2020}.} Furthermore, magnons \cite{TabuchiScience2015,LachanceScienceAdvan2017,LachanceQuirionScience2020} and phonons \cite{SatzingerNature2018,WhiteleyNphys2019,BienfaitScience2019} have been recently demonstrated as an active components in quantum information, with the implementation of surface acoustic wave (SAW) phononic circuits. In the perspective, we will first review on the fundamentals of magnetoelastic coupling, including material properties, excitation schematics and detection, for enabling magnon-phonon interconversion. Then, we will discuss the recent advances in coherent magnon-phonon interactions, with examples of angular momentum transfer, nonreciprocal acoustic phonon propagation and strong magnon-phonon coupling. Lastly, we will explore a few examples of surface acoustic wave resonator designs for realizing coherent interaction between acoustic phonons and quantum systems, and provide a future outlook of magnon-phonon coupled system for on-chip applications. With new physics and engineering of magnon-phonon coupling and new ideas in acoustic circuit design, we anticipate that the coherent magnon-phonon interaction will be a new avenue for empowering quantum information processing with magnons and acoustic phonons.

\section{Magnetoelastic coupling}

\begin{table*}[htb]
\centering
\begin{tabular}{>{\centering\arraybackslash}m{0.8in} >{\centering\arraybackslash}m{0.8in} >{\centering\arraybackslash}m{0.8in} >{\centering\arraybackslash}m{0.8in} >{\centering\arraybackslash}m{0.8in} >{\centering\arraybackslash}m{0.8in} >{\centering\arraybackslash}m{1.4in} }
\hline
\hline
Material & \multicolumn{3}{c}{Magnetoelastic coupling} &  Magnetization & Method & Reference  \\
& $b_1$ (MPa) & $b_1/M_s$ (T) & $\lambda_{100}$ ($\times 10^{-6}$)  & $\mu_0M_s$(T) & & \\
\hline
YIG &  0.74 & & & 0.175 & Bulk & Matthews \textit{et al.}\cite{MatthewsPRL1962}  \\
 &  0.59 & & &  & Bulk & Callen \textit{et al.}\cite{CallenPR1963}  \\
 &  0.35 & & &  & Bulk & Comstock \textit{et al.}\cite{ComstockProcIEEE1965}  \\
 &  0.35 & & $-1.4$ &  & FMR & Smith \textit{et al.}\cite{SmithJAP1963}  \\
\hline
Ni &   & 23 & &  0.59 & SAW-FMR & Dreher \textit{et al.} \cite{DreherPRB2012}  \\
& $^*9.5$  &  & $^*-38$ &   & cantilever & Klokholm \textit{et al.}\cite{KlokholmJAP1982}  \\
\hline
Fe & $^*1.72$  &  & $^*-6.8$ &   & cantilever & Klokholm \textit{et al.}\cite{KlokholmJAP1982}  \\
\hline
Co & $^*9.2$  &  & $^*-38$ &   & cantilever & Klokholm \textit{et al.}\cite{KlokholmJAP1982}  \\
\hline
Ni$_x$Fe$_{1-x}$ ($x$=0.62$\sim$0.87) &  &  & $^*15\sim-10$ &  0.95$\sim$1.33  & Optic & Bonin \textit{et al.} \cite{BoninJAP2005}  \\
\hline
Fe$_{1-x}$Ga$_x$ (galfenol, $x$=0.15$\sim$0.3) & -12$\sim$16  &  & 200$\sim$400 &   1.6  & Bulk & Clark \textit{et al.}\cite{ClarkJAP2003}  \\
\hline
Tb$_{0.3}$Dy$_{0.7}$Fe$_2$ (Terfenol-D) &   &  & $^*1100\sim1400$  &  1.0  & Bulk & Sandlund \textit{et al.}\cite{SandlundJAP1994} \\
\hline
Fe$_x$Co$_{1-x}$ ($x$=0.45$\sim$0.55) &   &  & $^*50\sim90$ &   & Optic, FMR & Cooke \textit{et al.}\cite{CookeJPD2000} \\
\hline
NiFe$_2$O$_4$ &   &  & $-44$ &  0.34  & FMR & Smith \textit{et al.}\cite{SmithJAP1963}  \\
\hline
NiZnAl ferrite&  2 & 15 & 10 &  0.15  & FMR & Emori \textit{et al.}\cite{EmoriAM2017}  \\
\hline
(Ga,Mn)As&   &   & $-5$ &  0.04  & Transport (4.2 K) & Glunk \textit{et al.}\cite{GlunkPRB2009}  \\
&   & 85  & &  0.025  & MOKE (1.6 K) & Scherbakov \textit{et al.} \cite{ScherbakovPRL2010}  \\
\hline
\hline
\end{tabular}
\caption{Parameters of a few typical magnetostrictive materials and their characterization approaches. *The samples are polycrystalline and the magnetoelastic constants represent an average of different crystalline orientations.}
\label{table1}
\end{table*}

The interaction between the magnetization and the strain of a magnetic materials, or the magnetoelastic coupling, is an intrinsic property which converts magnetic energy into kinetic energy or vice versa. This property has been found in a broad material category, including metallic ferromagnets, alloys and ferrites, and has been widely applied in building sensors and actuators. In a magnetic crystal, the dominating mechanism of magnetoelastic coupling is that the change of lattice by strain modifies the magnetocrystalline anisotropy \cite{KittelPhysRev1958}, which originates from the spin-orbit coupling. The coupling energy per volume can be expressed as \cite{SmithJAP1963,ComstockProcIEEE1965}:
\begin{align}\label{eq01}
{E \over V} = &b_1[\epsilon_{xx}m_x^2+\epsilon_{yy}m_y^2+\epsilon_{zz}m_z^2] \nonumber \\
+ &b_2[\epsilon_{xy}m_xm_y+\epsilon_{yz}m_ym_z+\epsilon_{zx}m_zm_x]
\end{align}
where $\epsilon_{ij}$ denotes the unitless strain and $m_i$ denotes the unit vector component of magnetization ($i,j \in \{x,y,z\}$). The magnetoelastic coupling constant $b_1$ and $b_2$ holds the unit of J/m$^3$ or Pascal. The value of $b$ can be also given as $b/M_s$, which denotes the effective magnetic field per unit of strain. Another commonly used term is the magnetostriction constants $\lambda_{100}$ $\lambda_{111}$ which represent the maximal shape distortion by saturating the magnetization. The conversions between $b_i$ and $\lambda_i$ are \cite{ComstockProcIEEE1965} $b_1=-(3/2)\lambda_{100}(c_{11}-c_{12})$ and $b_2=-3\lambda_{111}c_{44}$, where the $c_{ij}$ are elastic constants. Static magnetostrictive force can be used to engineer memory devices \cite{NovosadJAP2000}. Table \ref{table1} lists the reported magnetoelastic coupling of a few typical magnetic systems. Additional reviews of magnetoelastic materials can be found elsewhere \cite{ChuJPD2018,LiangSensors2020}. In particular, we focus on the magnetic systems with low dampings or linewidths, which means high quality factor in magnon resonance and efficient interaction with phonons.

\section{Magnetic excitation induced by acoustic waves}

\begin{figure}[htb]
 \centering
 \includegraphics[width=3.0 in]{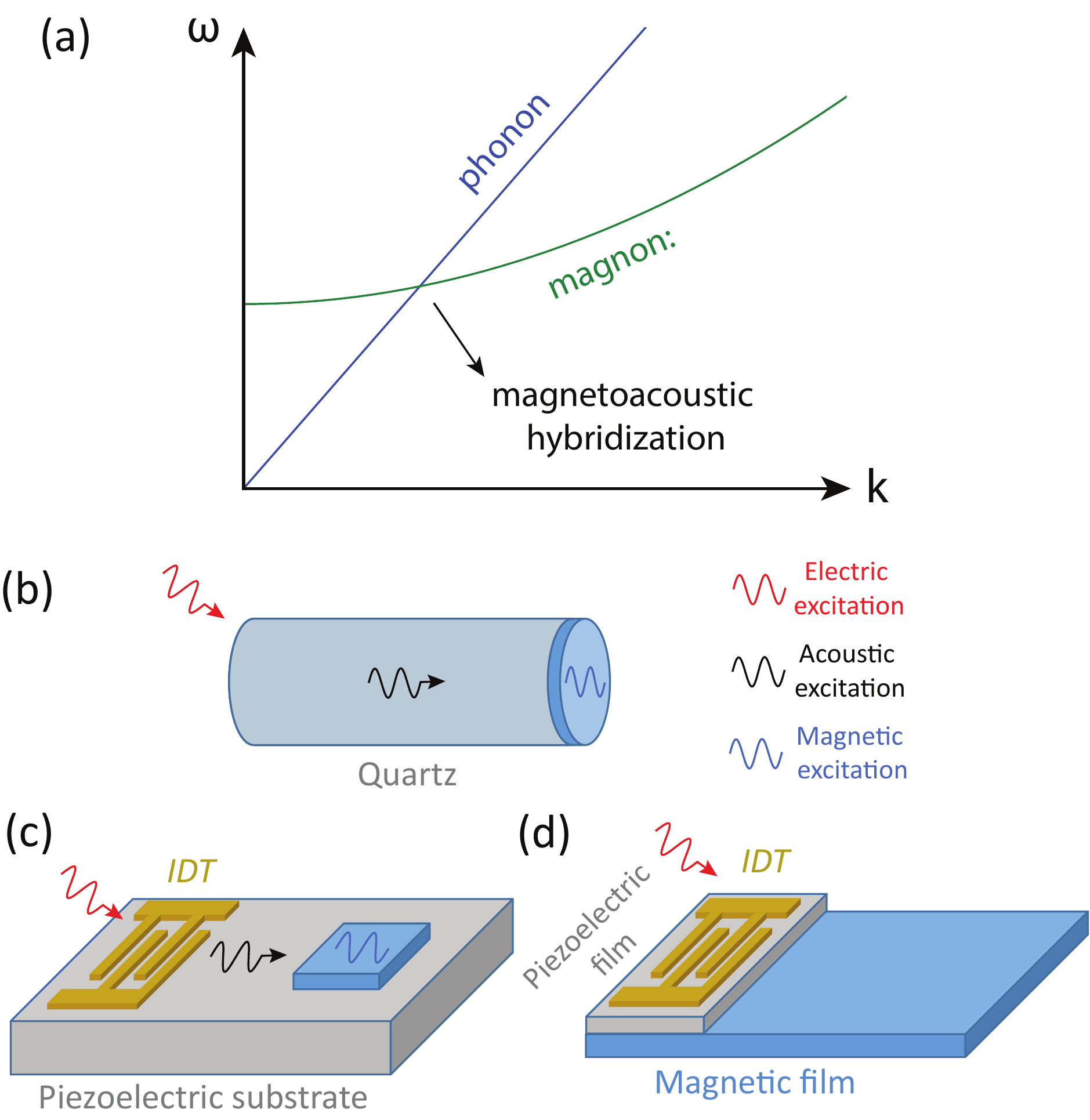}
 \caption{(a) Illustration of magnetoacoustic hybridization in $\omega$-$k$ space. (b-d) Different schematics of exciting magnons with acoustic waves. (b) Acoustic wave excitation and propagation in a quartz rod towards magnetic films grown on the far end. (c) Fabrications of an IDT and a magnetic device on a piezoelectric substrate. (d) Fabrication of an IDT on a magnetic film, with a layer of piezoelectric film deposited between the IDT and the magnetic film. }
 \label{fig1}
\end{figure}

In dynamics, the interaction of spin waves and acoustic waves has been extensively studied half a century ago \cite{KittelPhysRev1958,SpencerPRL1958,BommelPRL1959,PomerantzPRL1961,MatthewsPRL1962,SeaveyProcIEEE1965,ComstockProcIEEE1965,KobayashiPRB1973}. \textcolor{black}{In particular, the experiments were configured in the way that acoustic excitations were taken as a source to drive ferromagnetic resonance, or to be more accurate, the dynamics of magnetoacoustic modes \cite{KittelPhysRev1958,SchlomannJAP1960,SchlomannJAP1964,ComstockProcIEEE1965,LemanovSov1973,BelyaevaSov1992}. Fig. \ref{fig1}(a) shows the schematics of the dispersion relations for magnon [$\omega=\omega_0+(2\gamma A_{ex}/M_s)k^2$] and phonon ($\omega = v_c k$), where $\omega_0=\mu_0\gamma H_B$ is the uniform mode frequency determined by the biasing field $H_B$, $\gamma$ is the gyromagnetic ratio, $A_{ex}$ is the exchange stiffness, $M_s$ is the magnetization, and $v_c$ is the sound velocity of phonon. At the point where the dispersion curves for magnons and phonons intersect with each other, their coupling are enhanced due to the conservation of both energy and momentum, leading to the formation of hybrid magnetoacoustic modes and an enhanced energy absorption in acoustic-driven ferromagnetic resonance.}

As an example of different excitation schematics, magnetic thin films such as Ni \cite{BommelPRL1959} was deposited on one end face of a single-crystal quartz rod as a waveguide of acoustic excitations (Fig. \ref{fig1}b). The detection of ferromagnetic resonance in Ni films were realized by placing the Ni end in a microwave cavity for microwave transmission measurements \cite{PomerantzPRL1961}. Similar experiments have been also done in ferrites such as yttrium iron garnet (YIG) \cite{MatthewsPRL1962}, where magnetoelastic waves \cite{OkamotoPRB2020,VandervekenarXiv2020} were excited and a measurable rotation of polarization were measured in the shear elastic wave propagating in a YIG cylinder, a mechanical version of the optical Faraday effect.

\begin{figure}[htb]
 \centering
 \includegraphics[width=3.0 in]{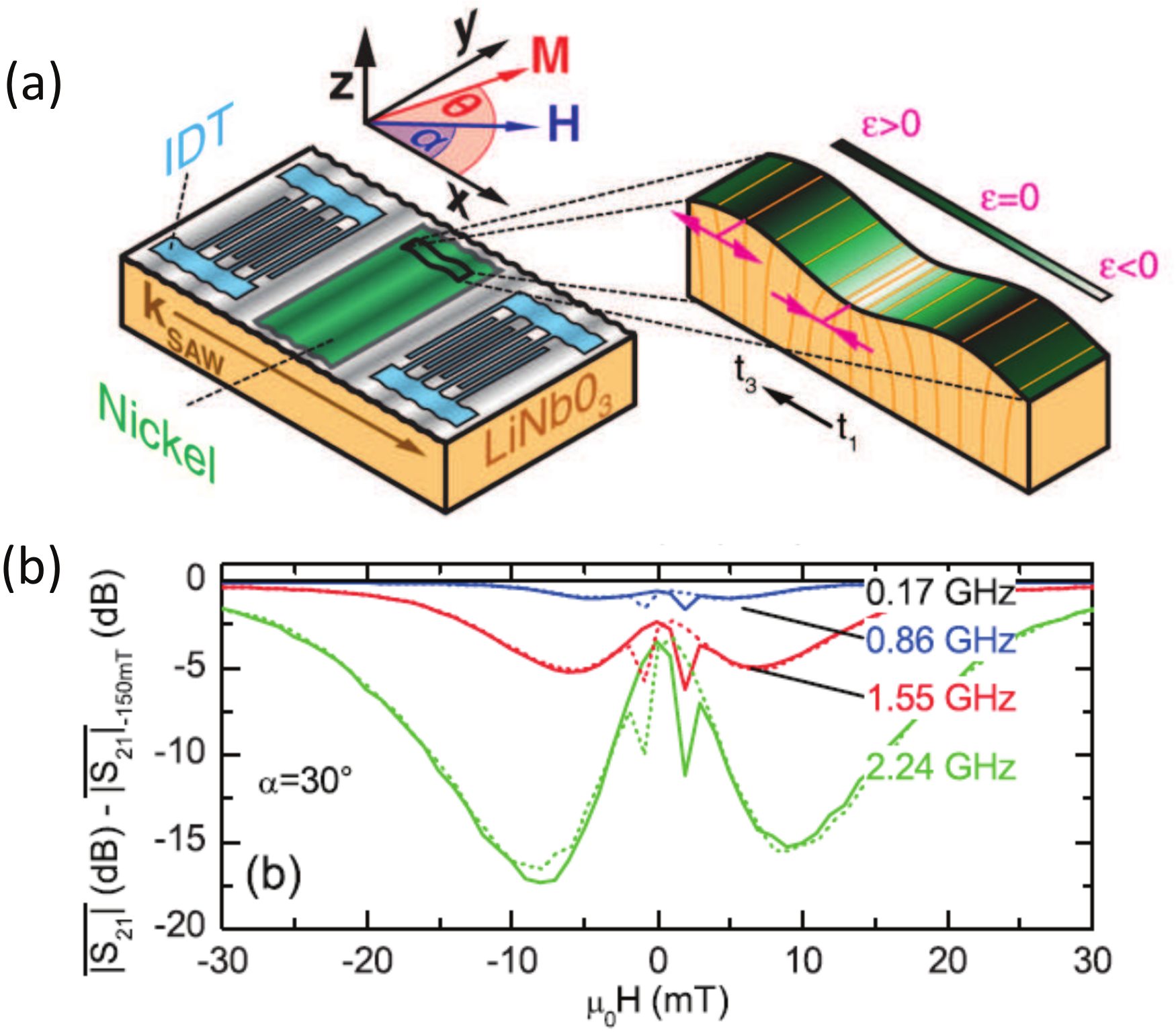}
 \caption{(a) Illustration of acoustic FMR in LiNbO$_3$/Ni system. (b) Experimental data of acoustic FMR measured for a Ni device. The excitation frequencies are harmonics of the base frequency 0.17 GHz determined by the period of the IDTs. Adapted from Ref. \cite{WeilerPRL2011}.}
 \label{fig1_exp}
\end{figure}

The exploration of magnon-phonon interaction has then moved to miniaturized micro-/nano-systems particularly with surface acoustic waves (SAWs) for their potentials in on-chip applications and integration with microwave circuits for advanced information processing. Shown in Fig. \ref{fig1}(c), the SAWs are electrically excited by interdigital transducers (IDTs) fabricated on a piezoelectric substrate, and interact with magnetic thin-film devices fabricated on the same substrate. The frequency and wavelength of the SAWs are determined by the period of the IDTs. \textcolor{black}{For IDT excitations of acoustic waves, the magnon dispersion curve can be shifted by $H_B$ to adapt the intersection point to the $k$ of the IDT.}  Higher harmonics of the SAW excitations are also conventionally used for obtaining higher frequencies. The detection of the magnon excitation induced by acoustic wave are conventionally realized by fabricating a second IDT on the pathway of acoustic wave. \textcolor{black}{One pioneering work by Weiler et al. \cite{WeilerPRL2011} has focused on the LiNbO$_3$/Ni systems to demonstrate electrical excitation and detection of acoustic-driven ferromagnetic resonance [Fig. \ref{fig1_exp}(a)]}. The magnetoelastic coupling of Ni converts the SAWs in LiNbO$_3$ to magnons in Ni, leading to a valley of microwave transmission between the two IDTs associated with Ni ferromagnetic resonance \cite{DreherPRB2012} [Fig. \ref{fig1_exp}(b)]. Another way to excite magnons with acoustic waves is to introduce a piezoelectric buffer layer, such as ZnO and Pb(Zr,Ti)O$_3$, between the IDT and the magnetic thin films, which is illustrated in Fig. \ref{fig1}(d). This technique is important for magnetic systems that usually need to be grown on specific substrates, with examples of YIG on GGG substrate \cite{HannaIEEE1988,HannaIEEE1990,ChumakPRB2010,UchidaAPL2011,KryshtalAPL2012} and (Ga,Mn)As on GaAs substrate \cite{ThevenardPRB2014,KuszewskiJPhys2018}. Note that the GaAs substrate is also a piezoelectric material for SAW excitation \cite{KuszewskiJPhys2018}.

It is worth noting that acoustic-driven ferromagnetic resonance can be also achieved by an ultrafast magneto-acoustic technique \cite{ScherbakovPRL2010,BombeckPRB2012,JagerPRB2015,KimPRL2012,KimPRB2017,JanusonisAPL2015,JanusonisPRB2016}
, where an ultrafast laser pulse illuminating the sample applies a strain pulse to \textcolor{black}{either the substrate \cite{ScherbakovPRL2010,BombeckPRB2012} or the ferromagnetic layer \cite{JagerPRB2015,KimPRL2012,KimPRB2017,JanusonisAPL2015,JanusonisPRB2016}}. This triggers resonant standing acoustic waves with the wavelength \textcolor{black}{typically} defined by the thickness of the substrate or the ferromagnetic layer; \textcolor{black}{in the latter case, either a phononic Bragg mirror \cite{JagerPRB2015} or a freestanding ferromagnetic film \cite{KimPRL2012} can be used to limit the standing wave within the layer.} \textcolor{black}{An in-plane acoustic wavevector can be also defined and tuned by a transient grating technique \cite{JanusonisAPL2015,JanusonisPRB2016}, where two interfering laser pulses illuminate the surface of the sample with a periodic pattern.} Magnetostrictive coupling in magnetic thin films grown on the substrate then converts the acoustic excitation to magnon excitations, which can be detected in the time domain by pump-probe systems that are typically paired with ultrafast lasers. \textcolor{black}{The main advantage of the ultrafast magneto-acoustic technique is that the acoustic pulse can be created without the need of piezoelectricity or electrodes. The capability of defining in-plane or perpendicular wavelengths can also greatly facilitate the coupling with various magnonic modes, including perpendicular standing spin waves and in-plane propagating spin waves.}

\section{Imaging magnon-phonon interaction}

\begin{figure}[htb]
 \centering
 \includegraphics[width=3.0 in]{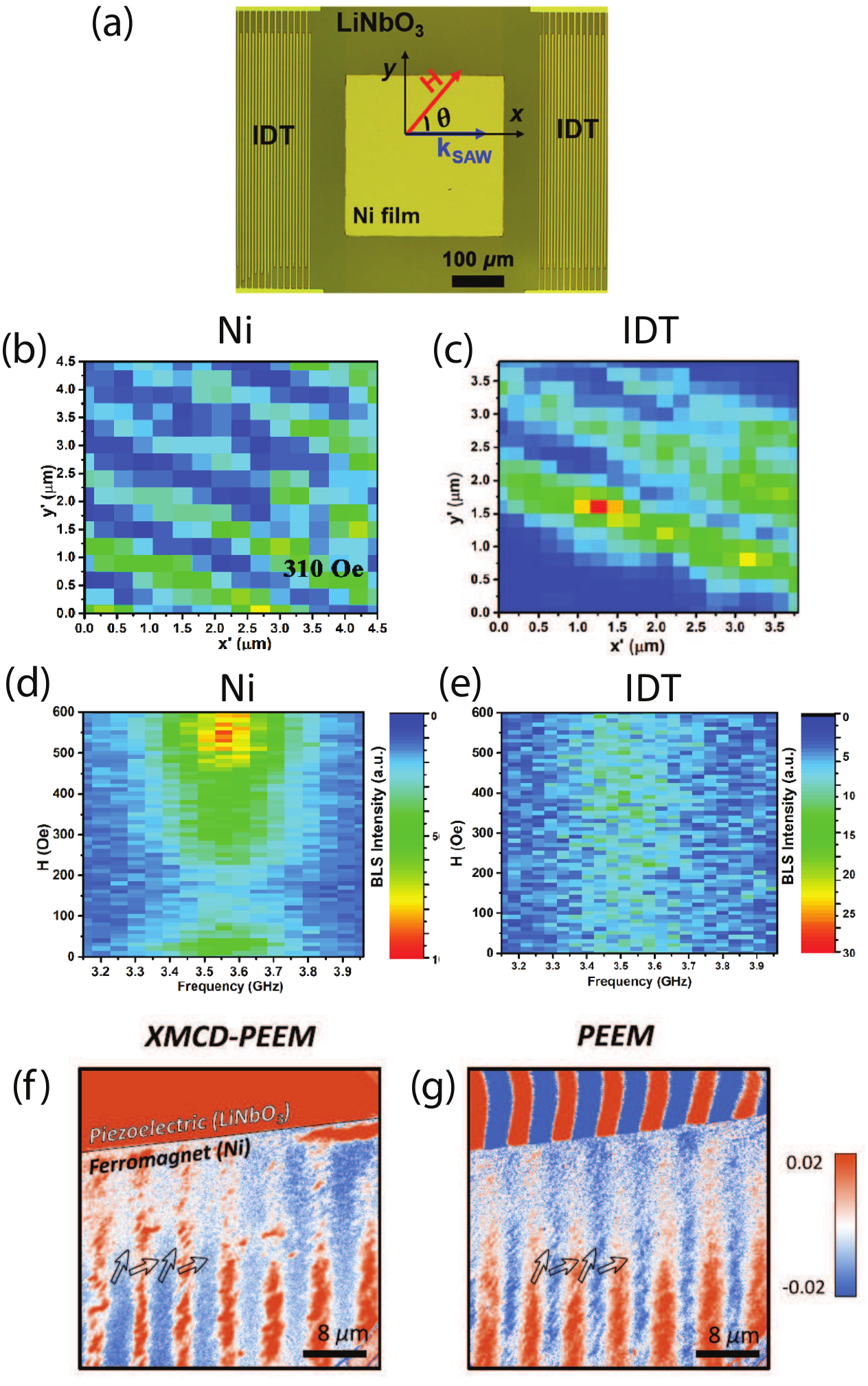}
 \caption{(a) Optical microscope image of a LiNbO$_3$/Ni(50 nm) system for the BLS measurements. (b-c) BLS microscopy of (b) the Ni device and (c) the IDT finger at $\omega/2\pi=3.56$~GHz, showing spatial wave profiles with the same wavelength. (d-e) BLS spectroscopy of (d) the Ni device and (e) the IDT finger by sweeping the source frequency and the biasing field. The frequency-domain peak at 3.56~GHz indicates the most efficient excitation with the IDT. The field-domain absorption valley at 150~Oe for Ni device is due to its FMR. No field-dependent absorption is observed for the IDT. Adapted from Ref. \cite{ZhaoPRApplied2021}. (f) XMCD and (g) PEEM microscopy of a LiNbO$_3$/Ni system at the Ni device edge. The former shows a clear contrast for magnons and the latter shows a clear contrast for SAWs. Adapted from Ref. \cite{CasalsPRL2020}.}
 \label{fig2}
\end{figure}

Imaging coupled magnon-phonon excitations are important for understanding their interaction and quantifying spatial propagation parameters. Recently, Zhao, et al. \cite{ZhaoPRApplied2021} have demonstrated visualization of acoustic ferromagnetic resonance (FMR) using micro-focused Brillouin light scattering ($\mu$-BLS). BLS is a powerful tool for investigating magnons with high spatial and frequency resolutions along with excellent sensitivity \cite{DemokritovPhysRep2001}. In a standard LiNbO$_3$/Ni system shown in Fig. \ref{fig2}(a), the BLS was applied on 1) the Ni thin film device and 2) the IDT. In both cases, a spatially wave excitation pattern is measured as shown in Figs. \ref{fig2}(b-c) with the same wavelength of 1.1 $\mu$m, suggesting the SAW-driven magnetization excitations in the Ni layer. Direct measurement of the LiNbO$_3$ surface yields a much weaker BLS signal because the LiNbO$_3$ substrate has a low reflecting rate of laser compared with the Au IDT. In addition, the field dependence of the Ni-film BLS spectrum shows a signal suppression at the FMR field of Ni, as shown in Fig. \ref{fig2}(d). This shows that in addition to the magnon excitation, the BLS signal from the Ni film contains a significant contribution from SAW \cite{VincentJPD2005}, which is damped at magnon resonance and leads to the BLS signal reduction. Thus the BLS provides a direct image of the magnetic field modulation of surface acoustic wave phonons by magnon-phonon coupling.

In another recent report, Casals, et al. \cite{CasalsPRL2020} have demonstrated independent imaging of both magnons and SAWs in the same LiNbO$_3$/Ni system with synchrotron X-ray source, with which photoemission electron microscopy (PEEM) was used to obtain electrical contract of SAWs and X-ray magnetic circular dichroism (XMCD) was used to achieve magnetic contrast of magnons. This technique has been demonstrated in a prior study of magnetic domain structures \cite{FoersterNComm2017}. As shown in Figs. \ref{fig2}(f-g), at the edge of Ni device, the XMCD signal shows a clear contrast of wave excitation on the Ni film but no contrast on the LiNbO$_3$ substrate. On the other hand, the PEEM signal at the same location shows a strong contrast of excitation on the LiNbO$_3$ substrate. The helicity-based magnetization resolution of XMCD has been also applied in optical measurements for characterizing magnetic excitations in novel materials down to the single atomic limit \cite{ZhangNMater2020,CenkerNPhys2020,JinarXiv2020}.

For the two techniques discussed above, BLS provides a convenient table-top detecting solution for magnon-phonon coupled dynamics and X-ray microscopy provides a powerful nano-imaging tool with clear distinction between strain and magnetization dynamics. The small wavelength of X-ray enables sub-100 nm spatial resolution \cite{BonettiNcomm2015,ChengJMMM2017,BonettiJPhys2017} and outperforms the resolution of $\mu$-BLS limited by the optical wavelength \cite{SebastianFrontPhys2015}, although a near-field BLS may provide a better spatial resolution \cite{JerschAPL2010}. Another notable advantage of BLS is its spectroscopy functionality, allowing broad-band excitation detection with fine steps [see Figs. \ref{fig2} (d-e)], whereas X-ray measurements need to be done only at multiple frequencies of the synchrotron repetition rate.

\section{Angular momentum transfer with magnon-phonon coupling}

Magnetic excitations naturally contain angular momentum from spins. When coupled with mechanical systems, the inverconversion from spin angular momentum to static lattice rotation has been experimentally demonstrated by Barnett \cite{BarnettPhysRev1915} and Einstein and de Haas \cite{EinsteindeHass1915} in bulk iron bars. However, for dynamic magnon systems and lattice vibration, it remains an interesting question if angular momentum can still be transferred on a microscopic level \cite{ZhangLifaPRL2014,GaraninPRB2015}.

\begin{figure}[htb]
 \centering
 \includegraphics[width=3.0 in]{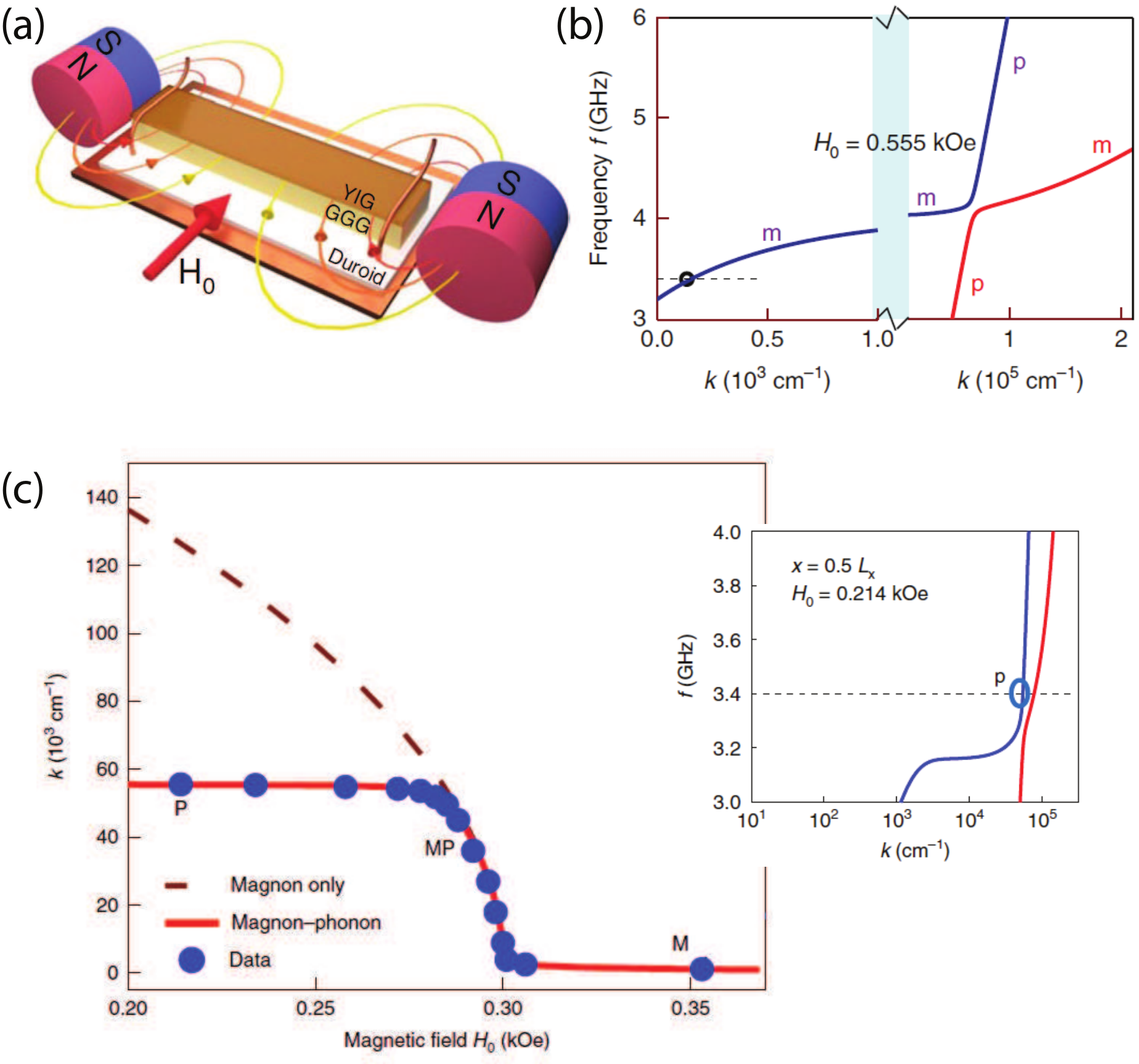}
 \caption{(a) YIG delay line with two permanent magnets placed at two ends creating a magnetic field gradient along the delay line. (b) Magnon-phonon hybridization in $\omega$-$k$ space. (c) Field peak position of the BLS signal measured at different $k$, with $k$ determined by the laser incident angle. The data are measured in the middle of the YIG thin film stripe where the magnons is adiabatically converted to phonon. The dashed curve shows the pure magnon dispersion and the red curve shows the dispersion of the magnetostatic excitation as shown in the inset. Adapted from Ref. \cite{HolandaNPhys2018}.}
 \label{fig3}
\end{figure}

Recently Holanda et al. \cite{HolandaNPhys2018} have shown that angular momentum can be transferred between magnon and phonon in a YIG-film delay line. In the YIG film, the $\omega$-$k$ dispersion curves of magnon and phonon modes intersect at $k\sim 10^5$ cm$^{-1}$ and form an avoided crossing of magnetoelastic waves [Fig. \ref{fig3}(b)]. By creating a strong non-uniform magnetic field on the YIG film by two adjacent permanent magnets, with the schematic shown in Fig. \ref{fig3}(a), the magnons propagating along the gradient field potential can be adiabatically converted to phonons along the upper (blue) branch of the magnetoelastic excitation. This will dramatically change the $k$ vector, which is measured by BLS and plotted in Fig. \ref{fig3}(c): when the biasing field is reduced for a given frequency of 3.4 GHz, $k$ saturates at around $55\times 10^3$ cm$^{-1}$. This is due to the constant group velocity of phonons in YIG, with $v_g = \omega/k = 3.9\times 10^{3}$ m/s. In addition, by measuring the regime where the magnetoelastic excitations are phonon-dominated, the BLS signals from linearly polarized light are shown to become circularly polarized, showing that the phonons carry angular momenta.

One consequence of phonons holding angular momenta is that the time-reversal symmetry is broken, which means that an asymmetry is created for phonons propagating forward or backward. As a macroscopic example of the above experiment, the polarization of a propagating transverse acoustic wave in a YIG cylinder has been observed to rotate \cite{MatthewsPRL1962}, which indicates the eigenstate of the magnetoelastic wave with circular polarization. Recently, the angular momenta of phonons have been applied in creating nonreciprocal SAW propagation \cite{SasakiPRB2017} and controlling magnetization state \cite{SasakiarXiv2020} in LiNbO$_3$/Ni structure. It is worth noting that another mechanism, spin-rotation coupling, can be also used to link lattice rotation with spin angular momenta of electrons even in a nonmagnetic materials \cite{MatsuoPRB2013,KobayashiPRL2017}. Thus utilizing angular momenta of phonons in couple with magnetic excitations provides a new avenue for engineering phonon propagation and creating nonreciprocity.

\section{Nonreciprocal SAW propagation induced by magnon-phonon coupling}

With the recent resurgence of acoustically driven FMR studies, integrating spin-wave manipulation into SAW systems provide a new way for making miniaturized radio frequency devices such as isolators and circulators while maintaining the tunability that is inherited from magnon dynamics. The key is to realize nonreciprocal magnon propagation in magnetic materials. While there are numerous proposals on how to implement nonreciprocal magnon propagation \cite{VerbaPRApplied2018,VerbaPRApplied2019,BauerPRB2020,YamamotoJPSJ2020}, we limit our discussion to two recent experimental works with clear demonstration of nonreciprocal SAW propagation that is coupled to magnetic thin-film devices \cite{ShahSciAdv2020,XuSciAdv2020}.

\begin{figure}[htb]
 \centering
 \includegraphics[width=3.0 in]{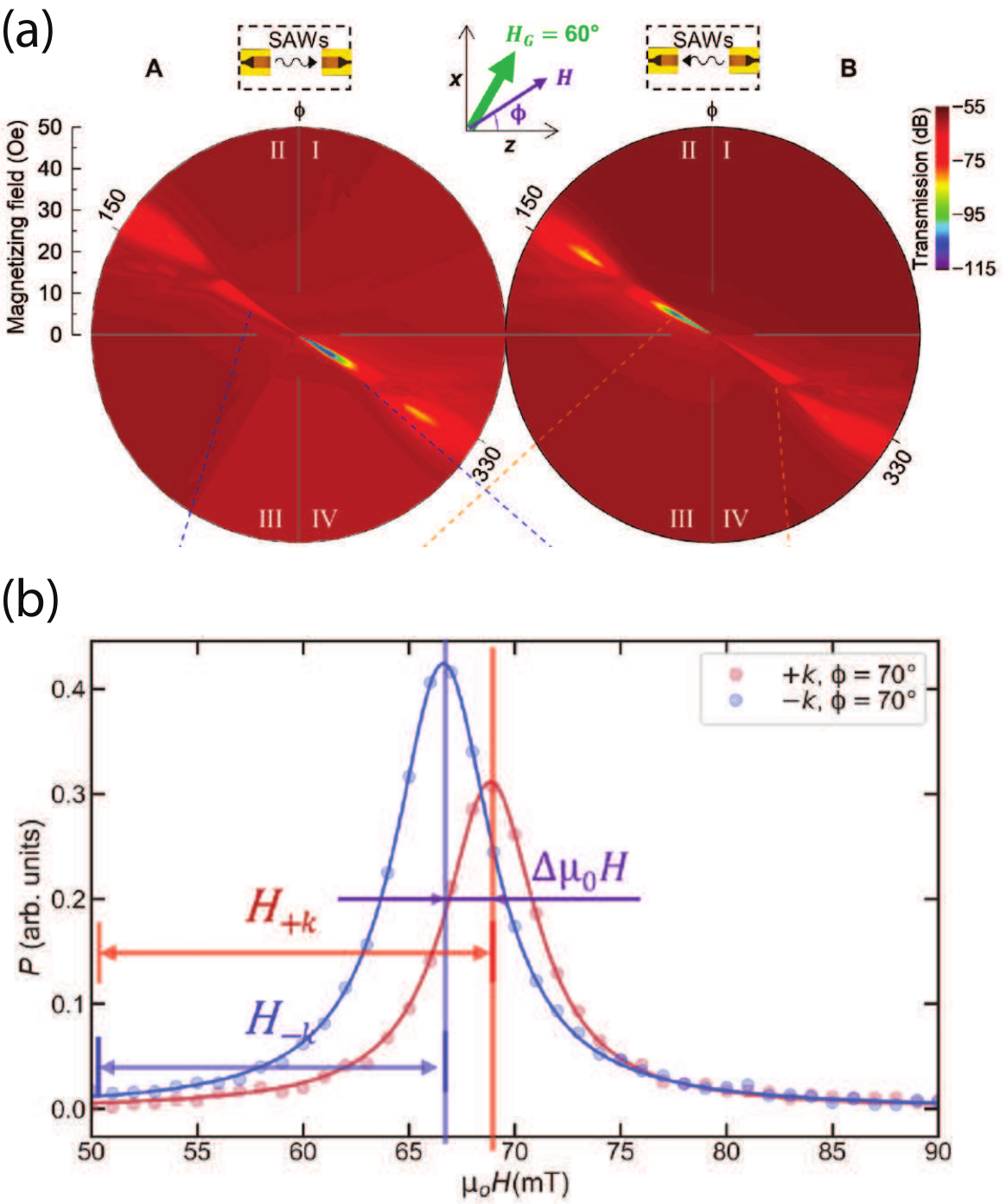}
 \caption{Magnon-induced nonreciprocal SAW propagation. (a) Nonreciprocal SAW transmission with FeGaB(20~nm)/Al$_2$O$_3$(5~nm)/FeGaB(20~nm) multilayer stack. Adapted from Ref. \cite{ShahSciAdv2020}. (b) Resonance field difference between acoustomagnetic waves in Ta(10~nm)/Co$_{20}$Fe$_{60}$B$_{20}$(1.6~nm)/MgO(2~nm) layers with SAW propagating in $+k$ and $-k$ direction. Both SAW systems and magnetic thin-film devices are fabricated on LiNbO$_3$ substrates. Adapted from Ref. \cite{XuSciAdv2020}.}
 \label{fig3_1}
\end{figure}

In the first example by Shah et al. \cite{ShahSciAdv2020}, the nonreciprocity is induced by different spin wave dispersions in a synthetic antiferromagnet (SAF), which has been predicted in theory \cite{VerbaPRApplied2019}. In order to realize nonreciprocal spin wave dispersion in a SAF, the magnetization state needs to be engineered \cite{VerbaPRApplied2019} such that 1) the static magnetization of the layers has a non-zero in-plane component and 2) the spin wave propagation in the SAF, which is determined by the SAW propagation, has a finite angle to the static magnetization. Thus the symmetry is broken by the magnetization vector, which lead to different spin wave dispersion characteristics \cite{DemokritovPhysRep2001}. In the experiment, a FeGaB/Al$_2$O$_3$/FeGaB trilayer has been used. This has yielded a SAW isolation ratio up to 48.4 dB (Fig.~\ref{fig3_1}a). Furthermore, high isolation is maintained in a finite field range from 0 to 20 Oe, meaning that the nonreciprocity is broad-band and does not depend on the resonance of a specific spin wave mode.

In the second example bu Xu et al. \cite{XuSciAdv2020}, the nonreciprocity is achieved from the Dzyaloshinskii-Moriya interaction (DMI), which create an effective field that depends on the propagating direction of spin waves \cite{VerbaPRApplied2018}. As a result, the forward and backward propagating spin waves will have different eigenfrequencies for a given field and wavenumber \cite{NembachNPhys2015}. Shown in Fig.~\ref{fig3_1}(b), a clear resonance field offset of 2.5 mT is measured between $+k$ and $-k$ spin wave propagations in a Ta/CoFeB/MgO thin film device, which is comparable to the resonance peak linewidth and can lead to sizable nonreciprocal SAW propagation. Note that the magnon-phonon coupling is dominated by the spin-rotation coupling \cite{KobayashiPRL2017} instead of magnetoelastic coupling. Similar results have been also reported in other multilayer systems \cite{KussPRL2020,HernandezPRApplied2020,GodejohannPRB2020}.

As a comparison between the two approaches, the SAF approach rely on the low-field canted magnetization states away from the $k$ vector in the two magnetic layers, while the DMI approach takes advantage of the effective field generated from the interfacal coupling. In bandwidth, the SAF approach shows potential for achieving broad isolation frequency band. However, due to unsaturated nature the system will only work for weak biasing field and thus low frequency for magnon band. Although in theory the unsaturated magnetization states may be challenging to control and can easily deviate from macrospin states with domain nucleation, the experimental results by Shah et al. \cite{ShahSciAdv2020} show promising flat transmissions from 0 to 10~Oe for both the passing band and the damped band along with very large isolation. For the DMI approach, a much broader isolation frequency tunability is available because the spin wave frequency can be freely modified by the biasing field in the saturated state. However, the isolation depth will be a challenge due to the fact that very thin ferromagnetic films such as CoFeB need to be used for significant DMI effective field. This will lead to limited magnon-SAW coupling and increased linewidth of spin waves for thin films. As a conclusion, the SAF approach wins in terms of isolation performance and the DMI appoach wins in terms of frequency range.

\section{Strong coupling between magnons and phonons}

One essential aspect of coherent information processing is the achievement of strong coupling, which is usually represented by the avoided crossing of two interacting excitations. In magnetoacoustic excitations, this happens when the magnon and phonon dispersion curves intersect in the $\omega$-$k$ space, leading to the formation of hybrid magnetoacoustic modes \cite{SchlomannJAP1960}. \textcolor{black}{They are also named as magnon polarons \cite{KamraPRB2015,FlebusPRB2017,CornelissenPRB2017,StreibPRL2018}, which are derived from the term ``polarons" describing the coupled excitations between electrons and lattice atom motions; here the spin excitations in magnons can be viewed as a special form of electron excitations.} Although the excitations of hybrid magnetoacoustic modes have been studied for long time \cite{BelyaevaSov1992}, recent experiments \cite{KikkawaPRL2016,BozhkoPRL2017,HayashiPRL2018,YahiroPRB2020} in magnetic thin-film systems have revealed how explicitly the degeneracy in both $\omega$- and $k$-spaces can enhance the excitations of magnon-phonon coupled modes. For example, Kikkawa et al. \cite{KikkawaPRL2016} have observed an enhanced peak in the spin Seebeck signals of a YIG/Pt thin-film bilayer by tuning the magnetic biasing field, which corresponds to the crossing of the magnon and phonon modes. Similar effects have been observed in the spin Peltier effects by Yahiro et al. \cite{YahiroPRB2020} and spin pumping by Hayashi et al. \cite{HayashiPRL2018}. Furthermore, Bozhko et al. \cite{BozhkoPRL2017} have directly observed the enhanced excitation of magnetoacoustic modes via wave-vector-resolved BLS measurements; similar phenomena have been also reported by other optical means \cite{OgawaPNAS2015,HashimotoNComm2017}. Here the coherent magnon-phonon interaction is different from the incoherent magnon-phonon scattering led by spin-orbit coupling, which is the major source of magnon damping process \cite{LiPRL2019_CoFe}.

\begin{figure*}[htb]
 \centering
 \includegraphics[width=6.0 in]{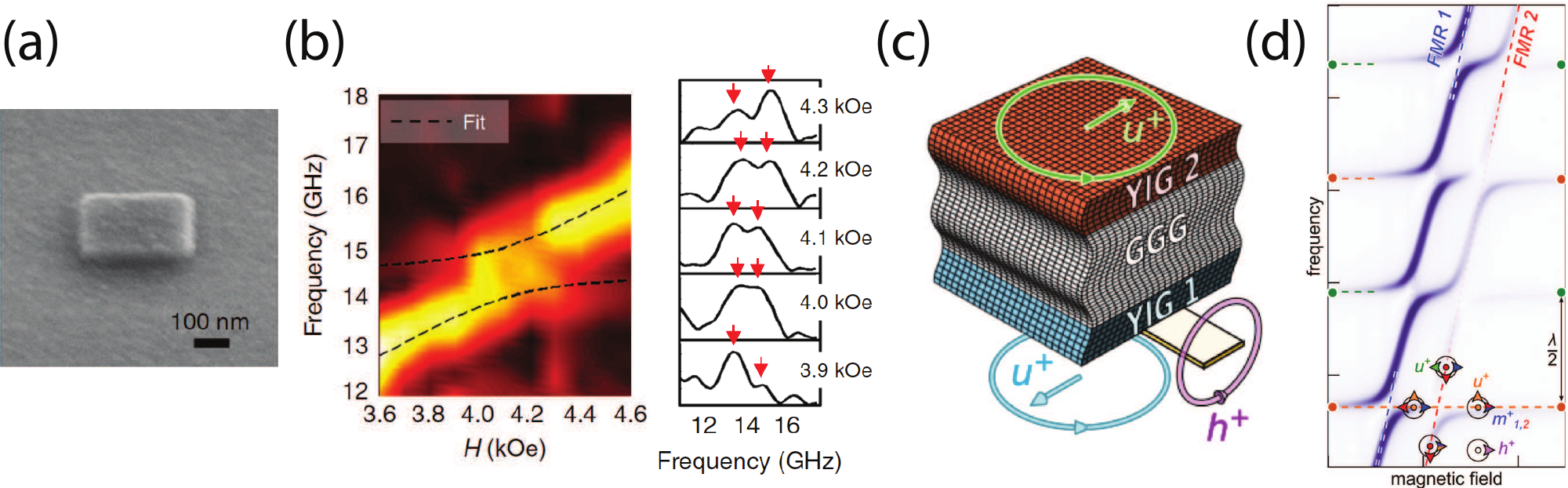}
 \caption{(a) Tilted SEM image of a $300\time 300$~nm$^2$ Ni nanodevice fabricated on a Si substrate. (b) Mode anticrossing between the magnon mode of Ni nanodevice and the (2,0) acoustic standing wave mode of the Ni square along the lateral direction. The signals are excited by a laser pulse and measured by time-resolved MOKE. Adapted from Ref. \cite{BerkNComm2019}. (c) Illustration of a YIG(200~nm)/GGG(500~$\mu$m)/YIG(200~nm) structure with microwave excitation by a coplanar waveguide underneath. (d) Mode anticrossing between the magnon modes of the two YIG films and the acoustic standing wave of GGG along the thickness direction. Adapted from Ref. \cite{AnPRB2020}.}
 \label{fig4}
\end{figure*}

\textcolor{black}{To fully demonstrate avoided crossing, the acoustic mode spectra need to be quantized and separated in order to allow for cavity-enhanced magnon-phonon coupling.} Two recent works \cite{BerkNComm2019,AnPRB2020} have experimentally achieved clear mode anticrossing with a geometrically confined phononic resonator. In the first work, Berk et al. \cite{BerkNComm2019} have fabricated a $330\times330\times30$ nm$^3$ Ni nanomagnet on a Si substrate [Fig. \ref{fig4}(a)]. The lateral dimension of the nanomagnet defines the wave vector of the mechanical standing wave, turning the continuous phonon spectrum of Ni into quantized modes and creating a mechanical resonator. The coupled dynamics was excited by a laser pulse which triggered the imbalance of both the magnon and phonon subsystems, and the detection was done by time-resolved magneto-optical Kerr effect (MOKE). Fig. \ref{fig4}(b) shows the mode anticrossing between the (2,0) mechanical mode and the magnon mode, with a maximal cooperativity of $C=1.65$ when the biasing field is 45$^\circ$ away from the $k$ vector of the mechanical mode. In the second work, An et al. \cite{AnPRB2020} have employed the perpendicular acoustic standing  wave of a 500-$\mu$m-thick GGG substrate for coupling with the magnons of the 200-nm-thick YIG films grown on the surface. The energy was injected into the magnon mode by a microwave antenna and an inductive FMR was used for the detection. As shown in Fig. \ref{fig4}(b), the magnon mode of YIG is chopped by many high-order phonon modes ($n\sim 1500$) indicated by the equally separated horizontal spectra. Clear avoided crossings are formed at each intersection, with an extracted cooporativity of $C=3$. The perpendicular acoustic standing waves were further shown to couple two remote magnonic systems of the top and bottom YIG layers, which has been also confirmed in theory \cite{RuckriegelPRL2020}.

Ignoring all the geometric factor, the magnon-phonon coupling strength can be expressed as \cite{KittelPhysRev1958,HolandaNPhys2018,AnPRB2020,VandervekenarXiv2020}:
\begin{equation}\label{eq03}
  \Omega = {b \sqrt{\gamma k_m k_{ph} \over \omega M_s \rho}}
\end{equation}
where $b$ is the magnetoelastic coupling, $k_m$ and $k_{ph}$ are the wave vectors of magnon and phonon, respectively, $\omega$ is the degenerated frequency, $M_s$ is the magnetization and $\rho$ is the mass density. Here the unitless prefactor and ellipticity of magnetization precession are omitted. In general $k_m=k_{ph}$ for optimal magnon-phonon coupling. Similar square-root average in coupling strength occurs for magnon-photon \cite{LachanceQuirionAPEx2019,LiJAPPerspective2020} and magnon-magnon coupling \cite{LiPRL2020_YIGPy}, which is characteristic for hybrid dynamic coupling. Eq. (\ref{eq03}) suggests that the geometric parameters do not change the coupling strength because the total energies of the magnons, phonons and magnon-phonon coupling are all proportional to the volume. The main difference of the works by Berk et al. \cite{BerkNComm2019} and An et al. \cite{AnPRB2020} compared with conventional magnetoacoustic wave excitations is that the phonon modes are quantized by the geometric confinement and the avoided crossing between magnon and phonon modes can be experimentally observed, while for continuous magnetic films the phonon modes are also continuous and fill up the entire $\omega$-$H$ space. For systems with separated magnon and phonon reservoirs such as in the case of YIG/GGG, the phonon wave vector $k_{ph}$ is replaced by the geometric factor $1/t_\text{GGG}$ \cite{AnPRB2020} (note that $k_m=k_{ph}$ is not required for perpendicular standing waves). Now the geometric parameters start to play a role in determining the coupling strength because the GGG substrate no longer takes part in the magnon-phonon coupling process. It is indicative that in order to achieve strong magnon-phonon coupling with SAW excitation, a smaller effective volume of the SAW resonator will lead to a stronger coupling strength $\Omega$ for improving coherent magnon-phonon interaction. It is also worth noting that another recent work \cite{GodejohannPRB2020} has demonstrated strong coupling between magnons and phonons with a cooperativity up to 8, where grating-defined epitaxial Fe$_{81}$Ga$_{19}$ films serve as both the magnon and SAW resonator and can outperform Ni \cite{BerkNComm2019} due to its strong magnetoelastic coupling along with reasonably low magnon damping.

\section{Enhancing magnetoelastic coupling via materials engineering}

\begin{figure}[htb]
 \centering
 \includegraphics[width=3.0 in]{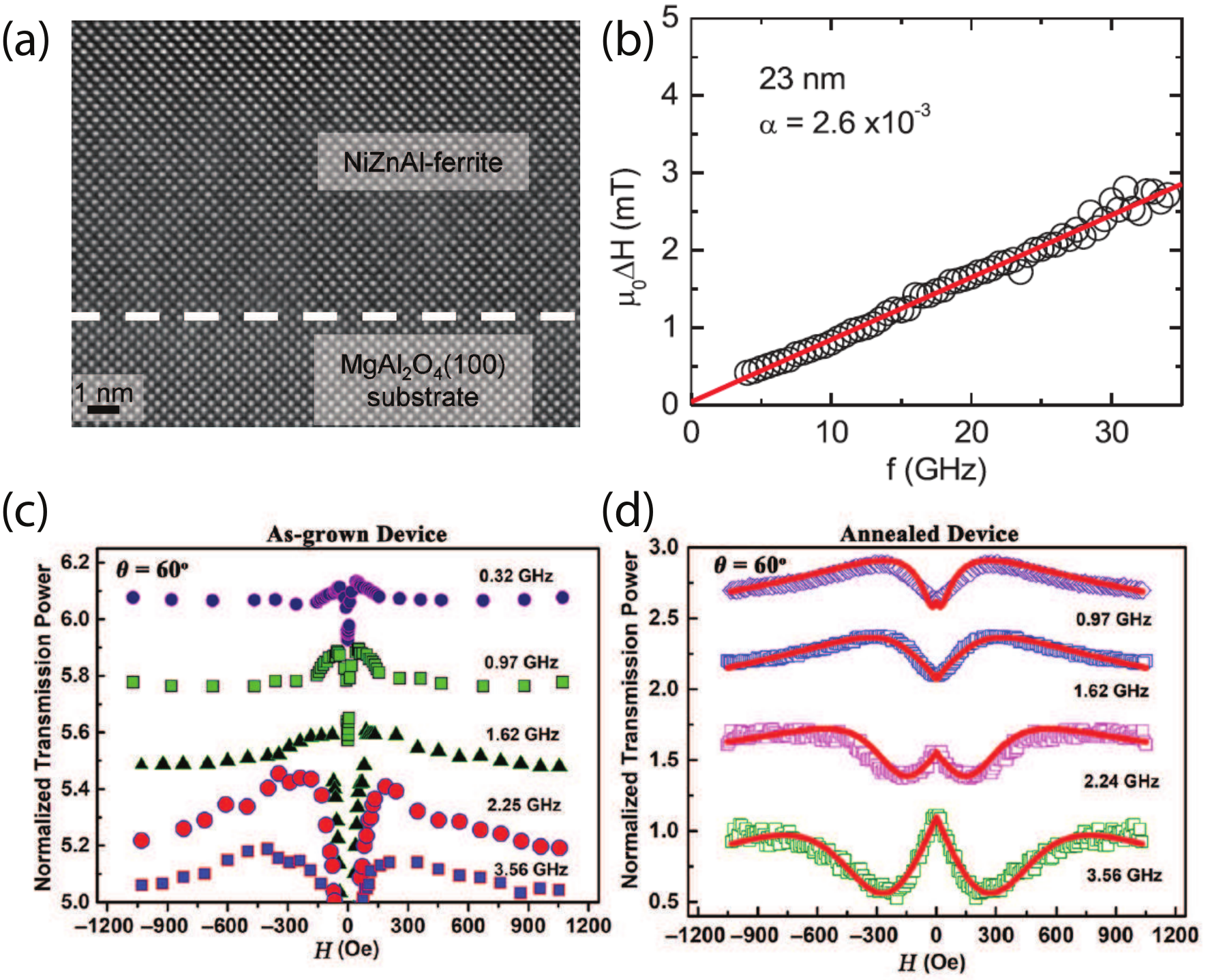}
 \caption{(a) A [100] High-resolution transmission electron microscopy image of epitaxial NiZnAl-ferrite thin films. (b) Ferromagnetic resonance linewidths of a 23-nm NiZnAl-ferrite thin film. (c-d) SAW transmission measured between two IDTs fabricated on a LiNbO3 substrate, with a 50-nm thick rectangular Ni thin film device fabricated between the two IDTs. (c) shows the transmission signals for the as-deposited Ni device and (d) shows the signals for the annealed Ni device. Adapted from refs. \cite{EmoriAM2017,ZhaoPRApplied2020}.}
 \label{fig_materials}
\end{figure}

Materialwise, it is desired to have both a strong magnetoelastic coupling and a low magnon damping rate for applications in coherent magnon-phonon interaction. However, since the magnon damping and magnetoelastic coupling shares the same spin-orbit coupling mechanism, balancing the two factors remains a challenge in material engineering. Recently Emori \textit{et al.} \cite{EmoriAM2017} has reported a spinel NiZnAl-ferrite epitaxial thin film system exhibiting both low damping and strong magnetoelastic coupling. Starting from a base NiFe$_2$O$_4$ spinel ferrite which exhibits a large magnetostriction coefficient ($\lambda_{100}=-4.4\times 10^{-5}$ )\cite{SmithJAP1963} but also a large ferromagnetic resonance linewidth (5 mT measured at 11 GHz \cite{SinghAM2017}), Emori \textit{et al.} partially substitute Ni$^{2+}$ cations with Zn$^{2+}$ cations which helps to reduce the damping by suppressing the spin-orbit stabilized uncompensated orbital angular momentum, as well as substitute Fe$^{3+}$ cations with Al$^{3+}$ cations which improves the lattice match with the MgAl$_2$O$_4$ substrate to minimize linewidth broadening induced by lattice disorder. The resulting composition Ni$_{0.65}$Zn$_{0.35}$Al$_{0.8}$Fe$_{1.2}$O$_4$ shows a large magnetoelastic coupling of $\lambda_{100}=-1\times 10^{-5}$ along with a small Gilbert damping of $\alpha=2.6\times 10^{-3}$ (a linewidth of 1 mT at 10 GHz) for a 23 nm thin film [Fig. \ref{fig_materials}(b)]. A nice lattice match between the film and the substrate as shown in Fig. \ref{fig_materials}(a) ensures minimal inhomogeneous broadening and low damping. The material chemistry provides a new pathway of engineering and optimizing the quality of magnon-phonon interactions for spin-mechanical applications.

In another recent report by Zhao \textit{et al.} \cite{ZhaoPRApplied2020}, it is shown that the magnon-phonon coupling efficiency can be greatly improved by post-annealing. A Ni thin-film device is fabricated on a LiNbO$_3$ substrate along with two gold IDTs fabricated on the two sides for SAW excitation measurements. As shown in Figs. \ref{fig_materials}(c) and (d), by annealing the Ni/LiNbO$_3$ sample at 400 $^\circ$C for 30 mins in vacuum, the SAW transmission responses between the two IDTs are significantly enhanced at low frequency ($\leq 1.62$ GHz). For higher frequencies, the lineshape of the annealed sample shows a suppression of the transmitted SAW power, indicating magnon mediated resonant energy absorption. From the line shape fitting, the effective magnon-phonon coupling coefficient $\eta\sim b^2/E_Y$ is found to enhance from 0.3 to 3.1 by post-annealing.

\section{Advanced SAW resonator design for studying coherent magnon-photon interaction}

\textcolor{black}{In order to allow for circuit integration of cavity-enhanced magnon-photon coupling and their mode anticrossing, it is important to bring in circuit SAW resonator designs, which have been widely applied for implementing strong coupling with quantum acoustic systems \cite{GustafssonScience2014,ManentiNComm2017,NoguchiPRL2017,SatzingerNature2018,ChuNature2018,MooresRPL2018,BolgarPRL2018,BienfaitScience2019,WhiteleyNphys2019,BienfaitPRX2020}. In addition, the use of superconducting coplanar resonators can significantly enhance the quality factor of SAW resonators by eliminating the Ohmic loss in IDTs; similar ideas have been applied for implementing strong magnon-photon coupling \cite{McKenziePRB2019,LiPRL2019_magnon,HouPRL2019}. Although the work of SAW resonators for studying magnon-photon coupling is scarce, we anticipate this direction to be the next breakthrough for engineering coherent magnon-phonon interactions.}

\begin{figure}[htb]
 \centering
 \includegraphics[width=3.0 in]{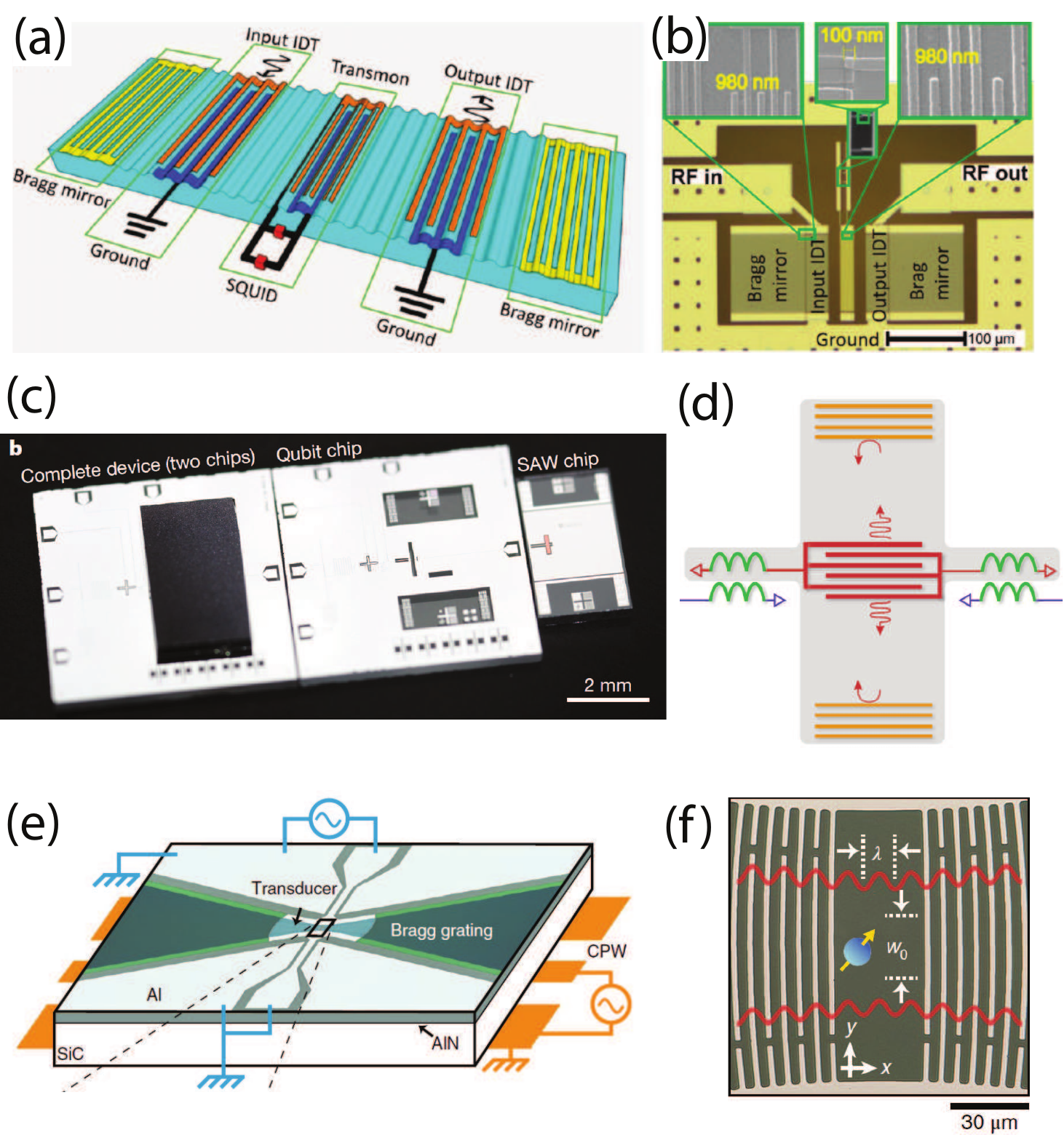}
 \caption{(a) Illustration of a SAW resonator with two Bragg mirrors and a pair of input and output IDTs in between. A transmon qubit IDT is fabricated in the middle of the SAW resonator for coherent SAW-qubit coupling. (b) Microscope image of the real device. Adapted from Ref. \cite{BolgarPRL2018}. (c) Picture of qubit-SAW hybrid chips. The middle and right are the individual qubit and SAW chip fabricated on sapphire and LiNbO$_3$ substrates, respectively. The left shows a qubit chip that is flip-covered by a SAW chip (black rectangle). (d) Illustration of the SAW resonator and a pair of IDTs in the middle which are connected to two different couplers for inductively coupling to two remote superconducting qubits on the qubit chip. Adapted from Ref. \cite{SatzingerNature2018} and Ref. \cite{BienfaitScience2019}. (e) Illustration of a Gaussian SAW resonator fabricated on a SiC substrate with an additional piezoelectric AlN thin film. (f) An optical micrograph of the resonator's acoustic focus, with a wavelength of $\lambda=12$~$\mu$m. Adapted from Ref. \cite{WhiteleyNphys2019}.}
 \label{fig5}
\end{figure}

\textcolor{black}{To serve as a guideline of SAW resonator design, we take three examples for studying SAW-qubit coupling in quantum acoustics. }
In the first example by Bolgar et al. \cite{BolgarPRL2018}, two Bragg mirrors are fabricated outside the input and output IDT pairs as shown in Figs. \ref{fig5}(a) and (b), forming a Fabry-P\'{e}rot SAW cavity. Each Bragg mirror consists of an array of equally spaced stripes with the period as half the wavelength of SAW, thus cancelling out the generated piezoelectric voltage. The period of the IDTs is 980 nm, which defines the excited SAW frequency of $\omega_c/2\pi=3.176$ GHz with a SAW group velocity of 3.16 km/s on quartz substrate. The full-width-half-maximum linewidth of the SAW cavity is $\Delta\omega_c=0.332$ MHz, corresponding to a quality factor of $\sim 10^4$. To couple with qubit which can be treated as a microwave resonator but with strong nonlinearity, the capacitive shunting of the qubit is also designed as an IDT with the same periodicity, allowing the SAW excitation to be transformed into electrical signal in the qubit. A SAW-qubit coupling strength of $g/2\pi= 15.7$ MHz is achieved, leading to a clear avoided crossing. Note that the Bragg mirror has a reflection bandwidth of 33 MHz which is determined by the number of repeated stripes. The hybrid modes need to stay within the mirror bandwidth. Otherwise the mirror can no longer effectively reflect the SAW and will cause the SAW coherence time to decrease. Similar demonstrations can be found in Refs. \cite{NoguchiPRL2017} and \cite{MooresRPL2018}.

In the second example by Satzinger et al. \cite{SatzingerNature2018}, the SAW resonator and the microwave superconducting circuit are fabricated on different chips. As shown in Fig. \ref{fig5}(c), the complete device (two chips) consists of a qubit chip fabricated on a sapphire substrate and a SAW chip fabricated on a LiNbO$_3$ substrate. To enable coupling, the SAW chip is flipped onto the qubit chip and the couplers on both chips are aligned upon flipping, providing an inductive coupling between the two couplers. The SAW resonator consists of a center IDT and two Bragg mirrors, similar to Fig. \ref{fig5}(a). The SAW resonator exhibits a wavelength of $\lambda=1 \mu$m and a quality factor of $Q>10^5$ at 10 mK. The same group has also demonstrated entanglement engineering of two superconducting qubits by such a flip-chip SAW resonator \cite{BienfaitScience2019}. By separating chip fabrications of the microwave and SAW circuits, the fabrication process can be simplified and optimized for each circuits and individual testing is allowed before assembly. This idea has also been adopted in a parallel work \cite{ChuNature2018} by utilizing bulk acoustic wave excitations.

In the third example by Whiteley et al. \cite{WhiteleyNphys2019}, the SAW resonator is fabricated with two Gaussian Bragg gratings, with the SAW focusing on a very small area in order to couple with spin defects in SiC. This design provides an even smaller effective volume for the SAW resonator and a higher sensitivity compared with the previous 1D geometry. A layer of aluminium nitride is sputtered on the SiC surface before the SAW resonator fabrication in order to amplify the piezoelectric response. A quality factor of $1.6\times 10^4$ is achieved at 30 K. By using the Gaussian SAW resonator, an avoided crossing with a coupling strength of $\Omega/2\pi=4$ MHz is achieved with diluted spin defects in an area of $30\times30$ $\mu$m$^2$.

To compare with the two phononic resonators used in Fig. \ref{fig4}, SAW resonators are much more flexible in design and fabrication, including wavelength definition and maintaining high quality factor. Furthermore, because the Rayleigh SAW is restricted to around one wavelength in the thickness direction, SAW resonators exhibit naturally small effective volumes. \textcolor{black}{By replacing qubits with magnonic devices, circuit inplementation of strong magnon-phonon coupling should be straightforward to be reached.} We also note that the interest in SAW phonons for quantum information comes from two unique properties: 1) phonons are potential for efficient conversion between microwave and optical photons \cite{SchuetzPRX2015,VainsencherAPL2016}, 2) phonons have low group velocity and can be used to create significant time delay for additional tunability \cite{GuoPRA2017,BienfaitPRX2020}. Similar properties are also shared by magnons in both microwave-to-optical transduction \cite{OsadaPRL2016,ZhangPRL2016,HaighPRL2016} and time delay \cite{SergaJPD2010}. With the recent demonstration of coupling a single magnon with a superconducting qubit \cite{TabuchiScience2015,LachanceScienceAdvan2017,LachanceQuirionScience2020} as well as on-chip magnon-photon hybridization with superconducting resonators \cite{LiPRL2019_magnon,HouPRL2019}, phonons and magnons are promising for playing active roles in quantum information.

\section{Conclusions}

The increasing efforts of coupling magnons and phonons have revealed novel discovery for the manipulation of microwave acoustic excitations empowered by magnetic materials, including new physics in spintronics and straintronics. In particular, with the capability of demonstrating strong magnon-phonon coupling and nonreciprocity, the interplay of magnons and phonons shows potential in coherent information processing with acoustic circuits. Applications may be also found in advanced information technology such as quantum computing by developing new functional modules in quantum acoustics with magnon engineering.

\section{Acknowledgement}

The preparation of this perspective article was supported by the U.S. Department of Energy, Office of Science, Basic Energy Sciences, Materials Sciences and Engineering Division. W. Z. acknowledges support from NSF-ECCS under grant no. 1941426. A. H. acknowledges support from Quantum Materials for Energy Efficient Neuromorphic Computing, an Energy Frontier Research Center funded by the U.S. DOE, Office of Science.

\section{Data Availability}

The data that support the findings of this study are available from the corresponding author upon reasonable request.


\begin{thebibliography}{137}%
\makeatletter
\providecommand \@ifxundefined [1]{%
 \@ifx{#1\undefined}
}%
\providecommand \@ifnum [1]{%
 \ifnum #1\expandafter \@firstoftwo
 \else \expandafter \@secondoftwo
 \fi
}%
\providecommand \@ifx [1]{%
 \ifx #1\expandafter \@firstoftwo
 \else \expandafter \@secondoftwo
 \fi
}%
\providecommand \natexlab [1]{#1}%
\providecommand \enquote  [1]{``#1''}%
\providecommand \bibnamefont  [1]{#1}%
\providecommand \bibfnamefont [1]{#1}%
\providecommand \citenamefont [1]{#1}%
\providecommand \href@noop [0]{\@secondoftwo}%
\providecommand \href [0]{\begingroup \@sanitize@url \@href}%
\providecommand \@href[1]{\@@startlink{#1}\@@href}%
\providecommand \@@href[1]{\endgroup#1\@@endlink}%
\providecommand \@sanitize@url [0]{\catcode `\\12\catcode `\$12\catcode
  `\&12\catcode `\#12\catcode `\^12\catcode `\_12\catcode `\%12\relax}%
\providecommand \@@startlink[1]{}%
\providecommand \@@endlink[0]{}%
\providecommand \url  [0]{\begingroup\@sanitize@url \@url }%
\providecommand \@url [1]{\endgroup\@href {#1}{\urlprefix }}%
\providecommand \urlprefix  [0]{URL }%
\providecommand \Eprint [0]{\href }%
\providecommand \doibase [0]{http://dx.doi.org/}%
\providecommand \selectlanguage [0]{\@gobble}%
\providecommand \bibinfo  [0]{\@secondoftwo}%
\providecommand \bibfield  [0]{\@secondoftwo}%
\providecommand \translation [1]{[#1]}%
\providecommand \BibitemOpen [0]{}%
\providecommand \bibitemStop [0]{}%
\providecommand \bibitemNoStop [0]{.\EOS\space}%
\providecommand \EOS [0]{\spacefactor3000\relax}%
\providecommand \BibitemShut  [1]{\csname bibitem#1\endcsname}%
\let\auto@bib@innerbib\@empty
\bibitem [{\citenamefont {Kurizki}\ \emph {et~al.}(2015)\citenamefont
  {Kurizki}, \citenamefont {Bertet}, \citenamefont {Kubo}, \citenamefont
  {M{\o}lmer}, \citenamefont {Petrosyan}, \citenamefont {Rabl},\ and\
  \citenamefont {Schmiedmayer}}]{KurizkiPNAS2015}%
  \BibitemOpen
  \bibfield  {author} {\bibinfo {author} {\bibfnamefont {G.}~\bibnamefont
  {Kurizki}}, \bibinfo {author} {\bibfnamefont {P.}~\bibnamefont {Bertet}},
  \bibinfo {author} {\bibfnamefont {Y.}~\bibnamefont {Kubo}}, \bibinfo {author}
  {\bibfnamefont {K.}~\bibnamefont {M{\o}lmer}}, \bibinfo {author}
  {\bibfnamefont {D.}~\bibnamefont {Petrosyan}}, \bibinfo {author}
  {\bibfnamefont {P.}~\bibnamefont {Rabl}}, \ and\ \bibinfo {author}
  {\bibfnamefont {J.}~\bibnamefont {Schmiedmayer}},\ }\href {\doibase
  10.1073/pnas.1419326112} {\bibfield  {journal} {\bibinfo  {journal} {Proc.
  Natl. Acad. Sci.}\ }\textbf {\bibinfo {volume} {112}},\ \bibinfo {pages}
  {3866} (\bibinfo {year} {2015})}\BibitemShut {NoStop}%
\bibitem [{\citenamefont {Clerk}\ \emph {et~al.}(2020)\citenamefont {Clerk},
  \citenamefont {Lehnert}, \citenamefont {Bertet}, \citenamefont {Petta},\ and\
  \citenamefont {Nakamura}}]{ClerkNphys2020}%
  \BibitemOpen
  \bibfield  {author} {\bibinfo {author} {\bibfnamefont {A.~A.}\ \bibnamefont
  {Clerk}}, \bibinfo {author} {\bibfnamefont {K.~W.}\ \bibnamefont {Lehnert}},
  \bibinfo {author} {\bibfnamefont {P.}~\bibnamefont {Bertet}}, \bibinfo
  {author} {\bibfnamefont {J.~R.}\ \bibnamefont {Petta}}, \ and\ \bibinfo
  {author} {\bibfnamefont {Y.}~\bibnamefont {Nakamura}},\ }\href@noop {}
  {\bibfield  {journal} {\bibinfo  {journal} {Nature Phys.}\ }\textbf {\bibinfo
  {volume} {16}},\ \bibinfo {pages} {257–267} (\bibinfo {year}
  {2020})}\BibitemShut {NoStop}%
\bibitem [{\citenamefont {Hu}(2016)}]{HuPhysCanada2016}%
  \BibitemOpen
  \bibfield  {author} {\bibinfo {author} {\bibfnamefont {C.-M.}\ \bibnamefont
  {Hu}},\ }\href@noop {} {\bibfield  {journal} {\bibinfo  {journal} {Phys.
  Canada}\ }\textbf {\bibinfo {volume} {72}},\ \bibinfo {pages} {76} (\bibinfo
  {year} {2016})}\BibitemShut {NoStop}%
\bibitem [{\citenamefont {Aspelmeyer}\ \emph {et~al.}(2014)\citenamefont
  {Aspelmeyer}, \citenamefont {Kippenberg},\ and\ \citenamefont
  {Marquardt}}]{AspelmeyerRMP2014}%
  \BibitemOpen
  \bibfield  {author} {\bibinfo {author} {\bibfnamefont {M.}~\bibnamefont
  {Aspelmeyer}}, \bibinfo {author} {\bibfnamefont {T.~J.}\ \bibnamefont
  {Kippenberg}}, \ and\ \bibinfo {author} {\bibfnamefont {F.}~\bibnamefont
  {Marquardt}},\ }\href {\doibase 10.1103/RevModPhys.86.1391} {\bibfield
  {journal} {\bibinfo  {journal} {Rev. Mod. Phys.}\ }\textbf {\bibinfo {volume}
  {86}},\ \bibinfo {pages} {1391} (\bibinfo {year} {2014})}\BibitemShut
  {NoStop}%
\bibitem [{\citenamefont {Viola~Kusminskiy}()}]{KusminskiyarXiv2019}%
  \BibitemOpen
  \bibfield  {author} {\bibinfo {author} {\bibfnamefont {S.}~\bibnamefont
  {Viola~Kusminskiy}},\ }\href@noop {} {\bibfield  {journal} {\bibinfo
  {journal} {arXiv}\ }}\bibinfo {note} {1911.11104}\BibitemShut {NoStop}%
\bibitem [{\citenamefont {Zhang}\ \emph
  {et~al.}(2016{\natexlab{a}})\citenamefont {Zhang}, \citenamefont {Zou},
  \citenamefont {Jiang},\ and\ \citenamefont {Tang}}]{ZhangScienceAdv2016}%
  \BibitemOpen
  \bibfield  {author} {\bibinfo {author} {\bibfnamefont {X.}~\bibnamefont
  {Zhang}}, \bibinfo {author} {\bibfnamefont {C.-L.}\ \bibnamefont {Zou}},
  \bibinfo {author} {\bibfnamefont {L.}~\bibnamefont {Jiang}}, \ and\ \bibinfo
  {author} {\bibfnamefont {H.~X.}\ \bibnamefont {Tang}},\ }\href {\doibase
  10.1126/sciadv.1501286} {\bibfield  {journal} {\bibinfo  {journal} {Science
  Advances}\ }\textbf {\bibinfo {volume} {2}} (\bibinfo {year}
  {2016}{\natexlab{a}}),\ 10.1126/sciadv.1501286}\BibitemShut {NoStop}%
\bibitem [{\citenamefont {Han}\ \emph {et~al.}(2020)\citenamefont {Han},
  \citenamefont {Fu}, \citenamefont {Zhong}, \citenamefont {Zou}, \citenamefont
  {Xu}, \citenamefont {Al~Sayem}, \citenamefont {Xu}, \citenamefont {Wang},
  \citenamefont {Cheng}, \citenamefont {Jiang},\ and\ \citenamefont
  {Tang}}]{HanNComm2020}%
  \BibitemOpen
  \bibfield  {author} {\bibinfo {author} {\bibfnamefont {X.}~\bibnamefont
  {Han}}, \bibinfo {author} {\bibfnamefont {W.}~\bibnamefont {Fu}}, \bibinfo
  {author} {\bibfnamefont {C.}~\bibnamefont {Zhong}}, \bibinfo {author}
  {\bibfnamefont {C.-L.}\ \bibnamefont {Zou}}, \bibinfo {author} {\bibfnamefont
  {Y.}~\bibnamefont {Xu}}, \bibinfo {author} {\bibfnamefont {A.}~\bibnamefont
  {Al~Sayem}}, \bibinfo {author} {\bibfnamefont {M.}~\bibnamefont {Xu}},
  \bibinfo {author} {\bibfnamefont {S.}~\bibnamefont {Wang}}, \bibinfo {author}
  {\bibfnamefont {R.}~\bibnamefont {Cheng}}, \bibinfo {author} {\bibfnamefont
  {L.}~\bibnamefont {Jiang}}, \ and\ \bibinfo {author} {\bibfnamefont {H.~X.}\
  \bibnamefont {Tang}},\ }\href@noop {} {\bibfield  {journal} {\bibinfo
  {journal} {Nature Commun.}\ }\textbf {\bibinfo {volume} {11}},\ \bibinfo
  {pages} {3237} (\bibinfo {year} {2020})}\BibitemShut {NoStop}%
\bibitem [{\citenamefont {Huebl}\ \emph {et~al.}(2013)\citenamefont {Huebl},
  \citenamefont {Zollitsch}, \citenamefont {Lotze}, \citenamefont {Hocke},
  \citenamefont {Greifenstein}, \citenamefont {Marx}, \citenamefont {Gross},\
  and\ \citenamefont {Goennenwein}}]{HueblPRL2013}%
  \BibitemOpen
  \bibfield  {author} {\bibinfo {author} {\bibfnamefont {H.}~\bibnamefont
  {Huebl}}, \bibinfo {author} {\bibfnamefont {C.~W.}\ \bibnamefont
  {Zollitsch}}, \bibinfo {author} {\bibfnamefont {J.}~\bibnamefont {Lotze}},
  \bibinfo {author} {\bibfnamefont {F.}~\bibnamefont {Hocke}}, \bibinfo
  {author} {\bibfnamefont {M.}~\bibnamefont {Greifenstein}}, \bibinfo {author}
  {\bibfnamefont {A.}~\bibnamefont {Marx}}, \bibinfo {author} {\bibfnamefont
  {R.}~\bibnamefont {Gross}}, \ and\ \bibinfo {author} {\bibfnamefont
  {S.~T.~B.}\ \bibnamefont {Goennenwein}},\ }\href {\doibase
  10.1103/PhysRevLett.111.127003} {\bibfield  {journal} {\bibinfo  {journal}
  {Phys. Rev. Lett.}\ }\textbf {\bibinfo {volume} {111}},\ \bibinfo {pages}
  {127003} (\bibinfo {year} {2013})}\BibitemShut {NoStop}%
\bibitem [{\citenamefont {Tabuchi}\ \emph {et~al.}(2014)\citenamefont
  {Tabuchi}, \citenamefont {Ishino}, \citenamefont {Ishikawa}, \citenamefont
  {Yamazaki}, \citenamefont {Usami},\ and\ \citenamefont
  {Nakamura}}]{TabuchiPRL2014}%
  \BibitemOpen
  \bibfield  {author} {\bibinfo {author} {\bibfnamefont {Y.}~\bibnamefont
  {Tabuchi}}, \bibinfo {author} {\bibfnamefont {S.}~\bibnamefont {Ishino}},
  \bibinfo {author} {\bibfnamefont {T.}~\bibnamefont {Ishikawa}}, \bibinfo
  {author} {\bibfnamefont {R.}~\bibnamefont {Yamazaki}}, \bibinfo {author}
  {\bibfnamefont {K.}~\bibnamefont {Usami}}, \ and\ \bibinfo {author}
  {\bibfnamefont {Y.}~\bibnamefont {Nakamura}},\ }\href {\doibase
  10.1103/PhysRevLett.113.083603} {\bibfield  {journal} {\bibinfo  {journal}
  {Phys. Rev. Lett.}\ }\textbf {\bibinfo {volume} {113}},\ \bibinfo {pages}
  {083603} (\bibinfo {year} {2014})}\BibitemShut {NoStop}%
\bibitem [{\citenamefont {Zhang}\ \emph {et~al.}(2014)\citenamefont {Zhang},
  \citenamefont {Zou}, \citenamefont {Jiang},\ and\ \citenamefont
  {Tang}}]{ZhangPRL2014}%
  \BibitemOpen
  \bibfield  {author} {\bibinfo {author} {\bibfnamefont {X.}~\bibnamefont
  {Zhang}}, \bibinfo {author} {\bibfnamefont {C.-L.}\ \bibnamefont {Zou}},
  \bibinfo {author} {\bibfnamefont {L.}~\bibnamefont {Jiang}}, \ and\ \bibinfo
  {author} {\bibfnamefont {H.~X.}\ \bibnamefont {Tang}},\ }\href {\doibase
  10.1103/PhysRevLett.113.156401} {\bibfield  {journal} {\bibinfo  {journal}
  {Phys. Rev. Lett.}\ }\textbf {\bibinfo {volume} {113}},\ \bibinfo {pages}
  {156401} (\bibinfo {year} {2014})}\BibitemShut {NoStop}%
\bibitem [{\citenamefont {Goryachev}\ \emph {et~al.}(2014)\citenamefont
  {Goryachev}, \citenamefont {Farr}, \citenamefont {Creedon}, \citenamefont
  {Fan}, \citenamefont {Kostylev},\ and\ \citenamefont
  {Tobar}}]{GoryachevPRApplied2014}%
  \BibitemOpen
  \bibfield  {author} {\bibinfo {author} {\bibfnamefont {M.}~\bibnamefont
  {Goryachev}}, \bibinfo {author} {\bibfnamefont {W.~G.}\ \bibnamefont {Farr}},
  \bibinfo {author} {\bibfnamefont {D.~L.}\ \bibnamefont {Creedon}}, \bibinfo
  {author} {\bibfnamefont {Y.}~\bibnamefont {Fan}}, \bibinfo {author}
  {\bibfnamefont {M.}~\bibnamefont {Kostylev}}, \ and\ \bibinfo {author}
  {\bibfnamefont {M.~E.}\ \bibnamefont {Tobar}},\ }\href {\doibase
  10.1103/PhysRevApplied.2.054002} {\bibfield  {journal} {\bibinfo  {journal}
  {Phys. Rev. Applied}\ }\textbf {\bibinfo {volume} {2}},\ \bibinfo {pages}
  {054002} (\bibinfo {year} {2014})}\BibitemShut {NoStop}%
\bibitem [{\citenamefont {Bhoi}\ \emph {et~al.}(2014)\citenamefont {Bhoi},
  \citenamefont {Cliff}, \citenamefont {Maksymov}, \citenamefont {Kostylev},
  \citenamefont {Aiyar}, \citenamefont {Venkataramani}, \citenamefont
  {Prasad},\ and\ \citenamefont {Stamps}}]{BhoiJAP2014}%
  \BibitemOpen
  \bibfield  {author} {\bibinfo {author} {\bibfnamefont {B.}~\bibnamefont
  {Bhoi}}, \bibinfo {author} {\bibfnamefont {T.}~\bibnamefont {Cliff}},
  \bibinfo {author} {\bibfnamefont {I.~S.}\ \bibnamefont {Maksymov}}, \bibinfo
  {author} {\bibfnamefont {M.}~\bibnamefont {Kostylev}}, \bibinfo {author}
  {\bibfnamefont {R.}~\bibnamefont {Aiyar}}, \bibinfo {author} {\bibfnamefont
  {N.}~\bibnamefont {Venkataramani}}, \bibinfo {author} {\bibfnamefont
  {S.}~\bibnamefont {Prasad}}, \ and\ \bibinfo {author} {\bibfnamefont {R.~L.}\
  \bibnamefont {Stamps}},\ }\href {\doibase 10.1063/1.4904857} {\bibfield
  {journal} {\bibinfo  {journal} {J. Appl. Phys.}\ }\textbf {\bibinfo {volume}
  {116}},\ \bibinfo {pages} {243906} (\bibinfo {year} {2014})}\BibitemShut
  {NoStop}%
\bibitem [{\citenamefont {Bai}\ \emph {et~al.}(2015)\citenamefont {Bai},
  \citenamefont {Harder}, \citenamefont {Chen}, \citenamefont {Fan},
  \citenamefont {Xiao},\ and\ \citenamefont {Hu}}]{BaiPRL2015}%
  \BibitemOpen
  \bibfield  {author} {\bibinfo {author} {\bibfnamefont {L.}~\bibnamefont
  {Bai}}, \bibinfo {author} {\bibfnamefont {M.}~\bibnamefont {Harder}},
  \bibinfo {author} {\bibfnamefont {Y.~P.}\ \bibnamefont {Chen}}, \bibinfo
  {author} {\bibfnamefont {X.}~\bibnamefont {Fan}}, \bibinfo {author}
  {\bibfnamefont {J.~Q.}\ \bibnamefont {Xiao}}, \ and\ \bibinfo {author}
  {\bibfnamefont {C.-M.}\ \bibnamefont {Hu}},\ }\href {\doibase
  10.1103/PhysRevLett.114.227201} {\bibfield  {journal} {\bibinfo  {journal}
  {Phys. Rev. Lett.}\ }\textbf {\bibinfo {volume} {114}},\ \bibinfo {pages}
  {227201} (\bibinfo {year} {2015})}\BibitemShut {NoStop}%
\bibitem [{\citenamefont {Li}\ \emph {et~al.}(2019{\natexlab{a}})\citenamefont
  {Li}, \citenamefont {Polakovic}, \citenamefont {Wang}, \citenamefont {Xu},
  \citenamefont {Lendinez}, \citenamefont {Zhang}, \citenamefont {Ding},
  \citenamefont {Khaire}, \citenamefont {Saglam}, \citenamefont {Divan},
  \citenamefont {Pearson}, \citenamefont {Kwok}, \citenamefont {Xiao},
  \citenamefont {Novosad}, \citenamefont {Hoffmann},\ and\ \citenamefont
  {Zhang}}]{LiPRL2019_magnon}%
  \BibitemOpen
  \bibfield  {author} {\bibinfo {author} {\bibfnamefont {Y.}~\bibnamefont
  {Li}}, \bibinfo {author} {\bibfnamefont {T.}~\bibnamefont {Polakovic}},
  \bibinfo {author} {\bibfnamefont {Y.-L.}\ \bibnamefont {Wang}}, \bibinfo
  {author} {\bibfnamefont {J.}~\bibnamefont {Xu}}, \bibinfo {author}
  {\bibfnamefont {S.}~\bibnamefont {Lendinez}}, \bibinfo {author}
  {\bibfnamefont {Z.}~\bibnamefont {Zhang}}, \bibinfo {author} {\bibfnamefont
  {J.}~\bibnamefont {Ding}}, \bibinfo {author} {\bibfnamefont {T.}~\bibnamefont
  {Khaire}}, \bibinfo {author} {\bibfnamefont {H.}~\bibnamefont {Saglam}},
  \bibinfo {author} {\bibfnamefont {R.}~\bibnamefont {Divan}}, \bibinfo
  {author} {\bibfnamefont {J.}~\bibnamefont {Pearson}}, \bibinfo {author}
  {\bibfnamefont {W.-K.}\ \bibnamefont {Kwok}}, \bibinfo {author}
  {\bibfnamefont {Z.}~\bibnamefont {Xiao}}, \bibinfo {author} {\bibfnamefont
  {V.}~\bibnamefont {Novosad}}, \bibinfo {author} {\bibfnamefont
  {A.}~\bibnamefont {Hoffmann}}, \ and\ \bibinfo {author} {\bibfnamefont
  {W.}~\bibnamefont {Zhang}},\ }\href {\doibase 10.1103/PhysRevLett.123.107701}
  {\bibfield  {journal} {\bibinfo  {journal} {Phys. Rev. Lett.}\ }\textbf
  {\bibinfo {volume} {123}},\ \bibinfo {pages} {107701} (\bibinfo {year}
  {2019}{\natexlab{a}})}\BibitemShut {NoStop}%
\bibitem [{\citenamefont {Hou}\ and\ \citenamefont {Liu}(2019)}]{HouPRL2019}%
  \BibitemOpen
  \bibfield  {author} {\bibinfo {author} {\bibfnamefont {J.~T.}\ \bibnamefont
  {Hou}}\ and\ \bibinfo {author} {\bibfnamefont {L.}~\bibnamefont {Liu}},\
  }\href {\doibase 10.1103/PhysRevLett.123.107702} {\bibfield  {journal}
  {\bibinfo  {journal} {Phys. Rev. Lett.}\ }\textbf {\bibinfo {volume} {123}},\
  \bibinfo {pages} {107702} (\bibinfo {year} {2019})}\BibitemShut {NoStop}%
\bibitem [{\citenamefont {Kittel}(1958)}]{KittelPhysRev1958}%
  \BibitemOpen
  \bibfield  {author} {\bibinfo {author} {\bibfnamefont {C.}~\bibnamefont
  {Kittel}},\ }\href@noop {} {\bibfield  {journal} {\bibinfo  {journal} {Phys.
  Rev.}\ }\textbf {\bibinfo {volume} {110}},\ \bibinfo {pages} {836} (\bibinfo
  {year} {1958})}\BibitemShut {NoStop}%
\bibitem [{\citenamefont {Spencer}\ and\ \citenamefont
  {LeCraw}(1958)}]{SpencerPRL1958}%
  \BibitemOpen
  \bibfield  {author} {\bibinfo {author} {\bibfnamefont {E.~G.}\ \bibnamefont
  {Spencer}}\ and\ \bibinfo {author} {\bibfnamefont {R.~C.}\ \bibnamefont
  {LeCraw}},\ }\href {\doibase 10.1103/PhysRevLett.1.241} {\bibfield  {journal}
  {\bibinfo  {journal} {Phys. Rev. Lett.}\ }\textbf {\bibinfo {volume} {1}},\
  \bibinfo {pages} {241} (\bibinfo {year} {1958})}\BibitemShut {NoStop}%
\bibitem [{\citenamefont {B\"ommel}\ and\ \citenamefont
  {Dransfeld}(1959)}]{BommelPRL1959}%
  \BibitemOpen
  \bibfield  {author} {\bibinfo {author} {\bibfnamefont {H.}~\bibnamefont
  {B\"ommel}}\ and\ \bibinfo {author} {\bibfnamefont {K.}~\bibnamefont
  {Dransfeld}},\ }\href@noop {} {\bibfield  {journal} {\bibinfo  {journal}
  {Phys. Rev. Lett.}\ }\textbf {\bibinfo {volume} {3}},\ \bibinfo {pages} {83}
  (\bibinfo {year} {1959})}\BibitemShut {NoStop}%
\bibitem [{\citenamefont {Schl\"{o}mann}(1960)}]{SchlomannJAP1960}%
  \BibitemOpen
  \bibfield  {author} {\bibinfo {author} {\bibfnamefont {E.}~\bibnamefont
  {Schl\"{o}mann}},\ }\href@noop {} {\bibfield  {journal} {\bibinfo  {journal}
  {J. Appl. Phys.}\ }\textbf {\bibinfo {volume} {31}},\ \bibinfo {pages} {1647}
  (\bibinfo {year} {1960})}\BibitemShut {NoStop}%
\bibitem [{\citenamefont {Pomerantz}(1961)}]{PomerantzPRL1961}%
  \BibitemOpen
  \bibfield  {author} {\bibinfo {author} {\bibfnamefont {M.}~\bibnamefont
  {Pomerantz}},\ }\href {\doibase 10.1103/PhysRevLett.7.312} {\bibfield
  {journal} {\bibinfo  {journal} {Phys. Rev. Lett.}\ }\textbf {\bibinfo
  {volume} {7}},\ \bibinfo {pages} {312} (\bibinfo {year} {1961})}\BibitemShut
  {NoStop}%
\bibitem [{\citenamefont {Matthews}\ and\ \citenamefont
  {LeCraw}(1962)}]{MatthewsPRL1962}%
  \BibitemOpen
  \bibfield  {author} {\bibinfo {author} {\bibfnamefont {H.}~\bibnamefont
  {Matthews}}\ and\ \bibinfo {author} {\bibfnamefont {R.~C.}\ \bibnamefont
  {LeCraw}},\ }\href {\doibase 10.1103/PhysRevLett.8.397} {\bibfield  {journal}
  {\bibinfo  {journal} {Phys. Rev. Lett.}\ }\textbf {\bibinfo {volume} {8}},\
  \bibinfo {pages} {397} (\bibinfo {year} {1962})}\BibitemShut {NoStop}%
\bibitem [{\citenamefont {Schl\"{o}mann}\ and\ \citenamefont
  {Joseph}(1964)}]{SchlomannJAP1964}%
  \BibitemOpen
  \bibfield  {author} {\bibinfo {author} {\bibfnamefont {E.}~\bibnamefont
  {Schl\"{o}mann}}\ and\ \bibinfo {author} {\bibfnamefont {R.~I.}\ \bibnamefont
  {Joseph}},\ }\href@noop {} {\bibfield  {journal} {\bibinfo  {journal} {J.
  Appl. Phys.}\ }\textbf {\bibinfo {volume} {35}},\ \bibinfo {pages} {2382}
  (\bibinfo {year} {1964})}\BibitemShut {NoStop}%
\bibitem [{\citenamefont {Seavey}(1965)}]{SeaveyProcIEEE1965}%
  \BibitemOpen
  \bibfield  {author} {\bibinfo {author} {\bibfnamefont {M.}~\bibnamefont
  {Seavey}},\ }\href@noop {} {\bibfield  {journal} {\bibinfo  {journal} {Proc.
  IEEE}\ }\textbf {\bibinfo {volume} {53}},\ \bibinfo {pages} {1387} (\bibinfo
  {year} {1965})}\BibitemShut {NoStop}%
\bibitem [{\citenamefont {Comstock}(1965)}]{ComstockProcIEEE1965}%
  \BibitemOpen
  \bibfield  {author} {\bibinfo {author} {\bibfnamefont {R.~L.}\ \bibnamefont
  {Comstock}},\ }\href@noop {} {\bibfield  {journal} {\bibinfo  {journal}
  {Proc. IEEE}\ }\textbf {\bibinfo {volume} {53}},\ \bibinfo {pages} {1508}
  (\bibinfo {year} {1965})}\BibitemShut {NoStop}%
\bibitem [{\citenamefont {Kobayashi}\ \emph {et~al.}(1973)\citenamefont
  {Kobayashi}, \citenamefont {Barker},\ and\ \citenamefont
  {Yelon}}]{KobayashiPRB1973}%
  \BibitemOpen
  \bibfield  {author} {\bibinfo {author} {\bibfnamefont {T.}~\bibnamefont
  {Kobayashi}}, \bibinfo {author} {\bibfnamefont {R.~C.}\ \bibnamefont
  {Barker}}, \ and\ \bibinfo {author} {\bibfnamefont {A.}~\bibnamefont
  {Yelon}},\ }\href@noop {} {\bibfield  {journal} {\bibinfo  {journal} {Phys.
  Rev. B}\ }\textbf {\bibinfo {volume} {7}},\ \bibinfo {pages} {3286} (\bibinfo
  {year} {1973})}\BibitemShut {NoStop}%
\bibitem [{\citenamefont {Belyaeva}\ \emph {et~al.}(1992)\citenamefont
  {Belyaeva}, \citenamefont {Karpachev},\ and\ \citenamefont
  {Zarembo}}]{BelyaevaSov1992}%
  \BibitemOpen
  \bibfield  {author} {\bibinfo {author} {\bibfnamefont {O.~Y.}\ \bibnamefont
  {Belyaeva}}, \bibinfo {author} {\bibfnamefont {S.~N.}\ \bibnamefont
  {Karpachev}}, \ and\ \bibinfo {author} {\bibfnamefont {L.~K.}\ \bibnamefont
  {Zarembo}},\ }\href@noop {} {\bibfield  {journal} {\bibinfo  {journal} {Sov.
  Phys. Usp.}\ }\textbf {\bibinfo {volume} {35}},\ \bibinfo {pages} {106}
  (\bibinfo {year} {1992})}\BibitemShut {NoStop}%
\bibitem [{\citenamefont {Weiler}\ \emph {et~al.}(2011)\citenamefont {Weiler},
  \citenamefont {Dreher}, \citenamefont {Heeg}, \citenamefont {Huebl},
  \citenamefont {Gross}, \citenamefont {Brandt},\ and\ \citenamefont
  {Goennenwein}}]{WeilerPRL2011}%
  \BibitemOpen
  \bibfield  {author} {\bibinfo {author} {\bibfnamefont {M.}~\bibnamefont
  {Weiler}}, \bibinfo {author} {\bibfnamefont {L.}~\bibnamefont {Dreher}},
  \bibinfo {author} {\bibfnamefont {C.}~\bibnamefont {Heeg}}, \bibinfo {author}
  {\bibfnamefont {H.}~\bibnamefont {Huebl}}, \bibinfo {author} {\bibfnamefont
  {R.}~\bibnamefont {Gross}}, \bibinfo {author} {\bibfnamefont {M.~S.}\
  \bibnamefont {Brandt}}, \ and\ \bibinfo {author} {\bibfnamefont {S.~T.~B.}\
  \bibnamefont {Goennenwein}},\ }\href@noop {} {\bibfield  {journal} {\bibinfo
  {journal} {Phys. Rev. Lett.}\ }\textbf {\bibinfo {volume} {106}},\ \bibinfo
  {pages} {117601} (\bibinfo {year} {2011})}\BibitemShut {NoStop}%
\bibitem [{\citenamefont {Dreher}\ \emph {et~al.}(2012)\citenamefont {Dreher},
  \citenamefont {Weiler}, \citenamefont {Pernpeintner}, \citenamefont {Huebl},
  \citenamefont {Gross}, \citenamefont {Brandt},\ and\ \citenamefont
  {Goennenwein}}]{DreherPRB2012}%
  \BibitemOpen
  \bibfield  {author} {\bibinfo {author} {\bibfnamefont {L.}~\bibnamefont
  {Dreher}}, \bibinfo {author} {\bibfnamefont {M.}~\bibnamefont {Weiler}},
  \bibinfo {author} {\bibfnamefont {M.}~\bibnamefont {Pernpeintner}}, \bibinfo
  {author} {\bibfnamefont {H.}~\bibnamefont {Huebl}}, \bibinfo {author}
  {\bibfnamefont {R.}~\bibnamefont {Gross}}, \bibinfo {author} {\bibfnamefont
  {M.~S.}\ \bibnamefont {Brandt}}, \ and\ \bibinfo {author} {\bibfnamefont
  {S.~T.~B.}\ \bibnamefont {Goennenwein}},\ }\href@noop {} {\bibfield
  {journal} {\bibinfo  {journal} {Phys. Rev. B}\ }\textbf {\bibinfo {volume}
  {86}},\ \bibinfo {pages} {134415} (\bibinfo {year} {2012})}\BibitemShut
  {NoStop}%
\bibitem [{\citenamefont {Chang}\ \emph {et~al.}(2018)\citenamefont {Chang},
  \citenamefont {Mieszczak}, \citenamefont {Zelent}, \citenamefont {Besse},
  \citenamefont {Martens}, \citenamefont {Tamming}, \citenamefont {Janusonis},
  \citenamefont {Graczyk}, \citenamefont {M\"unzenberg}, \citenamefont
  {K\l{}os},\ and\ \citenamefont {Tobey}}]{ChangPRApplied2018}%
  \BibitemOpen
  \bibfield  {author} {\bibinfo {author} {\bibfnamefont {C.~L.}\ \bibnamefont
  {Chang}}, \bibinfo {author} {\bibfnamefont {S.}~\bibnamefont {Mieszczak}},
  \bibinfo {author} {\bibfnamefont {M.}~\bibnamefont {Zelent}}, \bibinfo
  {author} {\bibfnamefont {V.}~\bibnamefont {Besse}}, \bibinfo {author}
  {\bibfnamefont {U.}~\bibnamefont {Martens}}, \bibinfo {author} {\bibfnamefont
  {R.}~\bibnamefont {Tamming}}, \bibinfo {author} {\bibfnamefont
  {J.}~\bibnamefont {Janusonis}}, \bibinfo {author} {\bibfnamefont
  {P.}~\bibnamefont {Graczyk}}, \bibinfo {author} {\bibfnamefont
  {M.}~\bibnamefont {M\"unzenberg}}, \bibinfo {author} {\bibfnamefont
  {J.}~\bibnamefont {K\l{}os}}, \ and\ \bibinfo {author} {\bibfnamefont
  {R.~I.}\ \bibnamefont {Tobey}},\ }\href {\doibase
  10.1103/PhysRevApplied.10.064051} {\bibfield  {journal} {\bibinfo  {journal}
  {Phys. Rev. Applied}\ }\textbf {\bibinfo {volume} {10}},\ \bibinfo {pages}
  {064051} (\bibinfo {year} {2018})}\BibitemShut {NoStop}%
\bibitem [{\citenamefont {Lisenkov}\ \emph {et~al.}(2019)\citenamefont
  {Lisenkov}, \citenamefont {Jander},\ and\ \citenamefont
  {Dhagat}}]{LisenkovPRB2019}%
  \BibitemOpen
  \bibfield  {author} {\bibinfo {author} {\bibfnamefont {I.}~\bibnamefont
  {Lisenkov}}, \bibinfo {author} {\bibfnamefont {A.}~\bibnamefont {Jander}}, \
  and\ \bibinfo {author} {\bibfnamefont {P.}~\bibnamefont {Dhagat}},\ }\href
  {\doibase 10.1103/PhysRevB.99.184433} {\bibfield  {journal} {\bibinfo
  {journal} {Phys. Rev. B}\ }\textbf {\bibinfo {volume} {99}},\ \bibinfo
  {pages} {184433} (\bibinfo {year} {2019})}\BibitemShut {NoStop}%
\bibitem [{\citenamefont {R\"uckriegel}\ \emph {et~al.}(2014)\citenamefont
  {R\"uckriegel}, \citenamefont {Kopietz}, \citenamefont {Bozhko},
  \citenamefont {Serga},\ and\ \citenamefont
  {Hillebrands}}]{RuckriegelPRB2014}%
  \BibitemOpen
  \bibfield  {author} {\bibinfo {author} {\bibfnamefont {A.}~\bibnamefont
  {R\"uckriegel}}, \bibinfo {author} {\bibfnamefont {P.}~\bibnamefont
  {Kopietz}}, \bibinfo {author} {\bibfnamefont {D.~A.}\ \bibnamefont {Bozhko}},
  \bibinfo {author} {\bibfnamefont {A.~A.}\ \bibnamefont {Serga}}, \ and\
  \bibinfo {author} {\bibfnamefont {B.}~\bibnamefont {Hillebrands}},\ }\href
  {\doibase 10.1103/PhysRevB.89.184413} {\bibfield  {journal} {\bibinfo
  {journal} {Phys. Rev. B}\ }\textbf {\bibinfo {volume} {89}},\ \bibinfo
  {pages} {184413} (\bibinfo {year} {2014})}\BibitemShut {NoStop}%
\bibitem [{\citenamefont {Li}\ \emph {et~al.}(2017)\citenamefont {Li},
  \citenamefont {Labanowski}, \citenamefont {Salahuddin},\ and\ \citenamefont
  {Lynch}}]{LiXJAP2017}%
  \BibitemOpen
  \bibfield  {author} {\bibinfo {author} {\bibfnamefont {X.}~\bibnamefont
  {Li}}, \bibinfo {author} {\bibfnamefont {D.}~\bibnamefont {Labanowski}},
  \bibinfo {author} {\bibfnamefont {S.}~\bibnamefont {Salahuddin}}, \ and\
  \bibinfo {author} {\bibfnamefont {C.~S.}\ \bibnamefont {Lynch}},\ }\href@noop
  {} {\bibfield  {journal} {\bibinfo  {journal} {J. Appl. Phys.}\ }\textbf
  {\bibinfo {volume} {122}},\ \bibinfo {pages} {043904} (\bibinfo {year}
  {2017})}\BibitemShut {NoStop}%
\bibitem [{\citenamefont {Puebla}\ \emph {et~al.}(2020)\citenamefont {Puebla},
  \citenamefont {Xu}, \citenamefont {Rana}, \citenamefont {Yamamoto},
  \citenamefont {Maekawa},\ and\ \citenamefont {Otani}}]{PueblaJPD2020}%
  \BibitemOpen
  \bibfield  {author} {\bibinfo {author} {\bibfnamefont {J.}~\bibnamefont
  {Puebla}}, \bibinfo {author} {\bibfnamefont {M.}~\bibnamefont {Xu}}, \bibinfo
  {author} {\bibfnamefont {B.}~\bibnamefont {Rana}}, \bibinfo {author}
  {\bibfnamefont {K.}~\bibnamefont {Yamamoto}}, \bibinfo {author}
  {\bibfnamefont {S.}~\bibnamefont {Maekawa}}, \ and\ \bibinfo {author}
  {\bibfnamefont {Y.}~\bibnamefont {Otani}},\ }\href@noop {} {\bibfield
  {journal} {\bibinfo  {journal} {J. Phys. D: Appl. Phys.}\ }\textbf {\bibinfo
  {volume} {53}},\ \bibinfo {pages} {264002} (\bibinfo {year}
  {2020})}\BibitemShut {NoStop}%
\bibitem [{\citenamefont {Hick}\ \emph {et~al.}(2012)\citenamefont {Hick},
  \citenamefont {Kloss},\ and\ \citenamefont {Kopietz}}]{HickPRB2012}%
  \BibitemOpen
  \bibfield  {author} {\bibinfo {author} {\bibfnamefont {J.}~\bibnamefont
  {Hick}}, \bibinfo {author} {\bibfnamefont {T.}~\bibnamefont {Kloss}}, \ and\
  \bibinfo {author} {\bibfnamefont {P.}~\bibnamefont {Kopietz}},\ }\href
  {\doibase 10.1103/PhysRevB.86.184417} {\bibfield  {journal} {\bibinfo
  {journal} {Phys. Rev. B}\ }\textbf {\bibinfo {volume} {86}},\ \bibinfo
  {pages} {184417} (\bibinfo {year} {2012})}\BibitemShut {NoStop}%
\bibitem [{\citenamefont {Bozhko}\ \emph {et~al.}(2017)\citenamefont {Bozhko},
  \citenamefont {Clausen}, \citenamefont {Melkov}, \citenamefont {L'vov},
  \citenamefont {Pomyalov}, \citenamefont {Vasyuchka}, \citenamefont {Chumak},
  \citenamefont {Hillebrands},\ and\ \citenamefont {Serga}}]{BozhkoPRL2017}%
  \BibitemOpen
  \bibfield  {author} {\bibinfo {author} {\bibfnamefont {D.~A.}\ \bibnamefont
  {Bozhko}}, \bibinfo {author} {\bibfnamefont {P.}~\bibnamefont {Clausen}},
  \bibinfo {author} {\bibfnamefont {G.~A.}\ \bibnamefont {Melkov}}, \bibinfo
  {author} {\bibfnamefont {V.~S.}\ \bibnamefont {L'vov}}, \bibinfo {author}
  {\bibfnamefont {A.}~\bibnamefont {Pomyalov}}, \bibinfo {author}
  {\bibfnamefont {V.~I.}\ \bibnamefont {Vasyuchka}}, \bibinfo {author}
  {\bibfnamefont {A.~V.}\ \bibnamefont {Chumak}}, \bibinfo {author}
  {\bibfnamefont {B.}~\bibnamefont {Hillebrands}}, \ and\ \bibinfo {author}
  {\bibfnamefont {A.~A.}\ \bibnamefont {Serga}},\ }\href {\doibase
  10.1103/PhysRevLett.118.237201} {\bibfield  {journal} {\bibinfo  {journal}
  {Phys. Rev. Lett.}\ }\textbf {\bibinfo {volume} {118}},\ \bibinfo {pages}
  {237201} (\bibinfo {year} {2017})}\BibitemShut {NoStop}%
\bibitem [{\citenamefont {Schneider}\ \emph {et~al.}(2020)\citenamefont
  {Schneider}, \citenamefont {Br\"{a}cher}, \citenamefont {Breitbach},
  \citenamefont {Lauer}, \citenamefont {Pirro}, \citenamefont {Bozhko},
  \citenamefont {Musiienko-Shmarova}, \citenamefont {Heinz}, \citenamefont
  {Wang}, \citenamefont {Meyer}, \citenamefont {Heussner}, \citenamefont
  {Keller}, \citenamefont {Papaioannou}, \citenamefont {L\"{a}gel},
  \citenamefont {L\"{o}ber}, \citenamefont {Dubs}, \citenamefont {Slavin},
  \citenamefont {Tiberkevich}, \citenamefont {Serga}, \citenamefont
  {Hillebrands},\ and\ \citenamefont {Chumak}}]{SchneiderNNano2020}%
  \BibitemOpen
  \bibfield  {author} {\bibinfo {author} {\bibfnamefont {M.}~\bibnamefont
  {Schneider}}, \bibinfo {author} {\bibfnamefont {T.}~\bibnamefont
  {Br\"{a}cher}}, \bibinfo {author} {\bibfnamefont {D.}~\bibnamefont
  {Breitbach}}, \bibinfo {author} {\bibfnamefont {V.}~\bibnamefont {Lauer}},
  \bibinfo {author} {\bibfnamefont {P.}~\bibnamefont {Pirro}}, \bibinfo
  {author} {\bibfnamefont {D.~A.}\ \bibnamefont {Bozhko}}, \bibinfo {author}
  {\bibfnamefont {H.~Y.}\ \bibnamefont {Musiienko-Shmarova}}, \bibinfo {author}
  {\bibfnamefont {B.}~\bibnamefont {Heinz}}, \bibinfo {author} {\bibfnamefont
  {Q.}~\bibnamefont {Wang}}, \bibinfo {author} {\bibfnamefont {T.}~\bibnamefont
  {Meyer}}, \bibinfo {author} {\bibfnamefont {F.}~\bibnamefont {Heussner}},
  \bibinfo {author} {\bibfnamefont {S.}~\bibnamefont {Keller}}, \bibinfo
  {author} {\bibfnamefont {E.~T.}\ \bibnamefont {Papaioannou}}, \bibinfo
  {author} {\bibfnamefont {B.}~\bibnamefont {L\"{a}gel}}, \bibinfo {author}
  {\bibfnamefont {T.}~\bibnamefont {L\"{o}ber}}, \bibinfo {author}
  {\bibfnamefont {C.}~\bibnamefont {Dubs}}, \bibinfo {author} {\bibfnamefont
  {A.~N.}\ \bibnamefont {Slavin}}, \bibinfo {author} {\bibfnamefont {V.~S.}\
  \bibnamefont {Tiberkevich}}, \bibinfo {author} {\bibfnamefont {A.~A.}\
  \bibnamefont {Serga}}, \bibinfo {author} {\bibfnamefont {B.}~\bibnamefont
  {Hillebrands}}, \ and\ \bibinfo {author} {\bibfnamefont {A.~V.}\ \bibnamefont
  {Chumak}},\ }\href@noop {} {\bibfield  {journal} {\bibinfo  {journal} {Nature
  Nanotech.}\ }\textbf {\bibinfo {volume} {15}},\ \bibinfo {pages} {457}
  (\bibinfo {year} {2020})}\BibitemShut {NoStop}%
\bibitem [{\citenamefont {Sasaki}\ \emph {et~al.}(2017)\citenamefont {Sasaki},
  \citenamefont {Nii}, \citenamefont {Iguchi},\ and\ \citenamefont
  {Onose}}]{SasakiPRB2017}%
  \BibitemOpen
  \bibfield  {author} {\bibinfo {author} {\bibfnamefont {R.}~\bibnamefont
  {Sasaki}}, \bibinfo {author} {\bibfnamefont {Y.}~\bibnamefont {Nii}},
  \bibinfo {author} {\bibfnamefont {Y.}~\bibnamefont {Iguchi}}, \ and\ \bibinfo
  {author} {\bibfnamefont {Y.}~\bibnamefont {Onose}},\ }\href@noop {}
  {\bibfield  {journal} {\bibinfo  {journal} {Phys. Rev. B}\ }\textbf {\bibinfo
  {volume} {95}},\ \bibinfo {pages} {020407} (\bibinfo {year}
  {2017})}\BibitemShut {NoStop}%
\bibitem [{\citenamefont {Holanda}\ \emph {et~al.}(2018)\citenamefont
  {Holanda}, \citenamefont {Maior}, \citenamefont {Azevedo},\ and\
  \citenamefont {Rezende}}]{HolandaNPhys2018}%
  \BibitemOpen
  \bibfield  {author} {\bibinfo {author} {\bibfnamefont {J.}~\bibnamefont
  {Holanda}}, \bibinfo {author} {\bibfnamefont {D.~S.}\ \bibnamefont {Maior}},
  \bibinfo {author} {\bibfnamefont {A.}~\bibnamefont {Azevedo}}, \ and\
  \bibinfo {author} {\bibfnamefont {S.~M.}\ \bibnamefont {Rezende}},\
  }\href@noop {} {\bibfield  {journal} {\bibinfo  {journal} {Nature Phys.}\
  }\textbf {\bibinfo {volume} {14}},\ \bibinfo {pages} {500} (\bibinfo {year}
  {2018})}\BibitemShut {NoStop}%
\bibitem [{\citenamefont {Bozhko}\ \emph {et~al.}(2020)\citenamefont {Bozhko},
  \citenamefont {Vasyuchka}, \citenamefont {Chumak},\ and\ \citenamefont
  {Serga}}]{BozhkoLTPhys2020}%
  \BibitemOpen
  \bibfield  {author} {\bibinfo {author} {\bibfnamefont {D.~A.}\ \bibnamefont
  {Bozhko}}, \bibinfo {author} {\bibfnamefont {V.~I.}\ \bibnamefont
  {Vasyuchka}}, \bibinfo {author} {\bibfnamefont {A.~V.}\ \bibnamefont
  {Chumak}}, \ and\ \bibinfo {author} {\bibfnamefont {A.~A.}\ \bibnamefont
  {Serga}},\ }\href@noop {} {\bibfield  {journal} {\bibinfo  {journal} {Low
  Temp. Phys.}\ }\textbf {\bibinfo {volume} {46}},\ \bibinfo {pages} {383}
  (\bibinfo {year} {2020})}\BibitemShut {NoStop}%
\bibitem [{\citenamefont {Kikkawa}\ \emph {et~al.}(2016)\citenamefont
  {Kikkawa}, \citenamefont {Shen}, \citenamefont {Flebus}, \citenamefont
  {Duine}, \citenamefont {Uchida}, \citenamefont {Qiu}, \citenamefont {Bauer},\
  and\ \citenamefont {Saitoh}}]{KikkawaPRL2016}%
  \BibitemOpen
  \bibfield  {author} {\bibinfo {author} {\bibfnamefont {T.}~\bibnamefont
  {Kikkawa}}, \bibinfo {author} {\bibfnamefont {K.}~\bibnamefont {Shen}},
  \bibinfo {author} {\bibfnamefont {B.}~\bibnamefont {Flebus}}, \bibinfo
  {author} {\bibfnamefont {R.~A.}\ \bibnamefont {Duine}}, \bibinfo {author}
  {\bibfnamefont {K.-i.}\ \bibnamefont {Uchida}}, \bibinfo {author}
  {\bibfnamefont {Z.}~\bibnamefont {Qiu}}, \bibinfo {author} {\bibfnamefont
  {G.~E.~W.}\ \bibnamefont {Bauer}}, \ and\ \bibinfo {author} {\bibfnamefont
  {E.}~\bibnamefont {Saitoh}},\ }\href {\doibase
  10.1103/PhysRevLett.117.207203} {\bibfield  {journal} {\bibinfo  {journal}
  {Phys. Rev. Lett.}\ }\textbf {\bibinfo {volume} {117}},\ \bibinfo {pages}
  {207203} (\bibinfo {year} {2016})}\BibitemShut {NoStop}%
\bibitem [{\citenamefont {Hayashi}\ and\ \citenamefont
  {Ando}(2018)}]{HayashiPRL2018}%
  \BibitemOpen
  \bibfield  {author} {\bibinfo {author} {\bibfnamefont {H.}~\bibnamefont
  {Hayashi}}\ and\ \bibinfo {author} {\bibfnamefont {K.}~\bibnamefont {Ando}},\
  }\href {\doibase 10.1103/PhysRevLett.121.237202} {\bibfield  {journal}
  {\bibinfo  {journal} {Phys. Rev. Lett.}\ }\textbf {\bibinfo {volume} {121}},\
  \bibinfo {pages} {237202} (\bibinfo {year} {2018})}\BibitemShut {NoStop}%
\bibitem [{\citenamefont {Berk}\ \emph {et~al.}(2019)\citenamefont {Berk},
  \citenamefont {Jaris}, \citenamefont {Yang}, \citenamefont {Dhuey},
  \citenamefont {Cabrini},\ and\ \citenamefont {Schmidt}}]{BerkNComm2019}%
  \BibitemOpen
  \bibfield  {author} {\bibinfo {author} {\bibfnamefont {C.}~\bibnamefont
  {Berk}}, \bibinfo {author} {\bibfnamefont {M.}~\bibnamefont {Jaris}},
  \bibinfo {author} {\bibfnamefont {W.}~\bibnamefont {Yang}}, \bibinfo {author}
  {\bibfnamefont {S.}~\bibnamefont {Dhuey}}, \bibinfo {author} {\bibfnamefont
  {S.}~\bibnamefont {Cabrini}}, \ and\ \bibinfo {author} {\bibfnamefont
  {H.}~\bibnamefont {Schmidt}},\ }\href@noop {} {\bibfield  {journal} {\bibinfo
   {journal} {Nature Commun.}\ }\textbf {\bibinfo {volume} {10}},\ \bibinfo
  {pages} {2652} (\bibinfo {year} {2019})}\BibitemShut {NoStop}%
\bibitem [{\citenamefont {An}\ \emph {et~al.}(2020)\citenamefont {An},
  \citenamefont {Litvinenko}, \citenamefont {Kohno}, \citenamefont {Fuad},
  \citenamefont {Naletov}, \citenamefont {Vila}, \citenamefont {Ebels},
  \citenamefont {de~Loubens}, \citenamefont {Hurdequint}, \citenamefont
  {Beaulieu}, \citenamefont {Ben~Youssef}, \citenamefont {Vukadinovic},
  \citenamefont {Bauer}, \citenamefont {Slavin}, \citenamefont {Tiberkevich},\
  and\ \citenamefont {Klein}}]{AnPRB2020}%
  \BibitemOpen
  \bibfield  {author} {\bibinfo {author} {\bibfnamefont {K.}~\bibnamefont
  {An}}, \bibinfo {author} {\bibfnamefont {A.~N.}\ \bibnamefont {Litvinenko}},
  \bibinfo {author} {\bibfnamefont {R.}~\bibnamefont {Kohno}}, \bibinfo
  {author} {\bibfnamefont {A.~A.}\ \bibnamefont {Fuad}}, \bibinfo {author}
  {\bibfnamefont {V.~V.}\ \bibnamefont {Naletov}}, \bibinfo {author}
  {\bibfnamefont {L.}~\bibnamefont {Vila}}, \bibinfo {author} {\bibfnamefont
  {U.}~\bibnamefont {Ebels}}, \bibinfo {author} {\bibfnamefont
  {G.}~\bibnamefont {de~Loubens}}, \bibinfo {author} {\bibfnamefont
  {H.}~\bibnamefont {Hurdequint}}, \bibinfo {author} {\bibfnamefont
  {N.}~\bibnamefont {Beaulieu}}, \bibinfo {author} {\bibfnamefont
  {J.}~\bibnamefont {Ben~Youssef}}, \bibinfo {author} {\bibfnamefont
  {N.}~\bibnamefont {Vukadinovic}}, \bibinfo {author} {\bibfnamefont
  {G.~E.~W.}\ \bibnamefont {Bauer}}, \bibinfo {author} {\bibfnamefont {A.~N.}\
  \bibnamefont {Slavin}}, \bibinfo {author} {\bibfnamefont {V.~S.}\
  \bibnamefont {Tiberkevich}}, \ and\ \bibinfo {author} {\bibfnamefont
  {O.}~\bibnamefont {Klein}},\ }\href {\doibase 10.1103/PhysRevB.101.060407}
  {\bibfield  {journal} {\bibinfo  {journal} {Phys. Rev. B}\ }\textbf {\bibinfo
  {volume} {101}},\ \bibinfo {pages} {060407} (\bibinfo {year}
  {2020})}\BibitemShut {NoStop}%
\bibitem [{\citenamefont {Yahiro}\ \emph {et~al.}(2020)\citenamefont {Yahiro},
  \citenamefont {Kikkawa}, \citenamefont {Ramos}, \citenamefont {Oyanagi},
  \citenamefont {Hioki}, \citenamefont {Daimon},\ and\ \citenamefont
  {Saitoh}}]{YahiroPRB2020}%
  \BibitemOpen
  \bibfield  {author} {\bibinfo {author} {\bibfnamefont {R.}~\bibnamefont
  {Yahiro}}, \bibinfo {author} {\bibfnamefont {T.}~\bibnamefont {Kikkawa}},
  \bibinfo {author} {\bibfnamefont {R.}~\bibnamefont {Ramos}}, \bibinfo
  {author} {\bibfnamefont {K.}~\bibnamefont {Oyanagi}}, \bibinfo {author}
  {\bibfnamefont {T.}~\bibnamefont {Hioki}}, \bibinfo {author} {\bibfnamefont
  {S.}~\bibnamefont {Daimon}}, \ and\ \bibinfo {author} {\bibfnamefont
  {E.}~\bibnamefont {Saitoh}},\ }\href {\doibase 10.1103/PhysRevB.101.024407}
  {\bibfield  {journal} {\bibinfo  {journal} {Phys. Rev. B}\ }\textbf {\bibinfo
  {volume} {101}},\ \bibinfo {pages} {024407} (\bibinfo {year}
  {2020})}\BibitemShut {NoStop}%
\bibitem [{\citenamefont {Godejohann}\ \emph {et~al.}(2020)\citenamefont
  {Godejohann}, \citenamefont {Scherbakov}, \citenamefont {Kukhtaruk},
  \citenamefont {Poddubny}, \citenamefont {Yaremkevich}, \citenamefont {Wang},
  \citenamefont {Nadzeyka}, \citenamefont {Yakovlev}, \citenamefont
  {Rushforth}, \citenamefont {Akimov},\ and\ \citenamefont
  {Bayer}}]{GodejohannPRB2020}%
  \BibitemOpen
  \bibfield  {author} {\bibinfo {author} {\bibfnamefont {F.}~\bibnamefont
  {Godejohann}}, \bibinfo {author} {\bibfnamefont {A.~V.}\ \bibnamefont
  {Scherbakov}}, \bibinfo {author} {\bibfnamefont {S.~M.}\ \bibnamefont
  {Kukhtaruk}}, \bibinfo {author} {\bibfnamefont {A.~N.}\ \bibnamefont
  {Poddubny}}, \bibinfo {author} {\bibfnamefont {D.~D.}\ \bibnamefont
  {Yaremkevich}}, \bibinfo {author} {\bibfnamefont {M.}~\bibnamefont {Wang}},
  \bibinfo {author} {\bibfnamefont {A.}~\bibnamefont {Nadzeyka}}, \bibinfo
  {author} {\bibfnamefont {D.~R.}\ \bibnamefont {Yakovlev}}, \bibinfo {author}
  {\bibfnamefont {A.~W.}\ \bibnamefont {Rushforth}}, \bibinfo {author}
  {\bibfnamefont {A.~V.}\ \bibnamefont {Akimov}}, \ and\ \bibinfo {author}
  {\bibfnamefont {M.}~\bibnamefont {Bayer}},\ }\href {\doibase
  10.1103/PhysRevB.102.144438} {\bibfield  {journal} {\bibinfo  {journal}
  {Phys. Rev. B}\ }\textbf {\bibinfo {volume} {102}},\ \bibinfo {pages}
  {144438} (\bibinfo {year} {2020})}\BibitemShut {NoStop}%
\bibitem [{\citenamefont {Tabuchi}\ \emph {et~al.}(2015)\citenamefont
  {Tabuchi}, \citenamefont {Ishino}, \citenamefont {Noguchi}, \citenamefont
  {Ishikawa}, \citenamefont {Yamazaki}, \citenamefont {Usami},\ and\
  \citenamefont {Nakamura}}]{TabuchiScience2015}%
  \BibitemOpen
  \bibfield  {author} {\bibinfo {author} {\bibfnamefont {Y.}~\bibnamefont
  {Tabuchi}}, \bibinfo {author} {\bibfnamefont {S.}~\bibnamefont {Ishino}},
  \bibinfo {author} {\bibfnamefont {A.}~\bibnamefont {Noguchi}}, \bibinfo
  {author} {\bibfnamefont {T.}~\bibnamefont {Ishikawa}}, \bibinfo {author}
  {\bibfnamefont {R.}~\bibnamefont {Yamazaki}}, \bibinfo {author}
  {\bibfnamefont {K.}~\bibnamefont {Usami}}, \ and\ \bibinfo {author}
  {\bibfnamefont {Y.}~\bibnamefont {Nakamura}},\ }\href@noop {} {\bibfield
  {journal} {\bibinfo  {journal} {Science}\ }\textbf {\bibinfo {volume}
  {349}},\ \bibinfo {pages} {405} (\bibinfo {year} {2015})}\BibitemShut
  {NoStop}%
\bibitem [{\citenamefont {Lachance-Quirion}\ \emph {et~al.}(2017)\citenamefont
  {Lachance-Quirion}, \citenamefont {Tabuchi}, \citenamefont {Ishino},
  \citenamefont {Noguchi}, \citenamefont {Ishikawa}, \citenamefont {Yamazaki},\
  and\ \citenamefont {Nakamura}}]{LachanceScienceAdvan2017}%
  \BibitemOpen
  \bibfield  {author} {\bibinfo {author} {\bibfnamefont {D.}~\bibnamefont
  {Lachance-Quirion}}, \bibinfo {author} {\bibfnamefont {Y.}~\bibnamefont
  {Tabuchi}}, \bibinfo {author} {\bibfnamefont {S.}~\bibnamefont {Ishino}},
  \bibinfo {author} {\bibfnamefont {A.}~\bibnamefont {Noguchi}}, \bibinfo
  {author} {\bibfnamefont {T.}~\bibnamefont {Ishikawa}}, \bibinfo {author}
  {\bibfnamefont {R.}~\bibnamefont {Yamazaki}}, \ and\ \bibinfo {author}
  {\bibfnamefont {Y.}~\bibnamefont {Nakamura}},\ }\href {\doibase
  10.1126/sciadv.1603150} {\bibfield  {journal} {\bibinfo  {journal} {Science
  Advances}\ }\textbf {\bibinfo {volume} {3}} (\bibinfo {year} {2017}),\
  10.1126/sciadv.1603150}\BibitemShut {NoStop}%
\bibitem [{\citenamefont {Lachance-Quirion}\ \emph {et~al.}(2020)\citenamefont
  {Lachance-Quirion}, \citenamefont {Wolski}, \citenamefont {Tabuchi},
  \citenamefont {Kono}, \citenamefont {Usami},\ and\ \citenamefont
  {Nakamura}}]{LachanceQuirionScience2020}%
  \BibitemOpen
  \bibfield  {author} {\bibinfo {author} {\bibfnamefont {D.}~\bibnamefont
  {Lachance-Quirion}}, \bibinfo {author} {\bibfnamefont {S.~P.}\ \bibnamefont
  {Wolski}}, \bibinfo {author} {\bibfnamefont {Y.}~\bibnamefont {Tabuchi}},
  \bibinfo {author} {\bibfnamefont {S.}~\bibnamefont {Kono}}, \bibinfo {author}
  {\bibfnamefont {K.}~\bibnamefont {Usami}}, \ and\ \bibinfo {author}
  {\bibfnamefont {Y.}~\bibnamefont {Nakamura}},\ }\href@noop {} {\bibfield
  {journal} {\bibinfo  {journal} {Science}\ }\textbf {\bibinfo {volume}
  {367}},\ \bibinfo {pages} {425} (\bibinfo {year} {2020})}\BibitemShut
  {NoStop}%
\bibitem [{\citenamefont {Satzinger}\ \emph {et~al.}(2018)\citenamefont
  {Satzinger}, \citenamefont {Zhong}, \citenamefont {Chang}, \citenamefont
  {Peairs}, \citenamefont {Bienfait}, \citenamefont {Chou}, \citenamefont
  {Cleland}, \citenamefont {Conner}, \citenamefont {Dumur}, \citenamefont
  {Grebel}, \citenamefont {Gutierrez}, \citenamefont {November}, \citenamefont
  {Povey}, \citenamefont {Whiteley}, \citenamefont {Awschalom}, \citenamefont
  {Schuster},\ and\ \citenamefont {Cleland}}]{SatzingerNature2018}%
  \BibitemOpen
  \bibfield  {author} {\bibinfo {author} {\bibfnamefont {K.~J.}\ \bibnamefont
  {Satzinger}}, \bibinfo {author} {\bibfnamefont {Y.~P.}\ \bibnamefont
  {Zhong}}, \bibinfo {author} {\bibfnamefont {H.-S.}\ \bibnamefont {Chang}},
  \bibinfo {author} {\bibfnamefont {G.~A.}\ \bibnamefont {Peairs}}, \bibinfo
  {author} {\bibfnamefont {A.}~\bibnamefont {Bienfait}}, \bibinfo {author}
  {\bibfnamefont {M.-H.}\ \bibnamefont {Chou}}, \bibinfo {author}
  {\bibfnamefont {A.~Y.}\ \bibnamefont {Cleland}}, \bibinfo {author}
  {\bibfnamefont {C.~R.}\ \bibnamefont {Conner}}, \bibinfo {author}
  {\bibfnamefont {E.}~\bibnamefont {Dumur}}, \bibinfo {author} {\bibfnamefont
  {J.}~\bibnamefont {Grebel}}, \bibinfo {author} {\bibfnamefont
  {I.}~\bibnamefont {Gutierrez}}, \bibinfo {author} {\bibfnamefont {B.~H.}\
  \bibnamefont {November}}, \bibinfo {author} {\bibfnamefont {R.~G.}\
  \bibnamefont {Povey}}, \bibinfo {author} {\bibfnamefont {S.~J.}\ \bibnamefont
  {Whiteley}}, \bibinfo {author} {\bibfnamefont {D.~D.}\ \bibnamefont
  {Awschalom}}, \bibinfo {author} {\bibfnamefont {D.~I.}\ \bibnamefont
  {Schuster}}, \ and\ \bibinfo {author} {\bibfnamefont {A.~N.}\ \bibnamefont
  {Cleland}},\ }\href@noop {} {\bibfield  {journal} {\bibinfo  {journal}
  {Nature}\ }\textbf {\bibinfo {volume} {563}},\ \bibinfo {pages} {661}
  (\bibinfo {year} {2018})}\BibitemShut {NoStop}%
\bibitem [{\citenamefont {Whiteley}\ \emph {et~al.}(2019)\citenamefont
  {Whiteley}, \citenamefont {Wolfowicz}, \citenamefont {Anderson},
  \citenamefont {Bourassa}, \citenamefont {Ma}, \citenamefont {Ye},
  \citenamefont {Koolstra}, \citenamefont {Satzinger}, \citenamefont {Holt},
  \citenamefont {Heremans}, \citenamefont {Cleland}, \citenamefont {Schuster},
  \citenamefont {Galli},\ and\ \citenamefont {Awschalom}}]{WhiteleyNphys2019}%
  \BibitemOpen
  \bibfield  {author} {\bibinfo {author} {\bibfnamefont {S.~J.}\ \bibnamefont
  {Whiteley}}, \bibinfo {author} {\bibfnamefont {G.}~\bibnamefont {Wolfowicz}},
  \bibinfo {author} {\bibfnamefont {C.~P.}\ \bibnamefont {Anderson}}, \bibinfo
  {author} {\bibfnamefont {A.}~\bibnamefont {Bourassa}}, \bibinfo {author}
  {\bibfnamefont {H.}~\bibnamefont {Ma}}, \bibinfo {author} {\bibfnamefont
  {M.}~\bibnamefont {Ye}}, \bibinfo {author} {\bibfnamefont {G.}~\bibnamefont
  {Koolstra}}, \bibinfo {author} {\bibfnamefont {K.~J.}\ \bibnamefont
  {Satzinger}}, \bibinfo {author} {\bibfnamefont {M.~V.}\ \bibnamefont {Holt}},
  \bibinfo {author} {\bibfnamefont {F.~J.}\ \bibnamefont {Heremans}}, \bibinfo
  {author} {\bibfnamefont {A.~N.}\ \bibnamefont {Cleland}}, \bibinfo {author}
  {\bibfnamefont {D.~I.}\ \bibnamefont {Schuster}}, \bibinfo {author}
  {\bibfnamefont {G.}~\bibnamefont {Galli}}, \ and\ \bibinfo {author}
  {\bibfnamefont {D.~D.}\ \bibnamefont {Awschalom}},\ }\href@noop {} {\bibfield
   {journal} {\bibinfo  {journal} {Nature Phys.}\ }\textbf {\bibinfo {volume}
  {15}},\ \bibinfo {pages} {490–495} (\bibinfo {year} {2019})}\BibitemShut
  {NoStop}%
\bibitem [{\citenamefont {Bienfait}\ \emph {et~al.}(2019)\citenamefont
  {Bienfait}, \citenamefont {Satzinger}, \citenamefont {Zhong}, \citenamefont
  {Chang}, \citenamefont {Chou}, \citenamefont {Conner}, \citenamefont {Dumur},
  \citenamefont {Grebel}, \citenamefont {Peairs}, \citenamefont {Povey},\ and\
  \citenamefont {Cleland}}]{BienfaitScience2019}%
  \BibitemOpen
  \bibfield  {author} {\bibinfo {author} {\bibfnamefont {A.}~\bibnamefont
  {Bienfait}}, \bibinfo {author} {\bibfnamefont {K.~J.}\ \bibnamefont
  {Satzinger}}, \bibinfo {author} {\bibfnamefont {Y.~P.}\ \bibnamefont
  {Zhong}}, \bibinfo {author} {\bibfnamefont {H.-S.}\ \bibnamefont {Chang}},
  \bibinfo {author} {\bibfnamefont {M.-H.}\ \bibnamefont {Chou}}, \bibinfo
  {author} {\bibfnamefont {C.~R.}\ \bibnamefont {Conner}}, \bibinfo {author}
  {\bibfnamefont {E.}~\bibnamefont {Dumur}}, \bibinfo {author} {\bibfnamefont
  {J.}~\bibnamefont {Grebel}}, \bibinfo {author} {\bibfnamefont {G.~A.}\
  \bibnamefont {Peairs}}, \bibinfo {author} {\bibfnamefont {R.~G.}\
  \bibnamefont {Povey}}, \ and\ \bibinfo {author} {\bibfnamefont {A.~N.}\
  \bibnamefont {Cleland}},\ }\href@noop {} {\bibfield  {journal} {\bibinfo
  {journal} {Science}\ }\textbf {\bibinfo {volume} {364}},\ \bibinfo {pages}
  {368} (\bibinfo {year} {2019})}\BibitemShut {NoStop}%
\bibitem [{\citenamefont {Callen}\ \emph {et~al.}(1963)\citenamefont {Callen},
  \citenamefont {Clark}, \citenamefont {DeSavage}, \citenamefont {Coleman},\
  and\ \citenamefont {Callen}}]{CallenPR1963}%
  \BibitemOpen
  \bibfield  {author} {\bibinfo {author} {\bibfnamefont {E.~R.}\ \bibnamefont
  {Callen}}, \bibinfo {author} {\bibfnamefont {A.~E.}\ \bibnamefont {Clark}},
  \bibinfo {author} {\bibfnamefont {B.}~\bibnamefont {DeSavage}}, \bibinfo
  {author} {\bibfnamefont {W.}~\bibnamefont {Coleman}}, \ and\ \bibinfo
  {author} {\bibfnamefont {H.~B.}\ \bibnamefont {Callen}},\ }\href@noop {}
  {\bibfield  {journal} {\bibinfo  {journal} {Phys. Rev.}\ }\textbf {\bibinfo
  {volume} {130}},\ \bibinfo {pages} {1735} (\bibinfo {year}
  {1963})}\BibitemShut {NoStop}%
\bibitem [{\citenamefont {Smith}\ and\ \citenamefont
  {Jones}(1963)}]{SmithJAP1963}%
  \BibitemOpen
  \bibfield  {author} {\bibinfo {author} {\bibfnamefont {A.~B.}\ \bibnamefont
  {Smith}}\ and\ \bibinfo {author} {\bibfnamefont {R.~V.}\ \bibnamefont
  {Jones}},\ }\href@noop {} {\bibfield  {journal} {\bibinfo  {journal} {J.
  Appl. Phys.}\ }\textbf {\bibinfo {volume} {34}},\ \bibinfo {pages} {1283}
  (\bibinfo {year} {1963})}\BibitemShut {NoStop}%
\bibitem [{\citenamefont {Klokholm}\ and\ \citenamefont
  {Aboaf}(1982)}]{KlokholmJAP1982}%
  \BibitemOpen
  \bibfield  {author} {\bibinfo {author} {\bibfnamefont {E.}~\bibnamefont
  {Klokholm}}\ and\ \bibinfo {author} {\bibfnamefont {J.}~\bibnamefont
  {Aboaf}},\ }\href@noop {} {\bibfield  {journal} {\bibinfo  {journal} {J.
  Appl. Phys.}\ }\textbf {\bibinfo {volume} {53}},\ \bibinfo {pages} {2661}
  (\bibinfo {year} {1982})}\BibitemShut {NoStop}%
\bibitem [{\citenamefont {Bonin}\ \emph {et~al.}(2005)\citenamefont {Bonin},
  \citenamefont {Schneider}, \citenamefont {Silva},\ and\ \citenamefont
  {Nibarger}}]{BoninJAP2005}%
  \BibitemOpen
  \bibfield  {author} {\bibinfo {author} {\bibfnamefont {R.}~\bibnamefont
  {Bonin}}, \bibinfo {author} {\bibfnamefont {M.~L.}\ \bibnamefont
  {Schneider}}, \bibinfo {author} {\bibfnamefont {T.~J.}\ \bibnamefont
  {Silva}}, \ and\ \bibinfo {author} {\bibfnamefont {J.~P.}\ \bibnamefont
  {Nibarger}},\ }\href@noop {} {\bibfield  {journal} {\bibinfo  {journal} {J.
  Appl. Phys.}\ }\textbf {\bibinfo {volume} {98}},\ \bibinfo {pages} {123904}
  (\bibinfo {year} {2005})}\BibitemShut {NoStop}%
\bibitem [{\citenamefont {Clark}\ \emph {et~al.}(2003)\citenamefont {Clark},
  \citenamefont {Hathaway}, \citenamefont {Wun-Fogle}, \citenamefont
  {Restorff}, \citenamefont {Lograsso}, \citenamefont {Keppens}, \citenamefont
  {Petculescu},\ and\ \citenamefont {Taylor}}]{ClarkJAP2003}%
  \BibitemOpen
  \bibfield  {author} {\bibinfo {author} {\bibfnamefont {A.~E.}\ \bibnamefont
  {Clark}}, \bibinfo {author} {\bibfnamefont {K.~B.}\ \bibnamefont {Hathaway}},
  \bibinfo {author} {\bibfnamefont {M.}~\bibnamefont {Wun-Fogle}}, \bibinfo
  {author} {\bibfnamefont {J.~B.}\ \bibnamefont {Restorff}}, \bibinfo {author}
  {\bibfnamefont {T.~A.}\ \bibnamefont {Lograsso}}, \bibinfo {author}
  {\bibfnamefont {V.~M.}\ \bibnamefont {Keppens}}, \bibinfo {author}
  {\bibfnamefont {G.}~\bibnamefont {Petculescu}}, \ and\ \bibinfo {author}
  {\bibfnamefont {R.~A.}\ \bibnamefont {Taylor}},\ }\href@noop {} {\bibfield
  {journal} {\bibinfo  {journal} {J. Appl. Phys.}\ }\textbf {\bibinfo {volume}
  {93}},\ \bibinfo {pages} {8621} (\bibinfo {year} {2003})}\BibitemShut
  {NoStop}%
\bibitem [{\citenamefont {Sandlund}\ \emph {et~al.}(1994)\citenamefont
  {Sandlund}, \citenamefont {Fahlander}, \citenamefont {Cedell}, \citenamefont
  {Clark}, \citenamefont {Restorff},\ and\ \citenamefont
  {Wun‐Fogle}}]{SandlundJAP1994}%
  \BibitemOpen
  \bibfield  {author} {\bibinfo {author} {\bibfnamefont {L.}~\bibnamefont
  {Sandlund}}, \bibinfo {author} {\bibfnamefont {M.}~\bibnamefont {Fahlander}},
  \bibinfo {author} {\bibfnamefont {T.}~\bibnamefont {Cedell}}, \bibinfo
  {author} {\bibfnamefont {A.~E.}\ \bibnamefont {Clark}}, \bibinfo {author}
  {\bibfnamefont {J.~B.}\ \bibnamefont {Restorff}}, \ and\ \bibinfo {author}
  {\bibfnamefont {M.}~\bibnamefont {Wun‐Fogle}},\ }\href@noop {} {\bibfield
  {journal} {\bibinfo  {journal} {J. Appl. Phys.}\ }\textbf {\bibinfo {volume}
  {75}},\ \bibinfo {pages} {5656} (\bibinfo {year} {1994})}\BibitemShut
  {NoStop}%
\bibitem [{\citenamefont {Cooke}\ \emph {et~al.}(2000)\citenamefont {Cooke},
  \citenamefont {Wang}, \citenamefont {Watts}, \citenamefont {Zuberek},
  \citenamefont {Heydon}, \citenamefont {Rainforth},\ and\ \citenamefont
  {Gehring}}]{CookeJPD2000}%
  \BibitemOpen
  \bibfield  {author} {\bibinfo {author} {\bibfnamefont {M.~D.}\ \bibnamefont
  {Cooke}}, \bibinfo {author} {\bibfnamefont {L.-C.}\ \bibnamefont {Wang}},
  \bibinfo {author} {\bibfnamefont {R.}~\bibnamefont {Watts}}, \bibinfo
  {author} {\bibfnamefont {R.}~\bibnamefont {Zuberek}}, \bibinfo {author}
  {\bibfnamefont {G.}~\bibnamefont {Heydon}}, \bibinfo {author} {\bibfnamefont
  {W.~M.}\ \bibnamefont {Rainforth}}, \ and\ \bibinfo {author} {\bibfnamefont
  {G.~A.}\ \bibnamefont {Gehring}},\ }\href@noop {} {\bibfield  {journal}
  {\bibinfo  {journal} {J. Phys. D: Appl. Phys.}\ }\textbf {\bibinfo {volume}
  {33}},\ \bibinfo {pages} {1450} (\bibinfo {year} {2000})}\BibitemShut
  {NoStop}%
\bibitem [{\citenamefont {Emori}\ \emph {et~al.}(2017)\citenamefont {Emori},
  \citenamefont {Gray}, \citenamefont {Jeon}, \citenamefont {Peoples},
  \citenamefont {Schmitt}, \citenamefont {Mahalingam}, \citenamefont {Hill},
  \citenamefont {McConney}, \citenamefont {Gray}, \citenamefont {Alaan},
  \citenamefont {Bornstein}, \citenamefont {Shafer}, \citenamefont {N'Diaye},
  \citenamefont {Arenholz}, \citenamefont {Haugstad}, \citenamefont {Meng},
  \citenamefont {Yang}, \citenamefont {Li}, \citenamefont {Mahat},
  \citenamefont {Cahill}, \citenamefont {Dhagat}, \citenamefont {Jander},
  \citenamefont {Sun}, \citenamefont {Suzuki},\ and\ \citenamefont
  {Howe}}]{EmoriAM2017}%
  \BibitemOpen
  \bibfield  {author} {\bibinfo {author} {\bibfnamefont {S.}~\bibnamefont
  {Emori}}, \bibinfo {author} {\bibfnamefont {B.~A.}\ \bibnamefont {Gray}},
  \bibinfo {author} {\bibfnamefont {H.}~\bibnamefont {Jeon}}, \bibinfo {author}
  {\bibfnamefont {J.}~\bibnamefont {Peoples}}, \bibinfo {author} {\bibfnamefont
  {M.}~\bibnamefont {Schmitt}}, \bibinfo {author} {\bibfnamefont
  {K.}~\bibnamefont {Mahalingam}}, \bibinfo {author} {\bibfnamefont
  {M.}~\bibnamefont {Hill}}, \bibinfo {author} {\bibfnamefont {M.~E.}\
  \bibnamefont {McConney}}, \bibinfo {author} {\bibfnamefont {M.~T.}\
  \bibnamefont {Gray}}, \bibinfo {author} {\bibfnamefont {U.~S.}\ \bibnamefont
  {Alaan}}, \bibinfo {author} {\bibfnamefont {A.~C.}\ \bibnamefont
  {Bornstein}}, \bibinfo {author} {\bibfnamefont {P.}~\bibnamefont {Shafer}},
  \bibinfo {author} {\bibfnamefont {A.~T.}\ \bibnamefont {N'Diaye}}, \bibinfo
  {author} {\bibfnamefont {E.}~\bibnamefont {Arenholz}}, \bibinfo {author}
  {\bibfnamefont {G.}~\bibnamefont {Haugstad}}, \bibinfo {author}
  {\bibfnamefont {K.}~\bibnamefont {Meng}}, \bibinfo {author} {\bibfnamefont
  {F.}~\bibnamefont {Yang}}, \bibinfo {author} {\bibfnamefont {D.}~\bibnamefont
  {Li}}, \bibinfo {author} {\bibfnamefont {S.}~\bibnamefont {Mahat}}, \bibinfo
  {author} {\bibfnamefont {D.~G.}\ \bibnamefont {Cahill}}, \bibinfo {author}
  {\bibfnamefont {P.}~\bibnamefont {Dhagat}}, \bibinfo {author} {\bibfnamefont
  {A.}~\bibnamefont {Jander}}, \bibinfo {author} {\bibfnamefont {N.~X.}\
  \bibnamefont {Sun}}, \bibinfo {author} {\bibfnamefont {Y.}~\bibnamefont
  {Suzuki}}, \ and\ \bibinfo {author} {\bibfnamefont {B.~M.}\ \bibnamefont
  {Howe}},\ }\href@noop {} {\bibfield  {journal} {\bibinfo  {journal} {Adv.
  Mater.}\ }\textbf {\bibinfo {volume} {29}},\ \bibinfo {pages} {1701130}
  (\bibinfo {year} {2017})}\BibitemShut {NoStop}%
\bibitem [{\citenamefont {Glunk}\ \emph {et~al.}(2009)\citenamefont {Glunk},
  \citenamefont {Daeubler}, \citenamefont {Dreher}, \citenamefont {Schwaiger},
  \citenamefont {Schoch}, \citenamefont {Sauer}, \citenamefont {Limmer},
  \citenamefont {Brandlmaier}, \citenamefont {Goennenwein}, \citenamefont
  {Bihler},\ and\ \citenamefont {Brandt}}]{GlunkPRB2009}%
  \BibitemOpen
  \bibfield  {author} {\bibinfo {author} {\bibfnamefont {M.}~\bibnamefont
  {Glunk}}, \bibinfo {author} {\bibfnamefont {J.}~\bibnamefont {Daeubler}},
  \bibinfo {author} {\bibfnamefont {L.}~\bibnamefont {Dreher}}, \bibinfo
  {author} {\bibfnamefont {S.}~\bibnamefont {Schwaiger}}, \bibinfo {author}
  {\bibfnamefont {W.}~\bibnamefont {Schoch}}, \bibinfo {author} {\bibfnamefont
  {R.}~\bibnamefont {Sauer}}, \bibinfo {author} {\bibfnamefont
  {W.}~\bibnamefont {Limmer}}, \bibinfo {author} {\bibfnamefont
  {A.}~\bibnamefont {Brandlmaier}}, \bibinfo {author} {\bibfnamefont
  {S.~T.~B.}\ \bibnamefont {Goennenwein}}, \bibinfo {author} {\bibfnamefont
  {C.}~\bibnamefont {Bihler}}, \ and\ \bibinfo {author} {\bibfnamefont {M.~S.}\
  \bibnamefont {Brandt}},\ }\href {\doibase 10.1103/PhysRevB.79.195206}
  {\bibfield  {journal} {\bibinfo  {journal} {Phys. Rev. B}\ }\textbf {\bibinfo
  {volume} {79}},\ \bibinfo {pages} {195206} (\bibinfo {year}
  {2009})}\BibitemShut {NoStop}%
\bibitem [{\citenamefont {Scherbakov}\ \emph {et~al.}(2010)\citenamefont
  {Scherbakov}, \citenamefont {Salasyuk}, \citenamefont {Akimov}, \citenamefont
  {Liu}, \citenamefont {Bombeck}, \citenamefont {Br\"uggemann}, \citenamefont
  {Yakovlev}, \citenamefont {Sapega}, \citenamefont {Furdyna},\ and\
  \citenamefont {Bayer}}]{ScherbakovPRL2010}%
  \BibitemOpen
  \bibfield  {author} {\bibinfo {author} {\bibfnamefont {A.~V.}\ \bibnamefont
  {Scherbakov}}, \bibinfo {author} {\bibfnamefont {A.~S.}\ \bibnamefont
  {Salasyuk}}, \bibinfo {author} {\bibfnamefont {A.~V.}\ \bibnamefont
  {Akimov}}, \bibinfo {author} {\bibfnamefont {X.}~\bibnamefont {Liu}},
  \bibinfo {author} {\bibfnamefont {M.}~\bibnamefont {Bombeck}}, \bibinfo
  {author} {\bibfnamefont {C.}~\bibnamefont {Br\"uggemann}}, \bibinfo {author}
  {\bibfnamefont {D.~R.}\ \bibnamefont {Yakovlev}}, \bibinfo {author}
  {\bibfnamefont {V.~F.}\ \bibnamefont {Sapega}}, \bibinfo {author}
  {\bibfnamefont {J.~K.}\ \bibnamefont {Furdyna}}, \ and\ \bibinfo {author}
  {\bibfnamefont {M.}~\bibnamefont {Bayer}},\ }\href {\doibase
  10.1103/PhysRevLett.105.117204} {\bibfield  {journal} {\bibinfo  {journal}
  {Phys. Rev. Lett.}\ }\textbf {\bibinfo {volume} {105}},\ \bibinfo {pages}
  {117204} (\bibinfo {year} {2010})}\BibitemShut {NoStop}%
\bibitem [{\citenamefont {Novosad}\ \emph {et~al.}(2000)\citenamefont
  {Novosad}, \citenamefont {Otani}, \citenamefont {Ohsawa}, \citenamefont
  {Kim}, \citenamefont {Fukamichi}, \citenamefont {Koike}, \citenamefont
  {Maruyama}, \citenamefont {Kitakami},\ and\ \citenamefont
  {Shimada}}]{NovosadJAP2000}%
  \BibitemOpen
  \bibfield  {author} {\bibinfo {author} {\bibfnamefont {V.}~\bibnamefont
  {Novosad}}, \bibinfo {author} {\bibfnamefont {Y.}~\bibnamefont {Otani}},
  \bibinfo {author} {\bibfnamefont {A.}~\bibnamefont {Ohsawa}}, \bibinfo
  {author} {\bibfnamefont {S.~G.}\ \bibnamefont {Kim}}, \bibinfo {author}
  {\bibfnamefont {K.}~\bibnamefont {Fukamichi}}, \bibinfo {author}
  {\bibfnamefont {J.}~\bibnamefont {Koike}}, \bibinfo {author} {\bibfnamefont
  {K.}~\bibnamefont {Maruyama}}, \bibinfo {author} {\bibfnamefont
  {O.}~\bibnamefont {Kitakami}}, \ and\ \bibinfo {author} {\bibfnamefont
  {Y.}~\bibnamefont {Shimada}},\ }\href@noop {} {\bibfield  {journal} {\bibinfo
   {journal} {J. Appl. Phys.}\ }\textbf {\bibinfo {volume} {87}},\ \bibinfo
  {pages} {6400} (\bibinfo {year} {2000})}\BibitemShut {NoStop}%
\bibitem [{\citenamefont {Chu}\ \emph {et~al.}(2018{\natexlab{a}})\citenamefont
  {Chu}, \citenamefont {PourhosseiniAsl},\ and\ \citenamefont
  {Dong}}]{ChuJPD2018}%
  \BibitemOpen
  \bibfield  {author} {\bibinfo {author} {\bibfnamefont {Z.}~\bibnamefont
  {Chu}}, \bibinfo {author} {\bibfnamefont {M.}~\bibnamefont
  {PourhosseiniAsl}}, \ and\ \bibinfo {author} {\bibfnamefont {S.}~\bibnamefont
  {Dong}},\ }\href@noop {} {\bibfield  {journal} {\bibinfo  {journal} {J. Phys.
  D: Appl. Phys.}\ }\textbf {\bibinfo {volume} {51}},\ \bibinfo {pages}
  {243001} (\bibinfo {year} {2018}{\natexlab{a}})}\BibitemShut {NoStop}%
\bibitem [{\citenamefont {Liang}\ \emph {et~al.}(2020)\citenamefont {Liang},
  \citenamefont {Dong}, \citenamefont {Chen}, \citenamefont {Wang},
  \citenamefont {Wei}, \citenamefont {Zaeimbashi}, \citenamefont {He},
  \citenamefont {Matyushov}, \citenamefont {Sun},\ and\ \citenamefont
  {Sun}}]{LiangSensors2020}%
  \BibitemOpen
  \bibfield  {author} {\bibinfo {author} {\bibfnamefont {X.}~\bibnamefont
  {Liang}}, \bibinfo {author} {\bibfnamefont {C.}~\bibnamefont {Dong}},
  \bibinfo {author} {\bibfnamefont {H.}~\bibnamefont {Chen}}, \bibinfo {author}
  {\bibfnamefont {J.}~\bibnamefont {Wang}}, \bibinfo {author} {\bibfnamefont
  {Y.}~\bibnamefont {Wei}}, \bibinfo {author} {\bibfnamefont {M.}~\bibnamefont
  {Zaeimbashi}}, \bibinfo {author} {\bibfnamefont {Y.}~\bibnamefont {He}},
  \bibinfo {author} {\bibfnamefont {A.}~\bibnamefont {Matyushov}}, \bibinfo
  {author} {\bibfnamefont {C.}~\bibnamefont {Sun}}, \ and\ \bibinfo {author}
  {\bibfnamefont {N.}~\bibnamefont {Sun}},\ }\href@noop {} {\bibfield
  {journal} {\bibinfo  {journal} {Sensors}\ }\textbf {\bibinfo {volume} {20}},\
  \bibinfo {pages} {1532} (\bibinfo {year} {2020})}\BibitemShut {NoStop}%
\bibitem [{\citenamefont {Lemanov}\ and\ \citenamefont
  {Smolenskii}(1973)}]{LemanovSov1973}%
  \BibitemOpen
  \bibfield  {author} {\bibinfo {author} {\bibfnamefont {V.~V.}\ \bibnamefont
  {Lemanov}}\ and\ \bibinfo {author} {\bibfnamefont {G.~A.}\ \bibnamefont
  {Smolenskii}},\ }\href@noop {} {\bibfield  {journal} {\bibinfo  {journal}
  {Sov. Phys. Usp.}\ }\textbf {\bibinfo {volume} {15}},\ \bibinfo {pages} {708}
  (\bibinfo {year} {1973})}\BibitemShut {NoStop}%
\bibitem [{\citenamefont {Okamoto}\ \emph {et~al.}(2020)\citenamefont
  {Okamoto}, \citenamefont {Murakami},\ and\ \citenamefont
  {Everschor-Sitte}}]{OkamotoPRB2020}%
  \BibitemOpen
  \bibfield  {author} {\bibinfo {author} {\bibfnamefont {A.}~\bibnamefont
  {Okamoto}}, \bibinfo {author} {\bibfnamefont {S.}~\bibnamefont {Murakami}}, \
  and\ \bibinfo {author} {\bibfnamefont {K.}~\bibnamefont {Everschor-Sitte}},\
  }\href {\doibase 10.1103/PhysRevB.101.064424} {\bibfield  {journal} {\bibinfo
   {journal} {Phys. Rev. B}\ }\textbf {\bibinfo {volume} {101}},\ \bibinfo
  {pages} {064424} (\bibinfo {year} {2020})}\BibitemShut {NoStop}%
\bibitem [{\citenamefont {Vanderveken}\ \emph {et~al.}()\citenamefont
  {Vanderveken}, \citenamefont {Ciubotaru},\ and\ \citenamefont
  {Adelmann}}]{VandervekenarXiv2020}%
  \BibitemOpen
  \bibfield  {author} {\bibinfo {author} {\bibfnamefont {F.}~\bibnamefont
  {Vanderveken}}, \bibinfo {author} {\bibfnamefont {F.}~\bibnamefont
  {Ciubotaru}}, \ and\ \bibinfo {author} {\bibfnamefont {C.}~\bibnamefont
  {Adelmann}},\ }\href@noop {} {\ }\bibinfo {note}
  {ArXiv:2003.12099}\BibitemShut {NoStop}%
\bibitem [{\citenamefont {Hanna}\ and\ \citenamefont
  {Murphy}(1988)}]{HannaIEEE1988}%
  \BibitemOpen
  \bibfield  {author} {\bibinfo {author} {\bibfnamefont {S.}~\bibnamefont
  {Hanna}}\ and\ \bibinfo {author} {\bibnamefont {Murphy}},\ }\href@noop {}
  {\bibfield  {journal} {\bibinfo  {journal} {IEEE Trans. Magn.}\ }\textbf
  {\bibinfo {volume} {24}},\ \bibinfo {pages} {2814} (\bibinfo {year}
  {1988})}\BibitemShut {NoStop}%
\bibitem [{\citenamefont {Hanna}\ \emph {et~al.}(1990)\citenamefont {Hanna},
  \citenamefont {Murphy}, \citenamefont {Sabetfakhri},\ and\ \citenamefont
  {Stratakis}}]{HannaIEEE1990}%
  \BibitemOpen
  \bibfield  {author} {\bibinfo {author} {\bibfnamefont {S.}~\bibnamefont
  {Hanna}}, \bibinfo {author} {\bibfnamefont {G.}~\bibnamefont {Murphy}},
  \bibinfo {author} {\bibfnamefont {K.}~\bibnamefont {Sabetfakhri}}, \ and\
  \bibinfo {author} {\bibfnamefont {K.}~\bibnamefont {Stratakis}},\ }\href@noop
  {} {\bibfield  {journal} {\bibinfo  {journal} {IEEE 1990 Ultrasonics
  Symposium Proceedings,}\ }\textbf {\bibinfo {volume} {1}},\ \bibinfo {pages}
  {209} (\bibinfo {year} {1990})}\BibitemShut {NoStop}%
\bibitem [{\citenamefont {Chumak}\ \emph {et~al.}(2010)\citenamefont {Chumak},
  \citenamefont {Dhagat}, \citenamefont {Jander}, \citenamefont {Serga},\ and\
  \citenamefont {Hillebrands}}]{ChumakPRB2010}%
  \BibitemOpen
  \bibfield  {author} {\bibinfo {author} {\bibfnamefont {A.~V.}\ \bibnamefont
  {Chumak}}, \bibinfo {author} {\bibfnamefont {P.}~\bibnamefont {Dhagat}},
  \bibinfo {author} {\bibfnamefont {A.}~\bibnamefont {Jander}}, \bibinfo
  {author} {\bibfnamefont {A.~A.}\ \bibnamefont {Serga}}, \ and\ \bibinfo
  {author} {\bibfnamefont {B.}~\bibnamefont {Hillebrands}},\ }\href {\doibase
  10.1103/PhysRevB.81.140404} {\bibfield  {journal} {\bibinfo  {journal} {Phys.
  Rev. B}\ }\textbf {\bibinfo {volume} {81}},\ \bibinfo {pages} {140404}
  (\bibinfo {year} {2010})}\BibitemShut {NoStop}%
\bibitem [{\citenamefont {Uchida}\ \emph {et~al.}(2011)\citenamefont {Uchida},
  \citenamefont {An}, \citenamefont {Kajiwara}, \citenamefont {Toda},\ and\
  \citenamefont {Saitoh}}]{UchidaAPL2011}%
  \BibitemOpen
  \bibfield  {author} {\bibinfo {author} {\bibfnamefont {K.-i.}\ \bibnamefont
  {Uchida}}, \bibinfo {author} {\bibfnamefont {T.}~\bibnamefont {An}}, \bibinfo
  {author} {\bibfnamefont {Y.}~\bibnamefont {Kajiwara}}, \bibinfo {author}
  {\bibfnamefont {M.}~\bibnamefont {Toda}}, \ and\ \bibinfo {author}
  {\bibfnamefont {E.}~\bibnamefont {Saitoh}},\ }\href@noop {} {\bibfield
  {journal} {\bibinfo  {journal} {Appl. Phys. Lett.}\ }\textbf {\bibinfo
  {volume} {99}},\ \bibinfo {pages} {212501} (\bibinfo {year}
  {2011})}\BibitemShut {NoStop}%
\bibitem [{\citenamefont {Kryshtal}\ and\ \citenamefont
  {Medved}(2012)}]{KryshtalAPL2012}%
  \BibitemOpen
  \bibfield  {author} {\bibinfo {author} {\bibfnamefont {R.~G.}\ \bibnamefont
  {Kryshtal}}\ and\ \bibinfo {author} {\bibfnamefont {A.~V.}\ \bibnamefont
  {Medved}},\ }\href@noop {} {\bibfield  {journal} {\bibinfo  {journal} {Appl.
  Phys. Lett.}\ }\textbf {\bibinfo {volume} {100}},\ \bibinfo {pages} {192410}
  (\bibinfo {year} {2012})}\BibitemShut {NoStop}%
\bibitem [{\citenamefont {Thevenard}\ \emph {et~al.}(2014)\citenamefont
  {Thevenard}, \citenamefont {Gourdon}, \citenamefont {Prieur}, \citenamefont
  {von Bardeleben}, \citenamefont {Vincent}, \citenamefont {Becerra},
  \citenamefont {Largeau},\ and\ \citenamefont {Duquesne}}]{ThevenardPRB2014}%
  \BibitemOpen
  \bibfield  {author} {\bibinfo {author} {\bibfnamefont {L.}~\bibnamefont
  {Thevenard}}, \bibinfo {author} {\bibfnamefont {C.}~\bibnamefont {Gourdon}},
  \bibinfo {author} {\bibfnamefont {J.~Y.}\ \bibnamefont {Prieur}}, \bibinfo
  {author} {\bibfnamefont {H.~J.}\ \bibnamefont {von Bardeleben}}, \bibinfo
  {author} {\bibfnamefont {S.}~\bibnamefont {Vincent}}, \bibinfo {author}
  {\bibfnamefont {L.}~\bibnamefont {Becerra}}, \bibinfo {author} {\bibfnamefont
  {L.}~\bibnamefont {Largeau}}, \ and\ \bibinfo {author} {\bibfnamefont
  {J.-Y.}\ \bibnamefont {Duquesne}},\ }\href {\doibase
  10.1103/PhysRevB.90.094401} {\bibfield  {journal} {\bibinfo  {journal} {Phys.
  Rev. B}\ }\textbf {\bibinfo {volume} {90}},\ \bibinfo {pages} {094401}
  (\bibinfo {year} {2014})}\BibitemShut {NoStop}%
\bibitem [{\citenamefont {Kuszewski}\ \emph {et~al.}(2018)\citenamefont
  {Kuszewski}, \citenamefont {Camara}, \citenamefont {Biarrotte}, \citenamefont
  {Becerra}, \citenamefont {von Bardeleben}, \citenamefont {Savero~Torres},
  \citenamefont {Lemaitre}, \citenamefont {Gourdon}, \citenamefont {Duquesne},\
  and\ \citenamefont {Thevenard}}]{KuszewskiJPhys2018}%
  \BibitemOpen
  \bibfield  {author} {\bibinfo {author} {\bibfnamefont {P.}~\bibnamefont
  {Kuszewski}}, \bibinfo {author} {\bibfnamefont {I.~S.}\ \bibnamefont
  {Camara}}, \bibinfo {author} {\bibfnamefont {N.}~\bibnamefont {Biarrotte}},
  \bibinfo {author} {\bibfnamefont {L.}~\bibnamefont {Becerra}}, \bibinfo
  {author} {\bibfnamefont {J.}~\bibnamefont {von Bardeleben}}, \bibinfo
  {author} {\bibfnamefont {W.}~\bibnamefont {Savero~Torres}}, \bibinfo {author}
  {\bibfnamefont {A.}~\bibnamefont {Lemaitre}}, \bibinfo {author}
  {\bibfnamefont {C.}~\bibnamefont {Gourdon}}, \bibinfo {author} {\bibfnamefont
  {J.-Y.}\ \bibnamefont {Duquesne}}, \ and\ \bibinfo {author} {\bibfnamefont
  {L.}~\bibnamefont {Thevenard}},\ }\href@noop {} {\bibfield  {journal}
  {\bibinfo  {journal} {J. Phys.: Condens. Matter}\ }\textbf {\bibinfo {volume}
  {30}},\ \bibinfo {pages} {244003} (\bibinfo {year} {2018})}\BibitemShut
  {NoStop}%
\bibitem [{\citenamefont {Bombeck}\ \emph {et~al.}(2012)\citenamefont
  {Bombeck}, \citenamefont {Salasyuk}, \citenamefont {Glavin}, \citenamefont
  {Scherbakov}, \citenamefont {Br\"uggemann}, \citenamefont {Yakovlev},
  \citenamefont {Sapega}, \citenamefont {Liu}, \citenamefont {Furdyna},
  \citenamefont {Akimov},\ and\ \citenamefont {Bayer}}]{BombeckPRB2012}%
  \BibitemOpen
  \bibfield  {author} {\bibinfo {author} {\bibfnamefont {M.}~\bibnamefont
  {Bombeck}}, \bibinfo {author} {\bibfnamefont {A.~S.}\ \bibnamefont
  {Salasyuk}}, \bibinfo {author} {\bibfnamefont {B.~A.}\ \bibnamefont
  {Glavin}}, \bibinfo {author} {\bibfnamefont {A.~V.}\ \bibnamefont
  {Scherbakov}}, \bibinfo {author} {\bibfnamefont {C.}~\bibnamefont
  {Br\"uggemann}}, \bibinfo {author} {\bibfnamefont {D.~R.}\ \bibnamefont
  {Yakovlev}}, \bibinfo {author} {\bibfnamefont {V.~F.}\ \bibnamefont
  {Sapega}}, \bibinfo {author} {\bibfnamefont {X.}~\bibnamefont {Liu}},
  \bibinfo {author} {\bibfnamefont {J.~K.}\ \bibnamefont {Furdyna}}, \bibinfo
  {author} {\bibfnamefont {A.~V.}\ \bibnamefont {Akimov}}, \ and\ \bibinfo
  {author} {\bibfnamefont {M.}~\bibnamefont {Bayer}},\ }\href {\doibase
  10.1103/PhysRevB.85.195324} {\bibfield  {journal} {\bibinfo  {journal} {Phys.
  Rev. B}\ }\textbf {\bibinfo {volume} {85}},\ \bibinfo {pages} {195324}
  (\bibinfo {year} {2012})}\BibitemShut {NoStop}%
\bibitem [{\citenamefont {J\"ager}\ \emph {et~al.}(2015)\citenamefont
  {J\"ager}, \citenamefont {Scherbakov}, \citenamefont {Glavin}, \citenamefont
  {Salasyuk}, \citenamefont {Campion}, \citenamefont {Rushforth}, \citenamefont
  {Yakovlev}, \citenamefont {Akimov},\ and\ \citenamefont
  {Bayer}}]{JagerPRB2015}%
  \BibitemOpen
  \bibfield  {author} {\bibinfo {author} {\bibfnamefont {J.~V.}\ \bibnamefont
  {J\"ager}}, \bibinfo {author} {\bibfnamefont {A.~V.}\ \bibnamefont
  {Scherbakov}}, \bibinfo {author} {\bibfnamefont {B.~A.}\ \bibnamefont
  {Glavin}}, \bibinfo {author} {\bibfnamefont {A.~S.}\ \bibnamefont
  {Salasyuk}}, \bibinfo {author} {\bibfnamefont {R.~P.}\ \bibnamefont
  {Campion}}, \bibinfo {author} {\bibfnamefont {A.~W.}\ \bibnamefont
  {Rushforth}}, \bibinfo {author} {\bibfnamefont {D.~R.}\ \bibnamefont
  {Yakovlev}}, \bibinfo {author} {\bibfnamefont {A.~V.}\ \bibnamefont
  {Akimov}}, \ and\ \bibinfo {author} {\bibfnamefont {M.}~\bibnamefont
  {Bayer}},\ }\href {\doibase 10.1103/PhysRevB.92.020404} {\bibfield  {journal}
  {\bibinfo  {journal} {Phys. Rev. B}\ }\textbf {\bibinfo {volume} {92}},\
  \bibinfo {pages} {020404} (\bibinfo {year} {2015})}\BibitemShut {NoStop}%
\bibitem [{\citenamefont {Kim}\ \emph {et~al.}(2012)\citenamefont {Kim},
  \citenamefont {Vomir},\ and\ \citenamefont {Bigot}}]{KimPRL2012}%
  \BibitemOpen
  \bibfield  {author} {\bibinfo {author} {\bibfnamefont {J.-W.}\ \bibnamefont
  {Kim}}, \bibinfo {author} {\bibfnamefont {M.}~\bibnamefont {Vomir}}, \ and\
  \bibinfo {author} {\bibfnamefont {J.-Y.}\ \bibnamefont {Bigot}},\ }\href
  {\doibase 10.1103/PhysRevLett.109.166601} {\bibfield  {journal} {\bibinfo
  {journal} {Phys. Rev. Lett.}\ }\textbf {\bibinfo {volume} {109}},\ \bibinfo
  {pages} {166601} (\bibinfo {year} {2012})}\BibitemShut {NoStop}%
\bibitem [{\citenamefont {Kim}\ and\ \citenamefont {Bigot}(2017)}]{KimPRB2017}%
  \BibitemOpen
  \bibfield  {author} {\bibinfo {author} {\bibfnamefont {J.-W.}\ \bibnamefont
  {Kim}}\ and\ \bibinfo {author} {\bibfnamefont {J.-Y.}\ \bibnamefont
  {Bigot}},\ }\href {\doibase 10.1103/PhysRevB.95.144422} {\bibfield  {journal}
  {\bibinfo  {journal} {Phys. Rev. B}\ }\textbf {\bibinfo {volume} {95}},\
  \bibinfo {pages} {144422} (\bibinfo {year} {2017})}\BibitemShut {NoStop}%
\bibitem [{\citenamefont {Janu\u{s}onis}\ \emph {et~al.}(2015)\citenamefont
  {Janu\u{s}onis}, \citenamefont {Chang}, \citenamefont {van Loosdrecht},\ and\
  \citenamefont {Tobey}}]{JanusonisAPL2015}%
  \BibitemOpen
  \bibfield  {author} {\bibinfo {author} {\bibfnamefont {J.}~\bibnamefont
  {Janu\u{s}onis}}, \bibinfo {author} {\bibfnamefont {C.~L.}\ \bibnamefont
  {Chang}}, \bibinfo {author} {\bibfnamefont {P.~H.~M.}\ \bibnamefont {van
  Loosdrecht}}, \ and\ \bibinfo {author} {\bibfnamefont {R.~I.}\ \bibnamefont
  {Tobey}},\ }\href@noop {} {\bibfield  {journal} {\bibinfo  {journal} {Appl.
  Phys. Lett.}\ }\textbf {\bibinfo {volume} {106}},\ \bibinfo {pages} {181601}
  (\bibinfo {year} {2015})}\BibitemShut {NoStop}%
\bibitem [{\citenamefont {Janu\u{s}onis}\ \emph {et~al.}(2016)\citenamefont
  {Janu\u{s}onis}, \citenamefont {Chang}, \citenamefont {Jansma}, \citenamefont
  {Gatilova}, \citenamefont {Vlasov}, \citenamefont {Lomonosov}, \citenamefont
  {Temnov},\ and\ \citenamefont {Tobey}}]{JanusonisPRB2016}%
  \BibitemOpen
  \bibfield  {author} {\bibinfo {author} {\bibfnamefont {J.}~\bibnamefont
  {Janu\u{s}onis}}, \bibinfo {author} {\bibfnamefont {C.~L.}\ \bibnamefont
  {Chang}}, \bibinfo {author} {\bibfnamefont {T.}~\bibnamefont {Jansma}},
  \bibinfo {author} {\bibfnamefont {A.}~\bibnamefont {Gatilova}}, \bibinfo
  {author} {\bibfnamefont {V.~S.}\ \bibnamefont {Vlasov}}, \bibinfo {author}
  {\bibfnamefont {A.~M.}\ \bibnamefont {Lomonosov}}, \bibinfo {author}
  {\bibfnamefont {V.~V.}\ \bibnamefont {Temnov}}, \ and\ \bibinfo {author}
  {\bibfnamefont {R.~I.}\ \bibnamefont {Tobey}},\ }\href {\doibase
  10.1103/PhysRevB.94.024415} {\bibfield  {journal} {\bibinfo  {journal} {Phys.
  Rev. B}\ }\textbf {\bibinfo {volume} {94}},\ \bibinfo {pages} {024415}
  (\bibinfo {year} {2016})}\BibitemShut {NoStop}%
\bibitem [{\citenamefont {Zhao}\ \emph {et~al.}(2021)\citenamefont {Zhao},
  \citenamefont {Zhang}, \citenamefont {Li}, \citenamefont {Zhang},
  \citenamefont {Pearson}, \citenamefont {Divan}, \citenamefont {Liu},
  \citenamefont {Novosad}, \citenamefont {Wang},\ and\ \citenamefont
  {Hoffmann}}]{ZhaoPRApplied2021}%
  \BibitemOpen
  \bibfield  {author} {\bibinfo {author} {\bibfnamefont {C.}~\bibnamefont
  {Zhao}}, \bibinfo {author} {\bibfnamefont {Z.}~\bibnamefont {Zhang}},
  \bibinfo {author} {\bibfnamefont {Y.}~\bibnamefont {Li}}, \bibinfo {author}
  {\bibfnamefont {W.}~\bibnamefont {Zhang}}, \bibinfo {author} {\bibfnamefont
  {J.~E.}\ \bibnamefont {Pearson}}, \bibinfo {author} {\bibfnamefont
  {R.}~\bibnamefont {Divan}}, \bibinfo {author} {\bibfnamefont
  {Q.}~\bibnamefont {Liu}}, \bibinfo {author} {\bibfnamefont {V.}~\bibnamefont
  {Novosad}}, \bibinfo {author} {\bibfnamefont {J.}~\bibnamefont {Wang}}, \
  and\ \bibinfo {author} {\bibfnamefont {A.}~\bibnamefont {Hoffmann}},\
  }\href@noop {} {\bibfield  {journal} {\bibinfo  {journal} {Phys. Rev.
  Applied}\ }\textbf {\bibinfo {volume} {15}},\ \bibinfo {pages} {014052}
  (\bibinfo {year} {2021})}\BibitemShut {NoStop}%
\bibitem [{\citenamefont {Casals}\ \emph {et~al.}(2020)\citenamefont {Casals},
  \citenamefont {Statuto}, \citenamefont {Foerster}, \citenamefont
  {Hern\'andez-M\'{\i}nguez}, \citenamefont {Cichelero}, \citenamefont
  {Manshausen}, \citenamefont {Mandziak}, \citenamefont {Aballe}, \citenamefont
  {Hern\`andez},\ and\ \citenamefont {Maci\`a}}]{CasalsPRL2020}%
  \BibitemOpen
  \bibfield  {author} {\bibinfo {author} {\bibfnamefont {B.}~\bibnamefont
  {Casals}}, \bibinfo {author} {\bibfnamefont {N.}~\bibnamefont {Statuto}},
  \bibinfo {author} {\bibfnamefont {M.}~\bibnamefont {Foerster}}, \bibinfo
  {author} {\bibfnamefont {A.}~\bibnamefont {Hern\'andez-M\'{\i}nguez}},
  \bibinfo {author} {\bibfnamefont {R.}~\bibnamefont {Cichelero}}, \bibinfo
  {author} {\bibfnamefont {P.}~\bibnamefont {Manshausen}}, \bibinfo {author}
  {\bibfnamefont {A.}~\bibnamefont {Mandziak}}, \bibinfo {author}
  {\bibfnamefont {L.}~\bibnamefont {Aballe}}, \bibinfo {author} {\bibfnamefont
  {J.~M.}\ \bibnamefont {Hern\`andez}}, \ and\ \bibinfo {author} {\bibfnamefont
  {F.}~\bibnamefont {Maci\`a}},\ }\href@noop {} {\bibfield  {journal} {\bibinfo
   {journal} {Phys. Rev. Lett.}\ }\textbf {\bibinfo {volume} {124}},\ \bibinfo
  {pages} {137202} (\bibinfo {year} {2020})}\BibitemShut {NoStop}%
\bibitem [{\citenamefont {Demokritov}\ \emph {et~al.}(2001)\citenamefont
  {Demokritov}, \citenamefont {Hillebrands},\ and\ \citenamefont
  {Slavin}}]{DemokritovPhysRep2001}%
  \BibitemOpen
  \bibfield  {author} {\bibinfo {author} {\bibfnamefont {S.~O.}\ \bibnamefont
  {Demokritov}}, \bibinfo {author} {\bibfnamefont {B.}~\bibnamefont
  {Hillebrands}}, \ and\ \bibinfo {author} {\bibfnamefont {A.~N.}\ \bibnamefont
  {Slavin}},\ }\href@noop {} {\bibfield  {journal} {\bibinfo  {journal} {Phys.
  Rep.}\ }\textbf {\bibinfo {volume} {348}},\ \bibinfo {pages} {441} (\bibinfo
  {year} {2001})}\BibitemShut {NoStop}%
\bibitem [{\citenamefont {Vincent}\ \emph {et~al.}(2005)\citenamefont
  {Vincent}, \citenamefont {Kruger}, \citenamefont {Elmazria}, \citenamefont
  {Bouvot}, \citenamefont {Mainka}, \citenamefont {Sanctuary}, \citenamefont
  {Rouxel},\ and\ \citenamefont {Alnot}}]{VincentJPD2005}%
  \BibitemOpen
  \bibfield  {author} {\bibinfo {author} {\bibfnamefont {B.}~\bibnamefont
  {Vincent}}, \bibinfo {author} {\bibfnamefont {J.~K.}\ \bibnamefont {Kruger}},
  \bibinfo {author} {\bibfnamefont {O.}~\bibnamefont {Elmazria}}, \bibinfo
  {author} {\bibfnamefont {L.}~\bibnamefont {Bouvot}}, \bibinfo {author}
  {\bibfnamefont {J.}~\bibnamefont {Mainka}}, \bibinfo {author} {\bibfnamefont
  {R.}~\bibnamefont {Sanctuary}}, \bibinfo {author} {\bibfnamefont
  {D.}~\bibnamefont {Rouxel}}, \ and\ \bibinfo {author} {\bibfnamefont
  {P.}~\bibnamefont {Alnot}},\ }\href@noop {} {\bibfield  {journal} {\bibinfo
  {journal} {J. Phys. D: Appl. Phys.}\ }\textbf {\bibinfo {volume} {38}},\
  \bibinfo {pages} {2026} (\bibinfo {year} {2005})}\BibitemShut {NoStop}%
\bibitem [{\citenamefont {Foerster}\ \emph {et~al.}(2017)\citenamefont
  {Foerster}, \citenamefont {Maci\`{a}}, \citenamefont {Statuto}, \citenamefont
  {Finizio}, \citenamefont {Hern\'{a}ndez-Minguez}, \citenamefont {Lendinez},
  \citenamefont {Santos}, \citenamefont {Fontcuberta}, \citenamefont
  {Manel~Hern\`{a}ndez}, \citenamefont {Kl\"{a}ui},\ and\ \citenamefont
  {Aballe}}]{FoersterNComm2017}%
  \BibitemOpen
  \bibfield  {author} {\bibinfo {author} {\bibfnamefont {M.}~\bibnamefont
  {Foerster}}, \bibinfo {author} {\bibfnamefont {F.}~\bibnamefont {Maci\`{a}}},
  \bibinfo {author} {\bibfnamefont {N.}~\bibnamefont {Statuto}}, \bibinfo
  {author} {\bibfnamefont {S.}~\bibnamefont {Finizio}}, \bibinfo {author}
  {\bibfnamefont {A.}~\bibnamefont {Hern\'{a}ndez-Minguez}}, \bibinfo {author}
  {\bibfnamefont {S.}~\bibnamefont {Lendinez}}, \bibinfo {author}
  {\bibfnamefont {P.~V.}\ \bibnamefont {Santos}}, \bibinfo {author}
  {\bibfnamefont {J.}~\bibnamefont {Fontcuberta}}, \bibinfo {author}
  {\bibfnamefont {J.}~\bibnamefont {Manel~Hern\`{a}ndez}}, \bibinfo {author}
  {\bibfnamefont {M.}~\bibnamefont {Kl\"{a}ui}}, \ and\ \bibinfo {author}
  {\bibfnamefont {L.}~\bibnamefont {Aballe}},\ }\href@noop {} {\bibfield
  {journal} {\bibinfo  {journal} {Nature Commun.}\ }\textbf {\bibinfo {volume}
  {8}},\ \bibinfo {pages} {407} (\bibinfo {year} {2017})}\BibitemShut {NoStop}%
\bibitem [{\citenamefont {Zhang}\ \emph
  {et~al.}(2020{\natexlab{a}})\citenamefont {Zhang}, \citenamefont {Li},
  \citenamefont {Weber}, \citenamefont {Goldberger}, \citenamefont {Mak},\ and\
  \citenamefont {Shan}}]{ZhangNMater2020}%
  \BibitemOpen
  \bibfield  {author} {\bibinfo {author} {\bibfnamefont {X.-X.}\ \bibnamefont
  {Zhang}}, \bibinfo {author} {\bibfnamefont {L.}~\bibnamefont {Li}}, \bibinfo
  {author} {\bibfnamefont {D.}~\bibnamefont {Weber}}, \bibinfo {author}
  {\bibfnamefont {J.}~\bibnamefont {Goldberger}}, \bibinfo {author}
  {\bibfnamefont {K.}~\bibnamefont {Mak}}, \ and\ \bibinfo {author}
  {\bibfnamefont {J.}~\bibnamefont {Shan}},\ }\href@noop {} {\bibfield
  {journal} {\bibinfo  {journal} {Nature Mater.}\ }\textbf {\bibinfo {volume}
  {19}},\ \bibinfo {pages} {838} (\bibinfo {year}
  {2020}{\natexlab{a}})}\BibitemShut {NoStop}%
\bibitem [{\citenamefont {Cenker}\ \emph {et~al.}(2020)\citenamefont {Cenker},
  \citenamefont {Huang}, \citenamefont {Suri}, \citenamefont {Thijssen},
  \citenamefont {Miller}, \citenamefont {Song}, \citenamefont {Taniguchi},
  \citenamefont {Watanabe}, \citenamefont {McGuire}, \citenamefont {Xiao},\
  and\ \citenamefont {Xu}}]{CenkerNPhys2020}%
  \BibitemOpen
  \bibfield  {author} {\bibinfo {author} {\bibfnamefont {J.}~\bibnamefont
  {Cenker}}, \bibinfo {author} {\bibfnamefont {B.}~\bibnamefont {Huang}},
  \bibinfo {author} {\bibfnamefont {N.}~\bibnamefont {Suri}}, \bibinfo {author}
  {\bibfnamefont {P.}~\bibnamefont {Thijssen}}, \bibinfo {author}
  {\bibfnamefont {A.}~\bibnamefont {Miller}}, \bibinfo {author} {\bibfnamefont
  {T.}~\bibnamefont {Song}}, \bibinfo {author} {\bibfnamefont {T.}~\bibnamefont
  {Taniguchi}}, \bibinfo {author} {\bibfnamefont {K.}~\bibnamefont {Watanabe}},
  \bibinfo {author} {\bibfnamefont {M.~A.}\ \bibnamefont {McGuire}}, \bibinfo
  {author} {\bibfnamefont {D.}~\bibnamefont {Xiao}}, \ and\ \bibinfo {author}
  {\bibfnamefont {X.}~\bibnamefont {Xu}},\ }\href@noop {} {\bibfield  {journal}
  {\bibinfo  {journal} {Nature Phys.}\ } (\bibinfo {year} {2020})}\BibitemShut
  {NoStop}%
\bibitem [{\citenamefont {Jin}\ \emph {et~al.}()\citenamefont {Jin},
  \citenamefont {Kim}, \citenamefont {Ye}, \citenamefont {Ye}, \citenamefont
  {Rojas}, \citenamefont {Luo}, \citenamefont {Yang}, \citenamefont {Yin},
  \citenamefont {Horng}, \citenamefont {Tian}, \citenamefont {Fu},
  \citenamefont {Deng}, \citenamefont {Lei}, \citenamefont {Sun}, \citenamefont
  {Tsen}, \citenamefont {He},\ and\ \citenamefont {Zhao}}]{JinarXiv2020}%
  \BibitemOpen
  \bibfield  {author} {\bibinfo {author} {\bibfnamefont {W.}~\bibnamefont
  {Jin}}, \bibinfo {author} {\bibfnamefont {H.}~\bibnamefont {Kim}}, \bibinfo
  {author} {\bibfnamefont {Z.}~\bibnamefont {Ye}}, \bibinfo {author}
  {\bibfnamefont {G.}~\bibnamefont {Ye}}, \bibinfo {author} {\bibfnamefont
  {L.}~\bibnamefont {Rojas}}, \bibinfo {author} {\bibfnamefont
  {X.}~\bibnamefont {Luo}}, \bibinfo {author} {\bibfnamefont {B.}~\bibnamefont
  {Yang}}, \bibinfo {author} {\bibfnamefont {F.}~\bibnamefont {Yin}}, \bibinfo
  {author} {\bibfnamefont {J.}~\bibnamefont {Horng}}, \bibinfo {author}
  {\bibfnamefont {S.}~\bibnamefont {Tian}}, \bibinfo {author} {\bibfnamefont
  {Y.}~\bibnamefont {Fu}}, \bibinfo {author} {\bibfnamefont {H.}~\bibnamefont
  {Deng}}, \bibinfo {author} {\bibfnamefont {H.}~\bibnamefont {Lei}}, \bibinfo
  {author} {\bibfnamefont {K.}~\bibnamefont {Sun}}, \bibinfo {author}
  {\bibfnamefont {A.~W.}\ \bibnamefont {Tsen}}, \bibinfo {author}
  {\bibfnamefont {R.}~\bibnamefont {He}}, \ and\ \bibinfo {author}
  {\bibfnamefont {L.}~\bibnamefont {Zhao}},\ }\href@noop {} {\bibfield
  {journal} {\bibinfo  {journal} {arXiv}\ }}\bibinfo {note}
  {2003.00176}\BibitemShut {NoStop}%
\bibitem [{\citenamefont {Bonetti}\ \emph {et~al.}(2015)\citenamefont
  {Bonetti}, \citenamefont {Kukreja}, \citenamefont {Chen}, \citenamefont
  {Maci\`{a}}, \citenamefont {Hern\`{a}ndez}, \citenamefont {Eklund},
  \citenamefont {Backes}, \citenamefont {Frisch}, \citenamefont {Katine},
  \citenamefont {Malm}, \citenamefont {Urazhdin}, \citenamefont {Kent},
  \citenamefont {St\"{o}hr}, \citenamefont {Ohldag},\ and\ \citenamefont
  {D\"{u}rr}}]{BonettiNcomm2015}%
  \BibitemOpen
  \bibfield  {author} {\bibinfo {author} {\bibfnamefont {S.}~\bibnamefont
  {Bonetti}}, \bibinfo {author} {\bibfnamefont {R.}~\bibnamefont {Kukreja}},
  \bibinfo {author} {\bibfnamefont {Z.}~\bibnamefont {Chen}}, \bibinfo {author}
  {\bibfnamefont {F.}~\bibnamefont {Maci\`{a}}}, \bibinfo {author}
  {\bibfnamefont {J.~M.}\ \bibnamefont {Hern\`{a}ndez}}, \bibinfo {author}
  {\bibfnamefont {A.}~\bibnamefont {Eklund}}, \bibinfo {author} {\bibfnamefont
  {D.}~\bibnamefont {Backes}}, \bibinfo {author} {\bibfnamefont
  {J.}~\bibnamefont {Frisch}}, \bibinfo {author} {\bibfnamefont
  {J.}~\bibnamefont {Katine}}, \bibinfo {author} {\bibfnamefont
  {G.}~\bibnamefont {Malm}}, \bibinfo {author} {\bibfnamefont {S.}~\bibnamefont
  {Urazhdin}}, \bibinfo {author} {\bibfnamefont {A.~D.}\ \bibnamefont {Kent}},
  \bibinfo {author} {\bibfnamefont {J.}~\bibnamefont {St\"{o}hr}}, \bibinfo
  {author} {\bibfnamefont {H.}~\bibnamefont {Ohldag}}, \ and\ \bibinfo {author}
  {\bibfnamefont {H.~A.}\ \bibnamefont {D\"{u}rr}},\ }\href@noop {} {\bibfield
  {journal} {\bibinfo  {journal} {Nature Commun.}\ }\textbf {\bibinfo {volume}
  {6}},\ \bibinfo {pages} {8889} (\bibinfo {year} {2015})}\BibitemShut
  {NoStop}%
\bibitem [{\citenamefont {Cheng}\ \emph {et~al.}(2017)\citenamefont {Cheng},
  \citenamefont {Cao},\ and\ \citenamefont {Bailey}}]{ChengJMMM2017}%
  \BibitemOpen
  \bibfield  {author} {\bibinfo {author} {\bibfnamefont {C.}~\bibnamefont
  {Cheng}}, \bibinfo {author} {\bibfnamefont {W.}~\bibnamefont {Cao}}, \ and\
  \bibinfo {author} {\bibfnamefont {W.~E.}\ \bibnamefont {Bailey}},\
  }\href@noop {} {\bibfield  {journal} {\bibinfo  {journal} {J. Magn. Magn.
  Mater.}\ }\textbf {\bibinfo {volume} {424}},\ \bibinfo {pages} {12} (\bibinfo
  {year} {2017})}\BibitemShut {NoStop}%
\bibitem [{\citenamefont {Bonetti}(2017)}]{BonettiJPhys2017}%
  \BibitemOpen
  \bibfield  {author} {\bibinfo {author} {\bibfnamefont {S.}~\bibnamefont
  {Bonetti}},\ }\href@noop {} {\bibfield  {journal} {\bibinfo  {journal} {J.
  Phys.: Condens. Matter}\ }\textbf {\bibinfo {volume} {29}},\ \bibinfo {pages}
  {133004} (\bibinfo {year} {2017})}\BibitemShut {NoStop}%
\bibitem [{\citenamefont {Sebastian}\ \emph {et~al.}(2015)\citenamefont
  {Sebastian}, \citenamefont {Schultheiss}, \citenamefont {Obry}, \citenamefont
  {Hillebrands},\ and\ \citenamefont {Schultheiss}}]{SebastianFrontPhys2015}%
  \BibitemOpen
  \bibfield  {author} {\bibinfo {author} {\bibfnamefont {T.}~\bibnamefont
  {Sebastian}}, \bibinfo {author} {\bibfnamefont {K.}~\bibnamefont
  {Schultheiss}}, \bibinfo {author} {\bibfnamefont {B.}~\bibnamefont {Obry}},
  \bibinfo {author} {\bibfnamefont {B.}~\bibnamefont {Hillebrands}}, \ and\
  \bibinfo {author} {\bibfnamefont {H.}~\bibnamefont {Schultheiss}},\
  }\href@noop {} {\bibfield  {journal} {\bibinfo  {journal} {Front. Phys.}\
  }\textbf {\bibinfo {volume} {3}},\ \bibinfo {pages} {35} (\bibinfo {year}
  {2015})}\BibitemShut {NoStop}%
\bibitem [{\citenamefont {Jersch}\ \emph {et~al.}(2010)\citenamefont {Jersch},
  \citenamefont {Demidov}, \citenamefont {Fuchs}, \citenamefont {Rott},
  \citenamefont {Krzysteczko}, \citenamefont {M\"{u}nchenberger}, \citenamefont
  {Reiss},\ and\ \citenamefont {Demokritov}}]{JerschAPL2010}%
  \BibitemOpen
  \bibfield  {author} {\bibinfo {author} {\bibfnamefont {J.}~\bibnamefont
  {Jersch}}, \bibinfo {author} {\bibfnamefont {V.~E.}\ \bibnamefont {Demidov}},
  \bibinfo {author} {\bibfnamefont {H.}~\bibnamefont {Fuchs}}, \bibinfo
  {author} {\bibfnamefont {K.}~\bibnamefont {Rott}}, \bibinfo {author}
  {\bibfnamefont {P.}~\bibnamefont {Krzysteczko}}, \bibinfo {author}
  {\bibfnamefont {J.}~\bibnamefont {M\"{u}nchenberger}}, \bibinfo {author}
  {\bibfnamefont {G.}~\bibnamefont {Reiss}}, \ and\ \bibinfo {author}
  {\bibfnamefont {S.~O.}\ \bibnamefont {Demokritov}},\ }\href@noop {}
  {\bibfield  {journal} {\bibinfo  {journal} {Appl. Phys. Lett.}\ }\textbf
  {\bibinfo {volume} {97}},\ \bibinfo {pages} {152502} (\bibinfo {year}
  {2010})}\BibitemShut {NoStop}%
\bibitem [{\citenamefont {Barnett}(1915)}]{BarnettPhysRev1915}%
  \BibitemOpen
  \bibfield  {author} {\bibinfo {author} {\bibfnamefont {S.~J.}\ \bibnamefont
  {Barnett}},\ }\href@noop {} {\bibfield  {journal} {\bibinfo  {journal} {Phys.
  Rev.}\ }\textbf {\bibinfo {volume} {6}},\ \bibinfo {pages} {239–270}
  (\bibinfo {year} {1915})}\BibitemShut {NoStop}%
\bibitem [{\citenamefont {Einstein}\ and\ \citenamefont
  {de~Haas}(1915)}]{EinsteindeHass1915}%
  \BibitemOpen
  \bibfield  {author} {\bibinfo {author} {\bibfnamefont {A.}~\bibnamefont
  {Einstein}}\ and\ \bibinfo {author} {\bibfnamefont {W.~J.}\ \bibnamefont
  {de~Haas}},\ }\href@noop {} {\bibfield  {journal} {\bibinfo  {journal} {Proc.
  Koninklijke Akademie van Wetenschappen te Amsterdam}\ }\textbf {\bibinfo
  {volume} {18}},\ \bibinfo {pages} {696–711} (\bibinfo {year}
  {1915})}\BibitemShut {NoStop}%
\bibitem [{\citenamefont {Zhang}\ and\ \citenamefont
  {Niu}(2014)}]{ZhangLifaPRL2014}%
  \BibitemOpen
  \bibfield  {author} {\bibinfo {author} {\bibfnamefont {L.}~\bibnamefont
  {Zhang}}\ and\ \bibinfo {author} {\bibfnamefont {Q.}~\bibnamefont {Niu}},\
  }\href {\doibase 10.1103/PhysRevLett.112.085503} {\bibfield  {journal}
  {\bibinfo  {journal} {Phys. Rev. Lett.}\ }\textbf {\bibinfo {volume} {112}},\
  \bibinfo {pages} {085503} (\bibinfo {year} {2014})}\BibitemShut {NoStop}%
\bibitem [{\citenamefont {Garanin}\ and\ \citenamefont
  {Chudnovsky}(2015)}]{GaraninPRB2015}%
  \BibitemOpen
  \bibfield  {author} {\bibinfo {author} {\bibfnamefont {D.~A.}\ \bibnamefont
  {Garanin}}\ and\ \bibinfo {author} {\bibfnamefont {E.~M.}\ \bibnamefont
  {Chudnovsky}},\ }\href {\doibase 10.1103/PhysRevB.92.024421} {\bibfield
  {journal} {\bibinfo  {journal} {Phys. Rev. B}\ }\textbf {\bibinfo {volume}
  {92}},\ \bibinfo {pages} {024421} (\bibinfo {year} {2015})}\BibitemShut
  {NoStop}%
\bibitem [{\citenamefont {Sasaki}\ \emph {et~al.}()\citenamefont {Sasaki},
  \citenamefont {Nii},\ and\ \citenamefont {Onose}}]{SasakiarXiv2020}%
  \BibitemOpen
  \bibfield  {author} {\bibinfo {author} {\bibfnamefont {R.}~\bibnamefont
  {Sasaki}}, \bibinfo {author} {\bibfnamefont {Y.}~\bibnamefont {Nii}}, \ and\
  \bibinfo {author} {\bibfnamefont {Y.}~\bibnamefont {Onose}},\ }\href@noop {}
  {\ }\bibinfo {note} {ArXiv:2007.03192}\BibitemShut {NoStop}%
\bibitem [{\citenamefont {Matsuo}\ \emph {et~al.}(2013)\citenamefont {Matsuo},
  \citenamefont {Ieda}, \citenamefont {Harii}, \citenamefont {Saitoh},\ and\
  \citenamefont {Maekawa}}]{MatsuoPRB2013}%
  \BibitemOpen
  \bibfield  {author} {\bibinfo {author} {\bibfnamefont {M.}~\bibnamefont
  {Matsuo}}, \bibinfo {author} {\bibfnamefont {J.}~\bibnamefont {Ieda}},
  \bibinfo {author} {\bibfnamefont {K.}~\bibnamefont {Harii}}, \bibinfo
  {author} {\bibfnamefont {E.}~\bibnamefont {Saitoh}}, \ and\ \bibinfo {author}
  {\bibfnamefont {S.}~\bibnamefont {Maekawa}},\ }\href@noop {} {\bibfield
  {journal} {\bibinfo  {journal} {Phys. Rev. B}\ }\textbf {\bibinfo {volume}
  {87}},\ \bibinfo {pages} {180402} (\bibinfo {year} {2013})}\BibitemShut
  {NoStop}%
\bibitem [{\citenamefont {Kobayashi}\ \emph {et~al.}(2017)\citenamefont
  {Kobayashi}, \citenamefont {Yoshikawa}, \citenamefont {Matsuo}, \citenamefont
  {Iguchi}, \citenamefont {Maekawa}, \citenamefont {Saitoh},\ and\
  \citenamefont {Nozaki}}]{KobayashiPRL2017}%
  \BibitemOpen
  \bibfield  {author} {\bibinfo {author} {\bibfnamefont {D.}~\bibnamefont
  {Kobayashi}}, \bibinfo {author} {\bibfnamefont {T.}~\bibnamefont
  {Yoshikawa}}, \bibinfo {author} {\bibfnamefont {M.}~\bibnamefont {Matsuo}},
  \bibinfo {author} {\bibfnamefont {R.}~\bibnamefont {Iguchi}}, \bibinfo
  {author} {\bibfnamefont {S.}~\bibnamefont {Maekawa}}, \bibinfo {author}
  {\bibfnamefont {E.}~\bibnamefont {Saitoh}}, \ and\ \bibinfo {author}
  {\bibfnamefont {Y.}~\bibnamefont {Nozaki}},\ }\href@noop {} {\bibfield
  {journal} {\bibinfo  {journal} {Phys. Rev. Lett.}\ }\textbf {\bibinfo
  {volume} {119}},\ \bibinfo {pages} {077202} (\bibinfo {year}
  {2017})}\BibitemShut {NoStop}%
\bibitem [{\citenamefont {Verba}\ \emph {et~al.}(2018)\citenamefont {Verba},
  \citenamefont {Lisenkov}, \citenamefont {Krivorotov}, \citenamefont
  {Tiberkevich},\ and\ \citenamefont {Slavin}}]{VerbaPRApplied2018}%
  \BibitemOpen
  \bibfield  {author} {\bibinfo {author} {\bibfnamefont {R.}~\bibnamefont
  {Verba}}, \bibinfo {author} {\bibfnamefont {I.}~\bibnamefont {Lisenkov}},
  \bibinfo {author} {\bibfnamefont {I.}~\bibnamefont {Krivorotov}}, \bibinfo
  {author} {\bibfnamefont {V.}~\bibnamefont {Tiberkevich}}, \ and\ \bibinfo
  {author} {\bibfnamefont {A.}~\bibnamefont {Slavin}},\ }\href@noop {}
  {\bibfield  {journal} {\bibinfo  {journal} {Phys. Rev. Applied}\ }\textbf
  {\bibinfo {volume} {9}},\ \bibinfo {pages} {064014} (\bibinfo {year}
  {2018})}\BibitemShut {NoStop}%
\bibitem [{\citenamefont {Verba}\ \emph {et~al.}(2019)\citenamefont {Verba},
  \citenamefont {Tiberkevich},\ and\ \citenamefont
  {Slavin}}]{VerbaPRApplied2019}%
  \BibitemOpen
  \bibfield  {author} {\bibinfo {author} {\bibfnamefont {R.}~\bibnamefont
  {Verba}}, \bibinfo {author} {\bibfnamefont {V.}~\bibnamefont {Tiberkevich}},
  \ and\ \bibinfo {author} {\bibfnamefont {A.}~\bibnamefont {Slavin}},\
  }\href@noop {} {\bibfield  {journal} {\bibinfo  {journal} {Phys. Rev.
  Applied}\ }\textbf {\bibinfo {volume} {12}},\ \bibinfo {pages} {054061}
  (\bibinfo {year} {2019})}\BibitemShut {NoStop}%
\bibitem [{\citenamefont {Zhang}\ \emph
  {et~al.}(2020{\natexlab{b}})\citenamefont {Zhang}, \citenamefont {Bauer},\
  and\ \citenamefont {Yu}}]{BauerPRB2020}%
  \BibitemOpen
  \bibfield  {author} {\bibinfo {author} {\bibfnamefont {X.}~\bibnamefont
  {Zhang}}, \bibinfo {author} {\bibfnamefont {G.~E.~W.}\ \bibnamefont {Bauer}},
  \ and\ \bibinfo {author} {\bibfnamefont {T.}~\bibnamefont {Yu}},\ }\href
  {\doibase 10.1103/PhysRevLett.125.077203} {\bibfield  {journal} {\bibinfo
  {journal} {Phys. Rev. Lett.}\ }\textbf {\bibinfo {volume} {125}},\ \bibinfo
  {pages} {077203} (\bibinfo {year} {2020}{\natexlab{b}})}\BibitemShut
  {NoStop}%
\bibitem [{\citenamefont {Yamamoto}\ \emph {et~al.}(2020)\citenamefont
  {Yamamoto}, \citenamefont {Yu}, \citenamefont {Yu}, \citenamefont {Puebla},
  \citenamefont {Xu}, \citenamefont {Maekawa},\ and\ \citenamefont
  {Bauer}}]{YamamotoJPSJ2020}%
  \BibitemOpen
  \bibfield  {author} {\bibinfo {author} {\bibfnamefont {K.}~\bibnamefont
  {Yamamoto}}, \bibinfo {author} {\bibfnamefont {W.}~\bibnamefont {Yu}},
  \bibinfo {author} {\bibfnamefont {T.}~\bibnamefont {Yu}}, \bibinfo {author}
  {\bibfnamefont {J.}~\bibnamefont {Puebla}}, \bibinfo {author} {\bibfnamefont
  {M.}~\bibnamefont {Xu}}, \bibinfo {author} {\bibfnamefont {S.}~\bibnamefont
  {Maekawa}}, \ and\ \bibinfo {author} {\bibfnamefont {G.}~\bibnamefont
  {Bauer}},\ }\href@noop {} {\bibfield  {journal} {\bibinfo  {journal} {Journal
  of the Physical Society of Japan}\ }\textbf {\bibinfo {volume} {89}},\
  \bibinfo {pages} {113702} (\bibinfo {year} {2020})}\BibitemShut {NoStop}%
\bibitem [{\citenamefont {Shah}\ \emph {et~al.}(2020)\citenamefont {Shah},
  \citenamefont {Bas}, \citenamefont {Lisenkov}, \citenamefont {Matyushov},
  \citenamefont {Sun},\ and\ \citenamefont {Page}}]{ShahSciAdv2020}%
  \BibitemOpen
  \bibfield  {author} {\bibinfo {author} {\bibfnamefont {P.~J.}\ \bibnamefont
  {Shah}}, \bibinfo {author} {\bibfnamefont {D.~A.}\ \bibnamefont {Bas}},
  \bibinfo {author} {\bibfnamefont {I.}~\bibnamefont {Lisenkov}}, \bibinfo
  {author} {\bibfnamefont {A.}~\bibnamefont {Matyushov}}, \bibinfo {author}
  {\bibfnamefont {N.~X.}\ \bibnamefont {Sun}}, \ and\ \bibinfo {author}
  {\bibfnamefont {M.~R.}\ \bibnamefont {Page}},\ }\href@noop {} {\bibfield
  {journal} {\bibinfo  {journal} {Sci. Adv.}\ }\textbf {\bibinfo {volume}
  {6}},\ \bibinfo {pages} {eabc5648} (\bibinfo {year} {2020})}\BibitemShut
  {NoStop}%
\bibitem [{\citenamefont {Xu}\ \emph {et~al.}(2020)\citenamefont {Xu},
  \citenamefont {Yamamoto}, \citenamefont {Puebla}, \citenamefont {Baumgaertl},
  \citenamefont {Rana}, \citenamefont {Miura}, \citenamefont {Takahashi},
  \citenamefont {Grundler}, \citenamefont {Maekawa},\ and\ \citenamefont
  {Otani}}]{XuSciAdv2020}%
  \BibitemOpen
  \bibfield  {author} {\bibinfo {author} {\bibfnamefont {M.}~\bibnamefont
  {Xu}}, \bibinfo {author} {\bibfnamefont {K.}~\bibnamefont {Yamamoto}},
  \bibinfo {author} {\bibfnamefont {J.}~\bibnamefont {Puebla}}, \bibinfo
  {author} {\bibfnamefont {K.}~\bibnamefont {Baumgaertl}}, \bibinfo {author}
  {\bibfnamefont {B.}~\bibnamefont {Rana}}, \bibinfo {author} {\bibfnamefont
  {K.}~\bibnamefont {Miura}}, \bibinfo {author} {\bibfnamefont
  {H.}~\bibnamefont {Takahashi}}, \bibinfo {author} {\bibfnamefont
  {D.}~\bibnamefont {Grundler}}, \bibinfo {author} {\bibfnamefont
  {S.}~\bibnamefont {Maekawa}}, \ and\ \bibinfo {author} {\bibfnamefont
  {Y.}~\bibnamefont {Otani}},\ }\href@noop {} {\bibfield  {journal} {\bibinfo
  {journal} {Sci. Adv.}\ }\textbf {\bibinfo {volume} {6}},\ \bibinfo {pages}
  {eabb1724} (\bibinfo {year} {2020})}\BibitemShut {NoStop}%
\bibitem [{\citenamefont {Nembach}\ \emph {et~al.}(2015)\citenamefont
  {Nembach}, \citenamefont {Shaw}, \citenamefont {Weiler}, \citenamefont
  {Ju\'{e}},\ and\ \citenamefont {Silva}}]{NembachNPhys2015}%
  \BibitemOpen
  \bibfield  {author} {\bibinfo {author} {\bibfnamefont {H.~T.}\ \bibnamefont
  {Nembach}}, \bibinfo {author} {\bibfnamefont {J.~M.}\ \bibnamefont {Shaw}},
  \bibinfo {author} {\bibfnamefont {M.}~\bibnamefont {Weiler}}, \bibinfo
  {author} {\bibfnamefont {E.}~\bibnamefont {Ju\'{e}}}, \ and\ \bibinfo
  {author} {\bibfnamefont {T.~J.}\ \bibnamefont {Silva}},\ }\href@noop {}
  {\bibfield  {journal} {\bibinfo  {journal} {Nature Phys.}\ }\textbf {\bibinfo
  {volume} {11}},\ \bibinfo {pages} {825} (\bibinfo {year} {2015})}\BibitemShut
  {NoStop}%
\bibitem [{\citenamefont {K\"u\ss{}}\ \emph {et~al.}(2020)\citenamefont
  {K\"u\ss{}}, \citenamefont {Heigl}, \citenamefont {Flacke}, \citenamefont
  {H\"orner}, \citenamefont {Weiler}, \citenamefont {Albrecht},\ and\
  \citenamefont {Wixforth}}]{KussPRL2020}%
  \BibitemOpen
  \bibfield  {author} {\bibinfo {author} {\bibfnamefont {M.}~\bibnamefont
  {K\"u\ss{}}}, \bibinfo {author} {\bibfnamefont {M.}~\bibnamefont {Heigl}},
  \bibinfo {author} {\bibfnamefont {L.}~\bibnamefont {Flacke}}, \bibinfo
  {author} {\bibfnamefont {A.}~\bibnamefont {H\"orner}}, \bibinfo {author}
  {\bibfnamefont {M.}~\bibnamefont {Weiler}}, \bibinfo {author} {\bibfnamefont
  {M.}~\bibnamefont {Albrecht}}, \ and\ \bibinfo {author} {\bibfnamefont
  {A.}~\bibnamefont {Wixforth}},\ }\href@noop {} {\bibfield  {journal}
  {\bibinfo  {journal} {Phys. Rev. Lett.}\ }\textbf {\bibinfo {volume} {125}},\
  \bibinfo {pages} {217203} (\bibinfo {year} {2020})}\BibitemShut {NoStop}%
\bibitem [{\citenamefont {Hern\'andez-M\'{\i}nguez}\ \emph
  {et~al.}(2020)\citenamefont {Hern\'andez-M\'{\i}nguez}, \citenamefont
  {Maci\`a}, \citenamefont {Hern\`andez}, \citenamefont {Herfort},\ and\
  \citenamefont {Santos}}]{HernandezPRApplied2020}%
  \BibitemOpen
  \bibfield  {author} {\bibinfo {author} {\bibfnamefont {A.}~\bibnamefont
  {Hern\'andez-M\'{\i}nguez}}, \bibinfo {author} {\bibfnamefont
  {F.}~\bibnamefont {Maci\`a}}, \bibinfo {author} {\bibfnamefont {J.~M.}\
  \bibnamefont {Hern\`andez}}, \bibinfo {author} {\bibfnamefont
  {J.}~\bibnamefont {Herfort}}, \ and\ \bibinfo {author} {\bibfnamefont
  {P.~V.}\ \bibnamefont {Santos}},\ }\href@noop {} {\bibfield  {journal}
  {\bibinfo  {journal} {Phys. Rev. Applied}\ }\textbf {\bibinfo {volume}
  {13}},\ \bibinfo {pages} {044018} (\bibinfo {year} {2020})}\BibitemShut
  {NoStop}%
\bibitem [{\citenamefont {Kamra}\ \emph {et~al.}(2015)\citenamefont {Kamra},
  \citenamefont {Keshtgar}, \citenamefont {Yan},\ and\ \citenamefont
  {Bauer}}]{KamraPRB2015}%
  \BibitemOpen
  \bibfield  {author} {\bibinfo {author} {\bibfnamefont {A.}~\bibnamefont
  {Kamra}}, \bibinfo {author} {\bibfnamefont {H.}~\bibnamefont {Keshtgar}},
  \bibinfo {author} {\bibfnamefont {P.}~\bibnamefont {Yan}}, \ and\ \bibinfo
  {author} {\bibfnamefont {G.~E.~W.}\ \bibnamefont {Bauer}},\ }\href {\doibase
  10.1103/PhysRevB.91.104409} {\bibfield  {journal} {\bibinfo  {journal} {Phys.
  Rev. B}\ }\textbf {\bibinfo {volume} {91}},\ \bibinfo {pages} {104409}
  (\bibinfo {year} {2015})}\BibitemShut {NoStop}%
\bibitem [{\citenamefont {Flebus}\ \emph {et~al.}(2017)\citenamefont {Flebus},
  \citenamefont {Shen}, \citenamefont {Kikkawa}, \citenamefont {Uchida},
  \citenamefont {Qiu}, \citenamefont {Saitoh}, \citenamefont {Duine},\ and\
  \citenamefont {Bauer}}]{FlebusPRB2017}%
  \BibitemOpen
  \bibfield  {author} {\bibinfo {author} {\bibfnamefont {B.}~\bibnamefont
  {Flebus}}, \bibinfo {author} {\bibfnamefont {K.}~\bibnamefont {Shen}},
  \bibinfo {author} {\bibfnamefont {T.}~\bibnamefont {Kikkawa}}, \bibinfo
  {author} {\bibfnamefont {K.-i.}\ \bibnamefont {Uchida}}, \bibinfo {author}
  {\bibfnamefont {Z.}~\bibnamefont {Qiu}}, \bibinfo {author} {\bibfnamefont
  {E.}~\bibnamefont {Saitoh}}, \bibinfo {author} {\bibfnamefont {R.~A.}\
  \bibnamefont {Duine}}, \ and\ \bibinfo {author} {\bibfnamefont {G.~E.~W.}\
  \bibnamefont {Bauer}},\ }\href {\doibase 10.1103/PhysRevB.95.144420}
  {\bibfield  {journal} {\bibinfo  {journal} {Phys. Rev. B}\ }\textbf {\bibinfo
  {volume} {95}},\ \bibinfo {pages} {144420} (\bibinfo {year}
  {2017})}\BibitemShut {NoStop}%
\bibitem [{\citenamefont {Cornelissen}\ \emph {et~al.}(2017)\citenamefont
  {Cornelissen}, \citenamefont {Oyanagi}, \citenamefont {Kikkawa},
  \citenamefont {Qiu}, \citenamefont {Kuschel}, \citenamefont {Bauer},
  \citenamefont {van Wees},\ and\ \citenamefont {Saitoh}}]{CornelissenPRB2017}%
  \BibitemOpen
  \bibfield  {author} {\bibinfo {author} {\bibfnamefont {L.~J.}\ \bibnamefont
  {Cornelissen}}, \bibinfo {author} {\bibfnamefont {K.}~\bibnamefont
  {Oyanagi}}, \bibinfo {author} {\bibfnamefont {T.}~\bibnamefont {Kikkawa}},
  \bibinfo {author} {\bibfnamefont {Z.}~\bibnamefont {Qiu}}, \bibinfo {author}
  {\bibfnamefont {T.}~\bibnamefont {Kuschel}}, \bibinfo {author} {\bibfnamefont
  {G.~E.~W.}\ \bibnamefont {Bauer}}, \bibinfo {author} {\bibfnamefont {B.~J.}\
  \bibnamefont {van Wees}}, \ and\ \bibinfo {author} {\bibfnamefont
  {E.}~\bibnamefont {Saitoh}},\ }\href@noop {} {\bibfield  {journal} {\bibinfo
  {journal} {Phys. Rev. B}\ }\textbf {\bibinfo {volume} {96}},\ \bibinfo
  {pages} {104441} (\bibinfo {year} {2017})}\BibitemShut {NoStop}%
\bibitem [{\citenamefont {Streib}\ \emph {et~al.}(2018)\citenamefont {Streib},
  \citenamefont {Keshtgar},\ and\ \citenamefont {Bauer}}]{StreibPRL2018}%
  \BibitemOpen
  \bibfield  {author} {\bibinfo {author} {\bibfnamefont {S.}~\bibnamefont
  {Streib}}, \bibinfo {author} {\bibfnamefont {H.}~\bibnamefont {Keshtgar}}, \
  and\ \bibinfo {author} {\bibfnamefont {G.~E.~W.}\ \bibnamefont {Bauer}},\
  }\href {\doibase 10.1103/PhysRevLett.121.027202} {\bibfield  {journal}
  {\bibinfo  {journal} {Phys. Rev. Lett.}\ }\textbf {\bibinfo {volume} {121}},\
  \bibinfo {pages} {027202} (\bibinfo {year} {2018})}\BibitemShut {NoStop}%
\bibitem [{\citenamefont {Ogawa}\ \emph {et~al.}(2015)\citenamefont {Ogawa},
  \citenamefont {Koshibae}, \citenamefont {Beekman}, \citenamefont {Nagaosa},
  \citenamefont {Kubota}, \citenamefont {Kawasaki},\ and\ \citenamefont
  {Tokura}}]{OgawaPNAS2015}%
  \BibitemOpen
  \bibfield  {author} {\bibinfo {author} {\bibfnamefont {N.}~\bibnamefont
  {Ogawa}}, \bibinfo {author} {\bibfnamefont {W.}~\bibnamefont {Koshibae}},
  \bibinfo {author} {\bibfnamefont {A.~J.}\ \bibnamefont {Beekman}}, \bibinfo
  {author} {\bibfnamefont {N.}~\bibnamefont {Nagaosa}}, \bibinfo {author}
  {\bibfnamefont {M.}~\bibnamefont {Kubota}}, \bibinfo {author} {\bibfnamefont
  {M.}~\bibnamefont {Kawasaki}}, \ and\ \bibinfo {author} {\bibfnamefont
  {Y.}~\bibnamefont {Tokura}},\ }\href@noop {} {\bibfield  {journal} {\bibinfo
  {journal} {Proc. Natl. Acad. Sci.}\ }\textbf {\bibinfo {volume} {112}},\
  \bibinfo {pages} {8977} (\bibinfo {year} {2015})}\BibitemShut {NoStop}%
\bibitem [{\citenamefont {Hashimoto}\ \emph {et~al.}(2017)\citenamefont
  {Hashimoto}, \citenamefont {Daimon}, \citenamefont {Iguchi}, \citenamefont
  {Oikawa}, \citenamefont {Sato}, \citenamefont {Bossini}, \citenamefont
  {Tabuchi}, \citenamefont {Satoh}, \citenamefont {Hillebrands}, \citenamefont
  {Bauer}, \citenamefont {Johansen}, \citenamefont {Kirilyuk}, \citenamefont
  {Rasing},\ and\ \citenamefont {Saitoh}}]{HashimotoNComm2017}%
  \BibitemOpen
  \bibfield  {author} {\bibinfo {author} {\bibfnamefont {Y.}~\bibnamefont
  {Hashimoto}}, \bibinfo {author} {\bibfnamefont {S.}~\bibnamefont {Daimon}},
  \bibinfo {author} {\bibfnamefont {R.}~\bibnamefont {Iguchi}}, \bibinfo
  {author} {\bibfnamefont {K.}~\bibnamefont {Oikawa}, \bibfnamefont
  {Y.~an~Shen}}, \bibinfo {author} {\bibfnamefont {K.}~\bibnamefont {Sato}},
  \bibinfo {author} {\bibfnamefont {D.}~\bibnamefont {Bossini}}, \bibinfo
  {author} {\bibfnamefont {Y.}~\bibnamefont {Tabuchi}}, \bibinfo {author}
  {\bibfnamefont {T.}~\bibnamefont {Satoh}}, \bibinfo {author} {\bibfnamefont
  {B.}~\bibnamefont {Hillebrands}}, \bibinfo {author} {\bibfnamefont
  {G.~E.~W.}\ \bibnamefont {Bauer}}, \bibinfo {author} {\bibfnamefont {T.~H.}\
  \bibnamefont {Johansen}}, \bibinfo {author} {\bibfnamefont {A.}~\bibnamefont
  {Kirilyuk}}, \bibinfo {author} {\bibfnamefont {T.}~\bibnamefont {Rasing}}, \
  and\ \bibinfo {author} {\bibfnamefont {E.}~\bibnamefont {Saitoh}},\
  }\href@noop {} {\bibfield  {journal} {\bibinfo  {journal} {Nature Commun.}\
  }\textbf {\bibinfo {volume} {8}},\ \bibinfo {pages} {15859} (\bibinfo {year}
  {2017})}\BibitemShut {NoStop}%
\bibitem [{\citenamefont {Li}\ \emph {et~al.}(2019{\natexlab{b}})\citenamefont
  {Li}, \citenamefont {Zeng}, \citenamefont {Zhang}, \citenamefont {Shin},
  \citenamefont {Saglam}, \citenamefont {Karakas}, \citenamefont {Ozatay},
  \citenamefont {Pearson}, \citenamefont {Heinonen}, \citenamefont {Wu},
  \citenamefont {Hoffmann},\ and\ \citenamefont {Zhang}}]{LiPRL2019_CoFe}%
  \BibitemOpen
  \bibfield  {author} {\bibinfo {author} {\bibfnamefont {Y.}~\bibnamefont
  {Li}}, \bibinfo {author} {\bibfnamefont {F.}~\bibnamefont {Zeng}}, \bibinfo
  {author} {\bibfnamefont {S.~S.-L.}\ \bibnamefont {Zhang}}, \bibinfo {author}
  {\bibfnamefont {H.}~\bibnamefont {Shin}}, \bibinfo {author} {\bibfnamefont
  {H.}~\bibnamefont {Saglam}}, \bibinfo {author} {\bibfnamefont
  {V.}~\bibnamefont {Karakas}}, \bibinfo {author} {\bibfnamefont
  {O.}~\bibnamefont {Ozatay}}, \bibinfo {author} {\bibfnamefont {J.~E.}\
  \bibnamefont {Pearson}}, \bibinfo {author} {\bibfnamefont {O.~G.}\
  \bibnamefont {Heinonen}}, \bibinfo {author} {\bibfnamefont {Y.}~\bibnamefont
  {Wu}}, \bibinfo {author} {\bibfnamefont {A.}~\bibnamefont {Hoffmann}}, \ and\
  \bibinfo {author} {\bibfnamefont {W.}~\bibnamefont {Zhang}},\ }\href@noop {}
  {\bibfield  {journal} {\bibinfo  {journal} {Phys. Rev. Lett.}\ }\textbf
  {\bibinfo {volume} {122}},\ \bibinfo {pages} {117203} (\bibinfo {year}
  {2019}{\natexlab{b}})}\BibitemShut {NoStop}%
\bibitem [{\citenamefont {R\"uckriegel}\ and\ \citenamefont
  {Duine}(2020)}]{RuckriegelPRL2020}%
  \BibitemOpen
  \bibfield  {author} {\bibinfo {author} {\bibfnamefont {A.}~\bibnamefont
  {R\"uckriegel}}\ and\ \bibinfo {author} {\bibfnamefont {R.~A.}\ \bibnamefont
  {Duine}},\ }\href@noop {} {\bibfield  {journal} {\bibinfo  {journal} {Phys.
  Rev. Lett.}\ }\textbf {\bibinfo {volume} {124}},\ \bibinfo {pages} {117201}
  (\bibinfo {year} {2020})}\BibitemShut {NoStop}%
\bibitem [{\citenamefont {Lachance-Quirion}\ \emph {et~al.}(2019)\citenamefont
  {Lachance-Quirion}, \citenamefont {Tabuchi}, \citenamefont {Gloppe},
  \citenamefont {Usami},\ and\ \citenamefont
  {Nakamura}}]{LachanceQuirionAPEx2019}%
  \BibitemOpen
  \bibfield  {author} {\bibinfo {author} {\bibfnamefont {D.}~\bibnamefont
  {Lachance-Quirion}}, \bibinfo {author} {\bibfnamefont {Y.}~\bibnamefont
  {Tabuchi}}, \bibinfo {author} {\bibfnamefont {A.}~\bibnamefont {Gloppe}},
  \bibinfo {author} {\bibfnamefont {K.}~\bibnamefont {Usami}}, \ and\ \bibinfo
  {author} {\bibfnamefont {Y.}~\bibnamefont {Nakamura}},\ }\href@noop {}
  {\bibfield  {journal} {\bibinfo  {journal} {Appl. Phys. Express}\ }\textbf
  {\bibinfo {volume} {12}},\ \bibinfo {pages} {070101} (\bibinfo {year}
  {2019})}\BibitemShut {NoStop}%
\bibitem [{\citenamefont {Li}\ \emph {et~al.}(2020{\natexlab{a}})\citenamefont
  {Li}, \citenamefont {Zhang}, \citenamefont {Tyberkevych}, \citenamefont
  {Kwok}, \citenamefont {Hoffmann},\ and\ \citenamefont
  {Novosad}}]{LiJAPPerspective2020}%
  \BibitemOpen
  \bibfield  {author} {\bibinfo {author} {\bibfnamefont {Y.}~\bibnamefont
  {Li}}, \bibinfo {author} {\bibfnamefont {W.}~\bibnamefont {Zhang}}, \bibinfo
  {author} {\bibfnamefont {V.}~\bibnamefont {Tyberkevych}}, \bibinfo {author}
  {\bibfnamefont {W.-K.}\ \bibnamefont {Kwok}}, \bibinfo {author}
  {\bibfnamefont {A.}~\bibnamefont {Hoffmann}}, \ and\ \bibinfo {author}
  {\bibfnamefont {V.}~\bibnamefont {Novosad}},\ }\href@noop {} {\bibfield
  {journal} {\bibinfo  {journal} {J. Appl. Phys.}\ }\textbf {\bibinfo {volume}
  {128}},\ \bibinfo {pages} {130902} (\bibinfo {year}
  {2020}{\natexlab{a}})}\BibitemShut {NoStop}%
\bibitem [{\citenamefont {Li}\ \emph {et~al.}(2020{\natexlab{b}})\citenamefont
  {Li}, \citenamefont {Cao}, \citenamefont {Amin}, \citenamefont {Zhang},
  \citenamefont {Gibbons}, \citenamefont {Sklenar}, \citenamefont {Pearson},
  \citenamefont {Haney}, \citenamefont {Stiles}, \citenamefont {Bailey},
  \citenamefont {Novosad}, \citenamefont {Hoffmann},\ and\ \citenamefont
  {Zhang}}]{LiPRL2020_YIGPy}%
  \BibitemOpen
  \bibfield  {author} {\bibinfo {author} {\bibfnamefont {Y.}~\bibnamefont
  {Li}}, \bibinfo {author} {\bibfnamefont {W.}~\bibnamefont {Cao}}, \bibinfo
  {author} {\bibfnamefont {V.~P.}\ \bibnamefont {Amin}}, \bibinfo {author}
  {\bibfnamefont {Z.}~\bibnamefont {Zhang}}, \bibinfo {author} {\bibfnamefont
  {J.}~\bibnamefont {Gibbons}}, \bibinfo {author} {\bibfnamefont
  {J.}~\bibnamefont {Sklenar}}, \bibinfo {author} {\bibfnamefont
  {J.}~\bibnamefont {Pearson}}, \bibinfo {author} {\bibfnamefont {P.~M.}\
  \bibnamefont {Haney}}, \bibinfo {author} {\bibfnamefont {M.~D.}\ \bibnamefont
  {Stiles}}, \bibinfo {author} {\bibfnamefont {W.~E.}\ \bibnamefont {Bailey}},
  \bibinfo {author} {\bibfnamefont {V.}~\bibnamefont {Novosad}}, \bibinfo
  {author} {\bibfnamefont {A.}~\bibnamefont {Hoffmann}}, \ and\ \bibinfo
  {author} {\bibfnamefont {W.}~\bibnamefont {Zhang}},\ }\href@noop {}
  {\bibfield  {journal} {\bibinfo  {journal} {Phys. Rev. Lett.}\ }\textbf
  {\bibinfo {volume} {124}},\ \bibinfo {pages} {117202} (\bibinfo {year}
  {2020}{\natexlab{b}})}\BibitemShut {NoStop}%
\bibitem [{\citenamefont {Singh}\ \emph {et~al.}(2017)\citenamefont {Singh},
  \citenamefont {Khodadadi}, \citenamefont {Mohammadi}, \citenamefont
  {Keshavarz}, \citenamefont {Mewes}, \citenamefont {Negi}, \citenamefont
  {Datta}, \citenamefont {Galazka}, \citenamefont {Uecker},\ and\ \citenamefont
  {Gupta}}]{SinghAM2017}%
  \BibitemOpen
  \bibfield  {author} {\bibinfo {author} {\bibfnamefont {A.~V.}\ \bibnamefont
  {Singh}}, \bibinfo {author} {\bibfnamefont {B.}~\bibnamefont {Khodadadi}},
  \bibinfo {author} {\bibfnamefont {J.~B.}\ \bibnamefont {Mohammadi}}, \bibinfo
  {author} {\bibfnamefont {S.}~\bibnamefont {Keshavarz}}, \bibinfo {author}
  {\bibfnamefont {T.}~\bibnamefont {Mewes}}, \bibinfo {author} {\bibfnamefont
  {D.~S.}\ \bibnamefont {Negi}}, \bibinfo {author} {\bibfnamefont
  {R.}~\bibnamefont {Datta}}, \bibinfo {author} {\bibfnamefont
  {Z.}~\bibnamefont {Galazka}}, \bibinfo {author} {\bibfnamefont
  {R.}~\bibnamefont {Uecker}}, \ and\ \bibinfo {author} {\bibfnamefont
  {A.}~\bibnamefont {Gupta}},\ }\href@noop {} {\bibfield  {journal} {\bibinfo
  {journal} {Adv. Mater.}\ }\textbf {\bibinfo {volume} {29}},\ \bibinfo {pages}
  {1701222} (\bibinfo {year} {2017})}\BibitemShut {NoStop}%
\bibitem [{\citenamefont {Zhao}\ \emph {et~al.}(2020)\citenamefont {Zhao},
  \citenamefont {Li}, \citenamefont {Zhang}, \citenamefont {Vogel},
  \citenamefont {Pearson}, \citenamefont {Wang}, \citenamefont {Zhang},
  \citenamefont {Novosad}, \citenamefont {Liu},\ and\ \citenamefont
  {Hoffmann}}]{ZhaoPRApplied2020}%
  \BibitemOpen
  \bibfield  {author} {\bibinfo {author} {\bibfnamefont {C.}~\bibnamefont
  {Zhao}}, \bibinfo {author} {\bibfnamefont {Y.}~\bibnamefont {Li}}, \bibinfo
  {author} {\bibfnamefont {Z.}~\bibnamefont {Zhang}}, \bibinfo {author}
  {\bibfnamefont {M.}~\bibnamefont {Vogel}}, \bibinfo {author} {\bibfnamefont
  {J.~E.}\ \bibnamefont {Pearson}}, \bibinfo {author} {\bibfnamefont
  {J.}~\bibnamefont {Wang}}, \bibinfo {author} {\bibfnamefont {W.}~\bibnamefont
  {Zhang}}, \bibinfo {author} {\bibfnamefont {V.}~\bibnamefont {Novosad}},
  \bibinfo {author} {\bibfnamefont {Q.}~\bibnamefont {Liu}}, \ and\ \bibinfo
  {author} {\bibfnamefont {A.}~\bibnamefont {Hoffmann}},\ }\href@noop {}
  {\bibfield  {journal} {\bibinfo  {journal} {Phys. Rev. Applied}\ }\textbf
  {\bibinfo {volume} {13}},\ \bibinfo {pages} {054032} (\bibinfo {year}
  {2020})}\BibitemShut {NoStop}%
\bibitem [{\citenamefont {Gustafsson}\ \emph {et~al.}(2014)\citenamefont
  {Gustafsson}, \citenamefont {Aref}, \citenamefont {Kockum}, \citenamefont
  {Ekstr\"{o}m}, \citenamefont {Johansson},\ and\ \citenamefont
  {Delsing}}]{GustafssonScience2014}%
  \BibitemOpen
  \bibfield  {author} {\bibinfo {author} {\bibfnamefont {M.~V.}\ \bibnamefont
  {Gustafsson}}, \bibinfo {author} {\bibfnamefont {T.}~\bibnamefont {Aref}},
  \bibinfo {author} {\bibfnamefont {A.~F.}\ \bibnamefont {Kockum}}, \bibinfo
  {author} {\bibfnamefont {M.~K.}\ \bibnamefont {Ekstr\"{o}m}}, \bibinfo
  {author} {\bibfnamefont {G.}~\bibnamefont {Johansson}}, \ and\ \bibinfo
  {author} {\bibfnamefont {P.}~\bibnamefont {Delsing}},\ }\href@noop {}
  {\bibfield  {journal} {\bibinfo  {journal} {Science}\ }\textbf {\bibinfo
  {volume} {346}},\ \bibinfo {pages} {207} (\bibinfo {year}
  {2014})}\BibitemShut {NoStop}%
\bibitem [{\citenamefont {Manenti}\ \emph {et~al.}(2017)\citenamefont
  {Manenti}, \citenamefont {Kockum}, \citenamefont {Patterson}, \citenamefont
  {Behrle}, \citenamefont {Rahamim}, \citenamefont {Tancredi}, \citenamefont
  {Nori},\ and\ \citenamefont {Leek}}]{ManentiNComm2017}%
  \BibitemOpen
  \bibfield  {author} {\bibinfo {author} {\bibfnamefont {R.}~\bibnamefont
  {Manenti}}, \bibinfo {author} {\bibfnamefont {A.~F.}\ \bibnamefont {Kockum}},
  \bibinfo {author} {\bibfnamefont {A.}~\bibnamefont {Patterson}}, \bibinfo
  {author} {\bibfnamefont {T.}~\bibnamefont {Behrle}}, \bibinfo {author}
  {\bibfnamefont {J.}~\bibnamefont {Rahamim}}, \bibinfo {author} {\bibfnamefont
  {G.}~\bibnamefont {Tancredi}}, \bibinfo {author} {\bibfnamefont
  {F.}~\bibnamefont {Nori}}, \ and\ \bibinfo {author} {\bibfnamefont {P.~J.}\
  \bibnamefont {Leek}},\ }\href@noop {} {\bibfield  {journal} {\bibinfo
  {journal} {Nature Commun.}\ }\textbf {\bibinfo {volume} {8}},\ \bibinfo
  {pages} {975} (\bibinfo {year} {2017})}\BibitemShut {NoStop}%
\bibitem [{\citenamefont {Noguchi}\ \emph {et~al.}(2017)\citenamefont
  {Noguchi}, \citenamefont {Yamazaki}, \citenamefont {Tabuchi},\ and\
  \citenamefont {Nakamura}}]{NoguchiPRL2017}%
  \BibitemOpen
  \bibfield  {author} {\bibinfo {author} {\bibfnamefont {A.}~\bibnamefont
  {Noguchi}}, \bibinfo {author} {\bibfnamefont {R.}~\bibnamefont {Yamazaki}},
  \bibinfo {author} {\bibfnamefont {Y.}~\bibnamefont {Tabuchi}}, \ and\
  \bibinfo {author} {\bibfnamefont {Y.}~\bibnamefont {Nakamura}},\ }\href
  {\doibase 10.1103/PhysRevLett.119.180505} {\bibfield  {journal} {\bibinfo
  {journal} {Phys. Rev. Lett.}\ }\textbf {\bibinfo {volume} {119}},\ \bibinfo
  {pages} {180505} (\bibinfo {year} {2017})}\BibitemShut {NoStop}%
\bibitem [{\citenamefont {Chu}\ \emph {et~al.}(2018{\natexlab{b}})\citenamefont
  {Chu}, \citenamefont {Kharel}, \citenamefont {Yoon}, \citenamefont {Frunzio},
  \citenamefont {Rakich},\ and\ \citenamefont {Schoelkopf}}]{ChuNature2018}%
  \BibitemOpen
  \bibfield  {author} {\bibinfo {author} {\bibfnamefont {Y.}~\bibnamefont
  {Chu}}, \bibinfo {author} {\bibfnamefont {P.}~\bibnamefont {Kharel}},
  \bibinfo {author} {\bibfnamefont {T.}~\bibnamefont {Yoon}}, \bibinfo {author}
  {\bibfnamefont {L.}~\bibnamefont {Frunzio}}, \bibinfo {author} {\bibfnamefont
  {P.~T.}\ \bibnamefont {Rakich}}, \ and\ \bibinfo {author} {\bibfnamefont
  {R.~J.}\ \bibnamefont {Schoelkopf}},\ }\href@noop {} {\bibfield  {journal}
  {\bibinfo  {journal} {Nature}\ }\textbf {\bibinfo {volume} {563}},\ \bibinfo
  {pages} {666} (\bibinfo {year} {2018}{\natexlab{b}})}\BibitemShut {NoStop}%
\bibitem [{\citenamefont {Moores}\ \emph {et~al.}(2018)\citenamefont {Moores},
  \citenamefont {Sletten}, \citenamefont {Viennot},\ and\ \citenamefont
  {Lehnert}}]{MooresRPL2018}%
  \BibitemOpen
  \bibfield  {author} {\bibinfo {author} {\bibfnamefont {B.~A.}\ \bibnamefont
  {Moores}}, \bibinfo {author} {\bibfnamefont {L.~R.}\ \bibnamefont {Sletten}},
  \bibinfo {author} {\bibfnamefont {J.~J.}\ \bibnamefont {Viennot}}, \ and\
  \bibinfo {author} {\bibfnamefont {K.~W.}\ \bibnamefont {Lehnert}},\ }\href
  {\doibase 10.1103/PhysRevLett.120.227701} {\bibfield  {journal} {\bibinfo
  {journal} {Phys. Rev. Lett.}\ }\textbf {\bibinfo {volume} {120}},\ \bibinfo
  {pages} {227701} (\bibinfo {year} {2018})}\BibitemShut {NoStop}%
\bibitem [{\citenamefont {Bolgar}\ \emph {et~al.}(2018)\citenamefont {Bolgar},
  \citenamefont {Zotova}, \citenamefont {Kirichenko}, \citenamefont {Besedin},
  \citenamefont {Semenov}, \citenamefont {Shaikhaidarov},\ and\ \citenamefont
  {Astafiev}}]{BolgarPRL2018}%
  \BibitemOpen
  \bibfield  {author} {\bibinfo {author} {\bibfnamefont {A.~N.}\ \bibnamefont
  {Bolgar}}, \bibinfo {author} {\bibfnamefont {J.~I.}\ \bibnamefont {Zotova}},
  \bibinfo {author} {\bibfnamefont {D.~D.}\ \bibnamefont {Kirichenko}},
  \bibinfo {author} {\bibfnamefont {I.~S.}\ \bibnamefont {Besedin}}, \bibinfo
  {author} {\bibfnamefont {A.~V.}\ \bibnamefont {Semenov}}, \bibinfo {author}
  {\bibfnamefont {R.~S.}\ \bibnamefont {Shaikhaidarov}}, \ and\ \bibinfo
  {author} {\bibfnamefont {O.~V.}\ \bibnamefont {Astafiev}},\ }\href {\doibase
  10.1103/PhysRevLett.120.223603} {\bibfield  {journal} {\bibinfo  {journal}
  {Phys. Rev. Lett.}\ }\textbf {\bibinfo {volume} {120}},\ \bibinfo {pages}
  {223603} (\bibinfo {year} {2018})}\BibitemShut {NoStop}%
\bibitem [{\citenamefont {Bienfait}\ \emph {et~al.}(2020)\citenamefont
  {Bienfait}, \citenamefont {Zhong}, \citenamefont {Chang}, \citenamefont
  {Chou}, \citenamefont {Conner}, \citenamefont {Dumur}, \citenamefont
  {Grebel}, \citenamefont {Peairs}, \citenamefont {Povey}, \citenamefont
  {Satzinger},\ and\ \citenamefont {Cleland}}]{BienfaitPRX2020}%
  \BibitemOpen
  \bibfield  {author} {\bibinfo {author} {\bibfnamefont {A.}~\bibnamefont
  {Bienfait}}, \bibinfo {author} {\bibfnamefont {Y.~P.}\ \bibnamefont {Zhong}},
  \bibinfo {author} {\bibfnamefont {H.-S.}\ \bibnamefont {Chang}}, \bibinfo
  {author} {\bibfnamefont {M.-H.}\ \bibnamefont {Chou}}, \bibinfo {author}
  {\bibfnamefont {C.~R.}\ \bibnamefont {Conner}}, \bibinfo {author}
  {\bibfnamefont {E.}~\bibnamefont {Dumur}}, \bibinfo {author} {\bibfnamefont
  {J.}~\bibnamefont {Grebel}}, \bibinfo {author} {\bibfnamefont {G.~A.}\
  \bibnamefont {Peairs}}, \bibinfo {author} {\bibfnamefont {R.~G.}\
  \bibnamefont {Povey}}, \bibinfo {author} {\bibfnamefont {K.~J.}\ \bibnamefont
  {Satzinger}}, \ and\ \bibinfo {author} {\bibfnamefont {A.~N.}\ \bibnamefont
  {Cleland}},\ }\href {\doibase 10.1103/PhysRevX.10.021055} {\bibfield
  {journal} {\bibinfo  {journal} {Phys. Rev. X}\ }\textbf {\bibinfo {volume}
  {10}},\ \bibinfo {pages} {021055} (\bibinfo {year} {2020})}\BibitemShut
  {NoStop}%
\bibitem [{\citenamefont {McKenzie-Sell}\ \emph {et~al.}(2019)\citenamefont
  {McKenzie-Sell}, \citenamefont {Xie}, \citenamefont {Lee}, \citenamefont
  {Robinson}, \citenamefont {Ciccarelli},\ and\ \citenamefont
  {Haigh}}]{McKenziePRB2019}%
  \BibitemOpen
  \bibfield  {author} {\bibinfo {author} {\bibfnamefont {L.}~\bibnamefont
  {McKenzie-Sell}}, \bibinfo {author} {\bibfnamefont {J.}~\bibnamefont {Xie}},
  \bibinfo {author} {\bibfnamefont {C.-M.}\ \bibnamefont {Lee}}, \bibinfo
  {author} {\bibfnamefont {J.~W.~A.}\ \bibnamefont {Robinson}}, \bibinfo
  {author} {\bibfnamefont {C.}~\bibnamefont {Ciccarelli}}, \ and\ \bibinfo
  {author} {\bibfnamefont {J.~A.}\ \bibnamefont {Haigh}},\ }\href {\doibase
  10.1103/PhysRevB.99.140414} {\bibfield  {journal} {\bibinfo  {journal} {Phys.
  Rev. B}\ }\textbf {\bibinfo {volume} {99}},\ \bibinfo {pages} {140414}
  (\bibinfo {year} {2019})}\BibitemShut {NoStop}%
\bibitem [{\citenamefont {Schuetz}\ \emph {et~al.}(2015)\citenamefont
  {Schuetz}, \citenamefont {Kessler}, \citenamefont {Giedke}, \citenamefont
  {Vandersypen}, \citenamefont {Lukin},\ and\ \citenamefont
  {Cirac}}]{SchuetzPRX2015}%
  \BibitemOpen
  \bibfield  {author} {\bibinfo {author} {\bibfnamefont {M.~J.~A.}\
  \bibnamefont {Schuetz}}, \bibinfo {author} {\bibfnamefont {E.~M.}\
  \bibnamefont {Kessler}}, \bibinfo {author} {\bibfnamefont {G.}~\bibnamefont
  {Giedke}}, \bibinfo {author} {\bibfnamefont {L.~M.~K.}\ \bibnamefont
  {Vandersypen}}, \bibinfo {author} {\bibfnamefont {M.~D.}\ \bibnamefont
  {Lukin}}, \ and\ \bibinfo {author} {\bibfnamefont {J.~I.}\ \bibnamefont
  {Cirac}},\ }\href {\doibase 10.1103/PhysRevX.5.031031} {\bibfield  {journal}
  {\bibinfo  {journal} {Phys. Rev. X}\ }\textbf {\bibinfo {volume} {5}},\
  \bibinfo {pages} {031031} (\bibinfo {year} {2015})}\BibitemShut {NoStop}%
\bibitem [{\citenamefont {Vainsencher}\ \emph {et~al.}(2016)\citenamefont
  {Vainsencher}, \citenamefont {Satzinger}, \citenamefont {Peairs},\ and\
  \citenamefont {Cleland}}]{VainsencherAPL2016}%
  \BibitemOpen
  \bibfield  {author} {\bibinfo {author} {\bibfnamefont {A.}~\bibnamefont
  {Vainsencher}}, \bibinfo {author} {\bibfnamefont {K.~J.}\ \bibnamefont
  {Satzinger}}, \bibinfo {author} {\bibfnamefont {G.~A.}\ \bibnamefont
  {Peairs}}, \ and\ \bibinfo {author} {\bibfnamefont {A.~N.}\ \bibnamefont
  {Cleland}},\ }\href@noop {} {\bibfield  {journal} {\bibinfo  {journal} {Appl.
  Phys. Lett.}\ }\textbf {\bibinfo {volume} {109}},\ \bibinfo {pages} {033107}
  (\bibinfo {year} {2016})}\BibitemShut {NoStop}%
\bibitem [{\citenamefont {Guo}\ \emph {et~al.}(2017)\citenamefont {Guo},
  \citenamefont {Grimsmo}, \citenamefont {Kockum}, \citenamefont {Pletyukhov},\
  and\ \citenamefont {Johansson}}]{GuoPRA2017}%
  \BibitemOpen
  \bibfield  {author} {\bibinfo {author} {\bibfnamefont {L.}~\bibnamefont
  {Guo}}, \bibinfo {author} {\bibfnamefont {A.}~\bibnamefont {Grimsmo}},
  \bibinfo {author} {\bibfnamefont {A.~F.}\ \bibnamefont {Kockum}}, \bibinfo
  {author} {\bibfnamefont {M.}~\bibnamefont {Pletyukhov}}, \ and\ \bibinfo
  {author} {\bibfnamefont {G.}~\bibnamefont {Johansson}},\ }\href {\doibase
  10.1103/PhysRevA.95.053821} {\bibfield  {journal} {\bibinfo  {journal} {Phys.
  Rev. A}\ }\textbf {\bibinfo {volume} {95}},\ \bibinfo {pages} {053821}
  (\bibinfo {year} {2017})}\BibitemShut {NoStop}%
\bibitem [{\citenamefont {Osada}\ \emph {et~al.}(2016)\citenamefont {Osada},
  \citenamefont {Hisatomi}, \citenamefont {Noguchi}, \citenamefont {Tabuchi},
  \citenamefont {Yamazaki}, \citenamefont {Usami}, \citenamefont {Sadgrove},
  \citenamefont {Yalla}, \citenamefont {Nomura},\ and\ \citenamefont
  {Nakamura}}]{OsadaPRL2016}%
  \BibitemOpen
  \bibfield  {author} {\bibinfo {author} {\bibfnamefont {A.}~\bibnamefont
  {Osada}}, \bibinfo {author} {\bibfnamefont {R.}~\bibnamefont {Hisatomi}},
  \bibinfo {author} {\bibfnamefont {A.}~\bibnamefont {Noguchi}}, \bibinfo
  {author} {\bibfnamefont {Y.}~\bibnamefont {Tabuchi}}, \bibinfo {author}
  {\bibfnamefont {R.}~\bibnamefont {Yamazaki}}, \bibinfo {author}
  {\bibfnamefont {K.}~\bibnamefont {Usami}}, \bibinfo {author} {\bibfnamefont
  {M.}~\bibnamefont {Sadgrove}}, \bibinfo {author} {\bibfnamefont
  {R.}~\bibnamefont {Yalla}}, \bibinfo {author} {\bibfnamefont
  {M.}~\bibnamefont {Nomura}}, \ and\ \bibinfo {author} {\bibfnamefont
  {Y.}~\bibnamefont {Nakamura}},\ }\href {\doibase
  10.1103/PhysRevLett.116.223601} {\bibfield  {journal} {\bibinfo  {journal}
  {Phys. Rev. Lett.}\ }\textbf {\bibinfo {volume} {116}},\ \bibinfo {pages}
  {223601} (\bibinfo {year} {2016})}\BibitemShut {NoStop}%
\bibitem [{\citenamefont {Zhang}\ \emph
  {et~al.}(2016{\natexlab{b}})\citenamefont {Zhang}, \citenamefont {Zhu},
  \citenamefont {Zou},\ and\ \citenamefont {Tang}}]{ZhangPRL2016}%
  \BibitemOpen
  \bibfield  {author} {\bibinfo {author} {\bibfnamefont {X.}~\bibnamefont
  {Zhang}}, \bibinfo {author} {\bibfnamefont {N.}~\bibnamefont {Zhu}}, \bibinfo
  {author} {\bibfnamefont {C.-L.}\ \bibnamefont {Zou}}, \ and\ \bibinfo
  {author} {\bibfnamefont {H.~X.}\ \bibnamefont {Tang}},\ }\href {\doibase
  10.1103/PhysRevLett.117.123605} {\bibfield  {journal} {\bibinfo  {journal}
  {Phys. Rev. Lett.}\ }\textbf {\bibinfo {volume} {117}},\ \bibinfo {pages}
  {123605} (\bibinfo {year} {2016}{\natexlab{b}})}\BibitemShut {NoStop}%
\bibitem [{\citenamefont {Haigh}\ \emph {et~al.}(2016)\citenamefont {Haigh},
  \citenamefont {Nunnenkamp}, \citenamefont {Ramsay},\ and\ \citenamefont
  {Ferguson}}]{HaighPRL2016}%
  \BibitemOpen
  \bibfield  {author} {\bibinfo {author} {\bibfnamefont {J.~A.}\ \bibnamefont
  {Haigh}}, \bibinfo {author} {\bibfnamefont {A.}~\bibnamefont {Nunnenkamp}},
  \bibinfo {author} {\bibfnamefont {A.~J.}\ \bibnamefont {Ramsay}}, \ and\
  \bibinfo {author} {\bibfnamefont {A.~J.}\ \bibnamefont {Ferguson}},\ }\href
  {\doibase 10.1103/PhysRevLett.117.133602} {\bibfield  {journal} {\bibinfo
  {journal} {Phys. Rev. Lett.}\ }\textbf {\bibinfo {volume} {117}},\ \bibinfo
  {pages} {133602} (\bibinfo {year} {2016})}\BibitemShut {NoStop}%
\bibitem [{\citenamefont {Serga}\ \emph {et~al.}(2010)\citenamefont {Serga},
  \citenamefont {Chumak},\ and\ \citenamefont {Hillebrands}}]{SergaJPD2010}%
  \BibitemOpen
  \bibfield  {author} {\bibinfo {author} {\bibfnamefont {A.~A.}\ \bibnamefont
  {Serga}}, \bibinfo {author} {\bibfnamefont {A.~V.}\ \bibnamefont {Chumak}}, \
  and\ \bibinfo {author} {\bibfnamefont {B.}~\bibnamefont {Hillebrands}},\
  }\href@noop {} {\bibfield  {journal} {\bibinfo  {journal} {J. Phys. D: Appl.
  Phys.}\ }\textbf {\bibinfo {volume} {43}},\ \bibinfo {pages} {264002}
  (\bibinfo {year} {2010})}\BibitemShut {NoStop}%
\end{thebibliography}

%

\end{document}